\documentclass[preprint,notoc]{JHEP3}

\usepackage{epsf,psfrag,graphicx,amssymb,amsmath}

\def\be{\begin{equation}}
\def\ee{\end{equation}}
\def\bea{\begin{eqnarray}}
\def\eea{\end{eqnarray}}
\def\Eqn#1{Equation~(\ref{#1})}

\def\eqn#1{eq.~(\ref{#1})}
\def\eqns#1#2{eqs.~(\ref{#1}) and~(\ref{#2})}
\def\fig#1{figure~{\ref{#1}}}
\def\Fig#1{Figure~{\ref{#1}}}

\def\sect#1{section~{\ref{#1}}}

\def\pol{\varepsilon}

\def\qb{{\bar{q}}}
\def\Qb{{\bar{Q}}}

\def\cg{c_\Gamma}
\def\e{\epsilon}
\def\eps{\epsilon}

\def\oneloop{{\rm 1\! -\! loop}}

\def\Ls{\mathrm{Ls}}
\def\Li{\mathrm{Li}}
\def\Ll{\mathrm{L}}

\def\treenum{{(0)}}
\def\oneloopnum{{(1)}}
\def\lloopnum{{(l)}}

\def\Ord{{\cal O}}

\def\lr{\leftrightarrow}

\def\alphas{\alpha_s}

\def\Psl{\not{\hbox{\kern-2.3pt $P$}}}
\def\psl{\not{\hbox{\kern-2.3pt $p$}}}
\def\Ksl{\not{\hbox{\kern-2.3pt $K$}}}
\def\ksl{\not{\hbox{\kern-2.3pt $k$}}}
\def\esl{\not{\hbox{\kern-2.3pt $\pol$}}}

\def\tr{\mathop{\rm tr}\nolimits}

\def\pol{\varepsilon}

\def\spa#1.#2{\left\langle#1\,#2\right\rangle}
\def\spb#1.#2{\left[#1\,#2\right]}
\def\lor#1.#2{\left(#1\,#2\right)}
\def\sand#1.#2.#3{%
\left\langle\smash{#1}{\vphantom1}^{-}\right|{#2}%
\left|\smash{#3}{\vphantom1}^{-}\right\rangle}
\def\sandp#1.#2.#3{%
\left\langle\smash{#1}{\vphantom1}^{-}\right|{#2}%
\left|\smash{#3}{\vphantom1}^{+}\right\rangle}
\def\sandpp#1.#2.#3{%
\left\langle\smash{#1}{\vphantom1}^{+}\right|{#2}%
\left|\smash{#3}{\vphantom1}^{+}\right\rangle}
\def\sandpm#1.#2.#3{%
\left\langle\smash{#1}{\vphantom1}^{+}\right|{#2}%
\left|\smash{#3}{\vphantom1}^{-}\right\rangle}
\def\sandmp#1.#2.#3{%
\left\langle\smash{#1}{\vphantom1}^{-}\right|{#2}%
\left|\smash{#3}{\vphantom1}^{+}\right\rangle}
\def\spab#1.#2.#3{\langle#1|#2|#3]}
\def\spba#1.#2.#3{[#1|#2|#3\rangle}
\newbox\charbox
\newbox\slabox
\def\s#1{{      
        \setbox\charbox=\hbox{$#1$}
        \setbox\slabox=\hbox{$/$}
        \dimen\charbox=\ht\slabox
        \advance\dimen\charbox by -\dp\slabox
        \advance\dimen\charbox by -\ht\charbox
        \advance\dimen\charbox by \dp\charbox
        \divide\dimen\charbox by 2
        \raise-\dimen\charbox\hbox to \wd\charbox{\hss/\hss}
        \llap{$#1$}
}}
\def\ksl{\s{k}}

\def\fl{{\rm f}}
\def\nn{\nonumber}
\def\hf{{\textstyle{1\over2}}}
\def\Res{\mathop{\rm Res}}
\def\Fact{{\cal F}}
\def\Ph{{{\hat P}_{r\ldots s}}}

\newcommand{\BlackHat}{{\sc BlackHat}}
\newcommand{\Maple}{{\sc Maple}}

\preprint{
  SLAC--PUB--13652\\
  May, 2009}

\title{Analytic one-loop amplitudes for a Higgs boson plus four partons
 \thanks{Research supported by the US Department of Energy under contract
  DE-AC02-76SF00515.}}

\author{Lance J. Dixon
    and Yorgos Sofianatos\\
    SLAC National Accelerator Laboratory, Stanford University,
    Stanford, CA 94309, USA\\
        E-mail: \email{lance@slac.stanford.edu, yorgos@slac.stanford.edu}}

\abstract{We compute the one-loop QCD amplitudes for the processes
$H\qb q\Qb Q$ and $H\qb qgg$, the latter restricted
to the case of opposite-helicity gluons.
Analytic expressions are presented for the color- and helicity-decomposed
amplitudes. The coupling of the Higgs boson to gluons is treated by
an effective interaction in the limit of large top quark mass.
The Higgs field is split into a complex field $\phi$ and its complex
conjugate $\phi^\dagger$.  The split is useful because amplitudes
involving $\phi$ have different analytic structure from those involving
$\phi^\dagger$. We compute the cut-containing pieces of the amplitudes using
generalized unitarity.  The remaining rational parts are obtained by
on-shell recursion.  Our results for $H\qb q\Qb Q$ agree with previous
semi-numerical computations.  We also show how to convert existing
semi-numerical results for the production of a scalar Higgs boson
into analogous results for a pseudoscalar Higgs boson.}


\dedicated{\centerline{Submitted to JHEP}}

\begin{document}


\section{Introduction}
\label{IntroSection}

The Large Hadron Collider (LHC) at CERN is now beginning
operation, and will be the major source of data from the energy
frontier for many years to come.  The main goal of the LHC
experimental program is to discover physics beyond the Standard
Model (SM), as well as the mechanism of electroweak symmetry
breaking. In the SM, this breaking is due to a single scalar
field, the Higgs field
\cite{Higgs1966ev,Englert1964et,Guralnik1964eu}. Similar fields
exist in most extensions of the SM, such as the Minimal
Supersymmetric Standard Model (MSSM).

The discovery and the measurement of the Higgs sector will be a
central piece of the experimental effort at the LHC. The
production of the Higgs boson is dominated by gluon fusion through
a top quark loop, for the whole relevant Higgs mass
range~\cite{Georgi1977gs}. The next-to-leading-order (NLO) corrections
to the gluon-fusion production cross section are very large, of
the order of 100\%~\cite{Djouadi1991tka,Dawson1990zj,%
Graudenz1992pv,Spira1995rr}. The second most important
contribution to the Higgs production cross section comes from the
vector boson fusion (VBF) process, which proceeds at tree-level
and receives much smaller QCD
corrections~\cite{Figy2003nv,Figy2004pt,Berger2004pca}.

Both of these channels participate in the phenomenologically
interesting signal of $pp\rightarrow H+2$~jets. The
jets coming from the two processes have different angular
distributions:  well-separated and forward jets in the VBF case,
in contrast to less separated jets 
and further central jet activity in the gluon fusion case. 
Thus, the cross-contamination can be reduced by
imposing appropriate experimental cuts. A good theoretical
understanding is also necessary to reduce the uncertainties coming
from the backgrounds and interference effects, and to allow us to
perform precision studies of the Higgs sector.

For the gluon-fusion contribution, in order to make the
computation more tractable, it is possible to make the
approximation of a large top quark mass
$m_t$~\cite{Wilczek1977zn,Shifman1978zn,Djouadi1991tka,Dawson1990zj}.
This approximation replaces the full one-loop coupling of the
Higgs boson to gluons via a top quark loop, by an effective local
operator $C(m_t) \, H\, G_{\mu\nu}\,G^{\mu\nu}$; thus it reduces
the problem by one loop order.  Formally, this approximation
requires the Higgs mass to obey $m_H \ll 2m_t \approx 350$~GeV.
However, for inclusive Higgs production,
the approximation works very well up to quite large Higgs masses,
$m_H \approx 2m_t$ or more,
if $C(m_t)$ is taken to have the exact $m_t$ dependence from one
loop~\cite{KLS98}. Here we are interested primarily in processes
where the Higgs boson is relatively light, but because of the
extra jet activity the partonic center-of-mass energy and
final-state invariant masses may be large.  It has been shown that
the large-$m_t$ approximation is still valid for such
configurations, as long as the jet transverse energies are smaller
than $m_t$~\cite{Hautmann2002tu,DelDuca2003ba}.

Several groups have computed various relevant quantities for the
gluon-fusion contribution to $pp\rightarrow H+2$~jets,
at LO, in the large $m_t$
limit~\cite{Dawson1991au,Kauffman1996ix,Kauffman1998yg} and with
the exact $m_t$ dependence~\cite{DDKOSZ}; and at NLO accuracy in
the large $m_t$ limit~\cite{Ellis2005qe,Campbell2006xx}. The real
corrections to this process, involving tree amplitudes for a Higgs
plus five partons, were studied in~\cite{DelDuca2004wt}. The
interference between gluon fusion and VBF production has been
computed, and found to be very
small~\cite{Andersen2007mp,Bredenstein2008tm}. Our results for the
$H\qb q \Qb Q$ amplitudes can also be used to calculate the
one-loop interference between the color-singlet pieces of these
two processes. We outline this calculation in
section~\ref{ResultsSection}.

New methods have been developed for computing one-loop amplitudes
for multi-leg processes.  Some are based on Feynman
diagrams~\cite{Anastasiou2004vj,Anastasiou2005cb,Anastasiou2007qb,%
Ellis2005zh,Denner2005nn, Ossola2006us,Ossola2007ax,Ossola2008xq,%
Mastrolia2008jb,Draggiotis2009yb,vanHameren2009dr} while others
exploit generalized
unitarity~\cite{Bern2007dw,Anastasiou2006jv,Anastasiou2006gt,%
Ellis2007br,Giele2008ve, Giele2008bc,Ellis2008ir} and recursion
relations~\cite{Berger2008sj}. The methods based on Feynman
diagrams have been employed to compute several quantities
involving the Higgs boson to
next-to-leading-order~\cite{Bredenstein2006ha,
Ciccolini2007jr,Ciccolini2007ec}.

In the large $m_t$ approximation, the NLO corrections to the
production of a Higgs boson plus various numbers of jets at a
hadron collider require one-loop amplitudes in QCD, with one
insertion of the effective operator $H\, G_{\mu\nu}\,G^{\mu\nu}$.
We will refer to these amplitudes as one-loop Higgs plus
multi-parton amplitudes. Because these virtual amplitudes contain
infrared divergences, they are invariably presented using
dimensional regularization (we take $D=4-2\e$), as a Laurent
expansion in $\e$ through the finite $\Ord(\e^0)$ terms.  The
complete set of such amplitudes for three external partons ($ggg$
or $q\bar{q}g$) was provided in ref.~\cite{Schmidt1997wr}. Results
for various numbers of legs and helicity configurations have
appeared more
recently~\cite{BadgerGlover,Berger2006sh,Badger2007si,Glover2008ffa}.
In particular, the amplitudes with four gluons, all of the same
helicity~\cite{BadgerGlover}, and those with two negative and two
positive helicities~\cite{Badger2007si,Glover2008ffa} have now
been computed analytically, using techniques very similar to what
we will employ here.

The full analytic results for the one-loop Higgs plus four-parton
amplitudes, for a complete set of parton helicities, have not yet
appeared.  However, the analytic expressions for the color- and
helicity-summed NLO interference of one-loop and tree amplitudes
have been presented for the four-quark case~\cite{Ellis2005qe},
along with numerical results for the two-quark-two-gluon and
four-gluon cases. These results have been incorporated into the
NLO gluon fusion contributions to $pp\rightarrow H+2$~jets
mentioned earlier~\cite{Campbell2006xx}. In this
paper we present analytic results at the amplitude level for the
four-quark case, $H\bar{q}q\bar{Q}Q$. We also give the
two-quark-two-gluon amplitudes, with the restriction that the two
gluons must have opposite helicity, $H\bar{q}qg^\pm g^\mp$.

In our method, the Higgs field $H$ is rewritten as the sum of a
complex field $\phi$, and its complex conjugate $\phi^\dagger$. This
has the advantage that the analytic structure of the two components
is much simpler than in the total amplitude. Furthermore, we only
need to compute the $\phi$-amplitudes because parity relates them to
the $\phi^\dagger$ ones. Our technique for calculating these
amplitudes consists of a unitarity-recursive bootstrap approach: by
performing appropriate unitarity cuts we obtain all cut-containing
terms of the amplitudes (logarithmic, polylogarithmic functions, and
associated terms), while the rational terms are computed using
on-shell recursion relations. In this process we only use on-shell
lower-point amplitudes as input in our calculation. This greatly
simplifies our task and allows us to obtain compact analytic answers
efficiently.

In slightly more detail, we employ quadruple
cuts~\cite{Britto2004nc} to determine the coefficients of scalar
box integrals in the amplitudes. The only scalar triangle
integrals appearing in the amplitudes described here have one or
two external massive legs, not three; the coefficients of such
integrals are fixed easily using the amplitudes' known infrared
poles.  The coefficients of bubble integrals are computed using
the method of spinor integration via residue
extraction~\cite{Britto2005ha,Britto2006sj}. As just mentioned,
the rational terms are computed using on-shell recursion
relations.  Certain spurious poles arising in this technique are
dealt with using an approach which is a {\it hybrid} between the
cut-completion method~\cite{Bern2005cq,Berger2006ci,Berger2006vq}
and the method of evaluating the spurious pole residue of the cut
part~\cite{Bern2008ef} which is implemented numerically in
\BlackHat~\cite{Berger2008sj}.

Our analytic expressions can easily be incorporated into one of
the NLO computer programs for computing cross sections. They
provide a fast evaluation of the amplitudes and are more stable
compared to (semi)numerical approaches.  In the computation of
Higgs amplitudes with yet one more external parton (five in all),
they can serve to check limiting cases, when two partons become
collinear, or one gluon becomes soft. In a numerical on-shell
recursive approach such as \BlackHat~\cite{Berger2008sj}, they
could provide a more important role: the four-parton amplitudes
could be used as a fast analytic input for some of the terms in
the on-shell recursion relations for the five-parton amplitudes.

By performing the appropriate sum over colors and helicities of
the interference between tree and one-loop amplitudes, we are able
to confirm (numerically) the expression for the virtual part of
the NLO cross section for $H\bar{q}q\bar{Q}Q$ presented in
ref.~\cite{Ellis2005qe}.  At the same time, because our results
are color decomposed, we can project them into a color-singlet
channel with respect to the $\bar{q}q$ (and $\bar{Q}Q$) quantum
numbers.  The color-singlet channel can interfere with the
electroweak VBF process~\cite{Andersen2006ag}.  We have verified
numerically that this interference, which is one of the two
virtual contributions to the full interference, agrees with
one~\cite{Andersen2007mp} of the two recent
computations~\cite{Andersen2007mp,Bredenstein2008tm} of this
(quite small) effect~\cite{ASPrivate}.

The color-singlet parts of the one-loop amplitudes may be of use
in determining how frequently the gluon-fusion process produces
events with large ``rapidity gaps'' that would survive typical
central jet vetoes proposed to select the VBF process.  Resummed
estimates of the efficacy of such vetoes have been performed
recently in ref.~\cite{Forshaw2009fz} for example; however, there
may also be important contributions at fixed order in $\alpha_s$.

The paper is organized as follows:  In \sect{NotationSection} we
introduce some basic notions that simplify the task of computing
the $H\qb q\Qb Q$ and $H\qb qgg$ amplitudes --- the
$\phi$-$\phi^\dagger$ and color decomposition of Higgs amplitudes,
and the spinor helicity formalism.  In \sect{BootstrapSection} we
outline the basics of the unitarity-recursive method at the core
of the calculation.  We also apply the technique to specific
examples.  In \sect{ResultsSection} we present the full analytic
answers, numerical results, and applications of our expressions.
We describe checks that we have used to verify their correctness.
In \sect{ConclusionSection} we present our conclusions and mention
possible future directions.


\section{Notation}
\label{NotationSection}

In this section we introduce the basic notation used in the rest
of this paper, as well as some prerequisite notions. In
particular, we decompose the Higgs amplitudes into $\phi$ and
$\phi^\dagger$ components with simpler analytic properties,
describe the color decomposition of scattering amplitudes in terms
of partial and primitive amplitudes, and recall the spinor-helicity formalism.


\subsection{The $\phi$-$\phi^\dagger$ decomposition of Higgs amplitudes}
\label{PhiDecompSubsection}

In the amplitudes we study in this paper, all external quarks are
taken to be massless, and the Higgs couples to them through
gluons. The coupling of the SM Higgs boson to gluons is dominated
by an intermediate top quark loop, because the top is much heavier
than the other quarks. In the limit of very large top mass, $m_t
\rightarrow \infty$, the top quark can be integrated out, giving
rise to the following effective interaction
\cite{Wilczek1977zn,Shifman1978zn},
\be \mathcal{L}_H^{\mathrm{int}} = \frac{C}{2} H\,\tr
G_{\mu\nu}\,G^{\mu\nu}\, , \label{Lint} \ee
with the coefficient $C$ given by $C = \alpha_s/(6\pi v) =
g^2/(24\pi^2 v)$, to leading order in $\alpha_s$. Here $v$ is the
vacuum expectation value of the Higgs field, $v = 246$ GeV. (The
value of $C$ is known to ${\cal
O}(\alpha_s^4)$~\cite{Chetyrkin1997un}.) Our convention for the
normalization of generators is $\tr T^a T^b = \delta^{ab}$, and
$G_{\mu\nu} = \sum_a T^a G^a_{\mu\nu}$, so that $\tr
G_{\mu\nu}\,G^{\mu\nu} = G^a_{\mu\nu}\,G^{a\,\mu\nu}$.

Tree-level amplitudes (not counting the top quark loop) for a
Higgs boson plus multiple partons were first computed analytically
for up to four partons in
refs.~\cite{Dawson1991au,Kauffman1996ix,Kauffman1998yg}, and up to
five partons in ref.~\cite{DFM}.  The structure of these
amplitudes could be simplified~\cite{Dixon2004za} by splitting the
effective interaction Lagrangian into two parts, a holomorphic
(self-dual) and an anti-holomorphic (anti-self-dual) part. In
fact, certain ``maximally helicity violating (MHV) rules'' that
had been developed for QCD tree amplitudes~\cite{Cachazo2004kj}
could be applied straightforwardly to Higgs amplitudes after
making this split~\cite{Dixon2004za,Badger2004ty}. Specifically,
we consider the Higgs boson to be the real part of a complex field
$\phi$, with
\be
\phi = \frac{1}{2} (H + iA)\, . \label{phi_definition}
\ee
The interaction Lagrangian can then be rewritten as
\bea
\mathcal{L}_{H,A}^{\mathrm{int}} &=& \frac{C}{2}\Big[ H\tr
G_{\mu\nu}\,G^{\mu\nu} + iA\tr G_{\mu\nu}\, {}^*G^{\mu\nu} \Big]\\
&=& C\Big[ \phi\tr
G_{{\scriptscriptstyle{SD}}\,\mu\nu}\,G_{\scriptscriptstyle{SD}}^{\mu\nu}
+ \phi^\dagger\tr
G_{{\scriptscriptstyle{ASD}}\,\mu\nu}\,G_{\scriptscriptstyle{ASD}}^{\mu\nu}
\Big]\, ,
\eea
where the gluon field strength has been divided into a self-dual
(SD) and an anti-self-dual (ASD) component, given by
\be G_{\scriptscriptstyle{SD}}^{\mu\nu} = \frac{1}{2}
(G^{\mu\nu}+{}^*G^{\mu\nu})\, , \quad
G_{\scriptscriptstyle{ASD}}^{\mu\nu} = \frac{1}{2}
(G^{\mu\nu}-{}^*G^{\mu\nu})\, , \quad {}^*G^{\mu\nu} \equiv
\frac{i}{2} \e^{\mu\nu\rho\sigma} G_{\rho\sigma}\, .
\ee

From \eqn{phi_definition} and its conjugate we can reconstruct the
scalar $H$ and pseudoscalar $A$ fields according to
\be
H = \phi + \phi^\dagger\, , 
\qquad 
A = \frac{1}{i} ( \phi - \phi^\dagger )\, .
\label{HA_reconstruct}
\ee
It follows from \eqn{HA_reconstruct} that the scattering amplitude
for a scalar Higgs boson plus any number of partons can be obtained,
at any loop order $l$, by the sum of the amplitudes with $\phi$ and
$\phi^\dagger$,
\be 
\mathcal{A}_n^\lloopnum(H,\ldots) =
\mathcal{A}_n^\lloopnum(\phi,\ldots) +
\mathcal{A}_n^\lloopnum(\phi^\dagger,\ldots)\,,
\label{Hreconstruct}
\ee
with
``$\ldots$'' indicating any arbitrary configuration of partons.

Similarly, the amplitudes for a pseudoscalar $A$ plus partons are
given by
\be
\mathcal{A}_n^\lloopnum(A,\ldots) = \frac{1}{i} \left[
\mathcal{A}_n^\lloopnum(\phi,\ldots) -
\mathcal{A}_n^\lloopnum(\phi^\dagger,\ldots) \right] \,,
\label{Areconstruct}
\ee
recognizing that the constant $C$ is different for the $A$
case~\cite{Berger2006sh}.
The relation between the $\phi$ and $\phi^\dagger$ amplitudes is
through parity, or complex conjugation of spinors,
\be
\mathcal{A}_n^\lloopnum(\phi^\dagger,1^{h_1},2^{h_2},\ldots,n^{h_n})
= (-1)^{n_{\qb q}} \Big[
\mathcal{A}_n^\lloopnum(\phi,1^{-h_1},2^{-h_2},\ldots,n^{-h_n})
\Big] \bigg|_{\spa{i}.{j} \leftrightarrow \spb{j}.{i} } \, ,
\label{phiphidaggerparity}
\ee
where the spinor products $\spa{i}.{j}$ and $\spb{j}.{i}$ are
defined in \eqns{defspa}{defspb}, and $n_{\qb q}$ denotes the number
of external antiquark-quark pairs~\cite{DFM}.
In other words, to generate an amplitude with $\phi^\dagger$ from an
amplitude with $\phi$, one reverses the helicities of all quarks and
gluons, and replaces all spinors $\spa{i}.{j}$ with $\spb{j}.{i}$.
Thus, it is possible to reconstruct the $H$ and $A$ helicity
amplitudes by computing only their $\phi$-components, using parity
to get the $\phi^\dagger$-components, and then assembling the two
ingredients together.

It is important to note that in the separation of any
Higgs amplitude into a $\phi$ and a $\phi^\dagger$ amplitude, all
color and kinematic information ({\it e.g.}, the momentum of the scalar
particle) remains the same in the original and the component
amplitudes. What separates is the self-duality properties of the
amplitudes, and consequently their analytic structure, resulting in
a simplification of the calculation.


\subsection{Color decomposition and color sums}
\label{ColorDecompSubsection}

An important tool for QCD calculations is the color decomposition
of amplitudes~\cite{TreeColor,Mangano1990by,Dixon1996wi}. It
allows us to write down a color-ordered expression for any
amplitude, which is a sum of products of color structures and
uncolored functions of the kinematic variables, called partial
amplitudes. The color and kinematic information is neatly
separated in this way, and one has to compute only the partial
amplitudes, which have simpler analytic properties.  The partial
amplitudes can be expressed in terms of yet simpler building
blocks, called primitive amplitudes.   Primitive amplitudes are
color ordered; that is, they receive contributions only from
planar one-loop Feynman diagrams with a fixed cyclic ordering of
the external legs. Furthermore, a given primitive amplitude either
contains a closed fermion loop (f), or it does not.  If it does not,
then the primitive amplitude is further characterized by how the
external fermion lines are routed as they enter the loop: whether
they turn left ($L$) or right ($R$) in the case of one fermion
line; and according to leading-color (lc) and subleading-color
(slc) designations in the case of two separate fermion lines.
Computing the primitive amplitudes will be the focus of this
paper.  First, we describe the decomposition of the $\phi\qb q\Qb
Q$ and $\phi\qb qgg$ amplitudes in terms of partial, and then
primitive, amplitudes.

\subsubsection{$\phi\qb q\Qb Q$}

Because the Higgs boson, and $\phi$ and $\phi^\dagger$ fields, are
color-neutral, they play no role in the color structure of the
amplitude.  The color decomposition of the $H\qb q\Qb Q$ or
$\phi\qb q\Qb Q$ amplitude is identical to the decomposition of
the four-quark $\qb q\Qb Q$ amplitude \cite{Bern1996ka}. There are
two independent color structures, corresponding to the two
independent ways we can contract the color indices of the quarks.
At tree level, the coefficients of the two color structures are
simply related,
\be
{\cal A}_4^\treenum(\phi,1_\qb,2_q,3_\Qb,4_Q)
 = Cg^2 \, A_4^\treenum(\phi,1_\qb,2_q,3_\Qb,4_Q) \biggl[
\delta_{i_4}^{\,\,\,\bar{\imath}_1} \delta_{i_2}^{\,\,\,\bar{\imath}_3}
- {1\over N_c}
\delta_{i_2}^{\,\,\,\bar{\imath}_1}\delta_{i_4}^{\,\,\,\bar{\imath}_3}
\biggr] \,,
\label{qqQQtreedecomp}
\ee
where the tree helicity amplitudes $A_4^\treenum(\phi,1_\qb,2_q,3_\Qb,4_Q)$
are given in \eqn{qqQQtrees}, and $N_c$ denotes the number of colors,
$N_c=3$ in QCD.

At one loop, the color decomposition is
\bea
{\cal A}_4^\oneloopnum(\phi,1_\qb,2_q,3_\Qb,4_Q) &=& Cg^4 \, \cg \biggl[
N_c \, \delta_{i_4}^{\,\,\,\bar{\imath}_1} \delta_{i_2}^{\,\,\,\bar{\imath}_3}
A_{4;1}(\phi,1_\qb,2_q,3_\Qb,4_Q)
\nonumber\\ && \hskip1.3cm \null
+ \delta_{i_2}^{\,\,\,\bar{\imath}_1} \delta_{i_4}^{\,\,\,\bar{\imath}_3}
A_{4;2}(\phi,1_\qb,2_q,3_\Qb,4_Q)\biggr] \,,
\label{qqQQloopdecomp}
\eea
where we have also extracted a factor from the loop integrals,
\be
\cg \equiv  {1\over(4\pi)^{2-\eps}}
   {\Gamma(1+\eps)\Gamma^2(1-\eps)\over\Gamma(1-2\eps)} \,.
\label{cgdefn}
\ee
%


\FIGURE[t]{
\resizebox{0.65\textwidth}{!}{\includegraphics{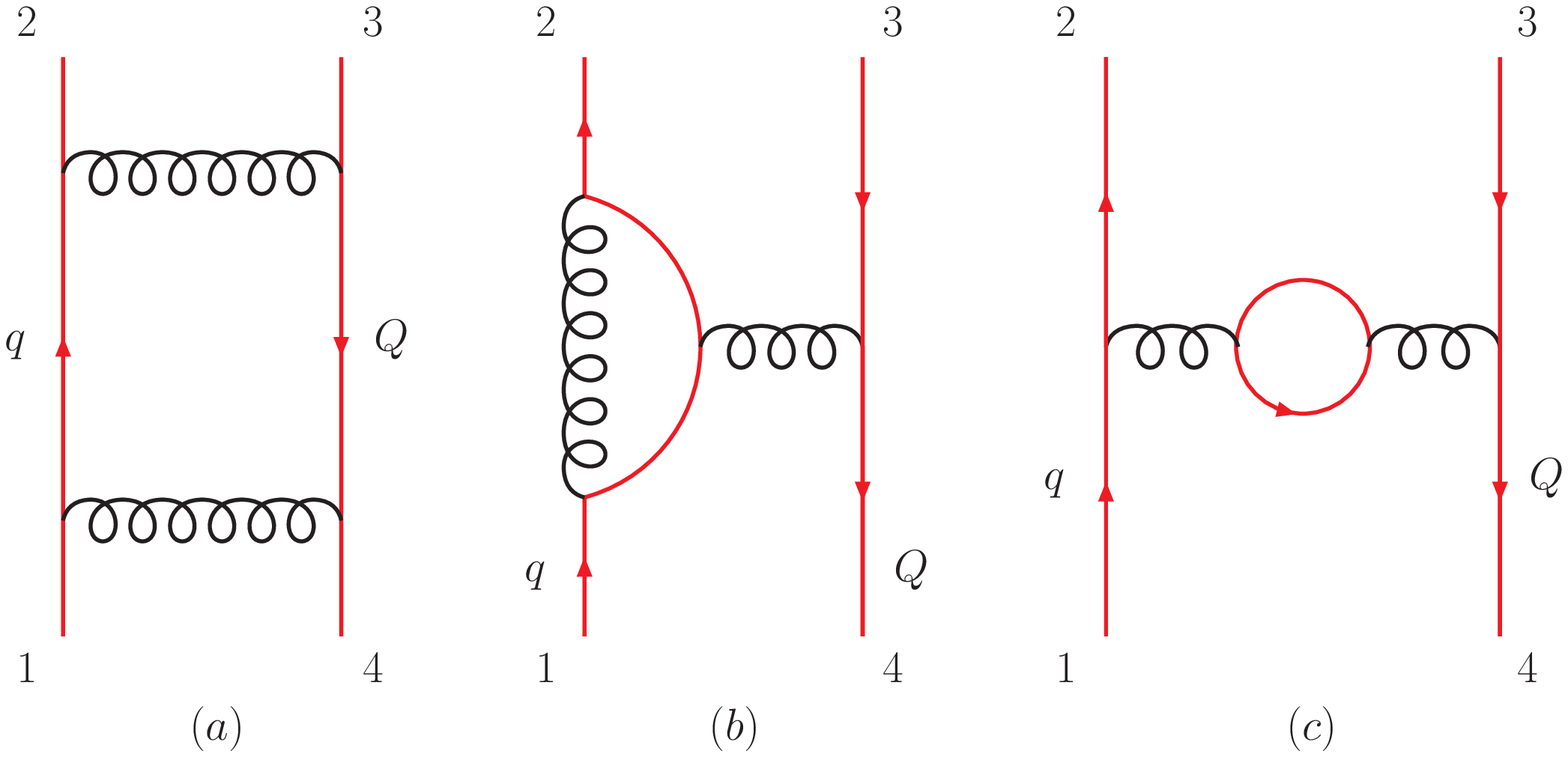}}
\caption{ Sample diagrams corresponding to the (a) leading color or ``lc'',
(b) subleading-color or ``slc'' and (c) fermion loop or ``$\fl$''
primitive amplitudes.
The Higgs field can attach to any gluon line in these diagrams.}
\label{primitive_amplitudes} }

Formulas~(\ref{qqQQtreedecomp}) and (\ref{qqQQloopdecomp}) apply
to the case of different quark
flavors, $q\neq Q$.  The amplitude for identical flavors, $q=Q$,
is found from the unequal-flavor formula by subtracting the
same formula with the labels for $q$ and $Q$ exchanged,
\be
{\cal A}_4^{(l)}(\phi,1_\qb,2_q,3_\qb,4_q)
= {\cal A}_4^{(l)}(\phi,1_\qb,2_q,3_\Qb,4_Q)
 - {\cal A}_4^{(l)}(\phi,1_\qb,4_q,3_\Qb,2_Q).
\label{identicalquarks}
\ee
Note that the helicities of $q$ and $Q$ must be the same in order to get
a nonvanishing exchange term.
Of course if there are two identical quarks in the final state, there
is also an identical-particle factor of $\hf$ in the phase-space
measure.

The partial amplitudes $A_{4;1}$ and $A_{4;2}$ can in turn be
expressed in terms of primitive amplitudes, using the results of
refs.~\cite{Bern1996ka} for the analogous amplitudes,
$e^+e^-\rightarrow \qb q \Qb Q$. Because the $e^+e^-$ pair and the
scalar $\phi$ are both colorless, the color structure is identical
to our case.  The results could be given in a helicity-independent
form; however, we list the two different helicity cases
separately, so that we can use relations among the primitive
amplitudes in order to minimize the number that appear:
\bea
A_{4;1}(\phi,1_\qb^-,2_q^+,3_\Qb^+,4_Q^-) &=&
A_4^{\mathrm{lc}}(\phi,1_\qb^-,2_q^+,3_\Qb^+,4_Q^-)
\label{mppm1}
\\ && \hskip0.0cm \null
- \frac{2}{N_c^2}
\left[A_4^{\mathrm{lc}}(\phi,1_\qb^-,2_q^+,3_\Qb^+,4_Q^-) +
A_4^{\mathrm{lc}}(\phi,1_\qb^-,2_q^+,4_\Qb^-,3_Q^+)\right]
\nonumber\\ && \hskip0.0cm \null
- \frac{1}{N_c^2} A_4^{\mathrm{slc}}(\phi,1_\qb^-,2_q^+,3_\Qb^+,4_Q^-)
+ \frac{n_f}{N_c}
A_4^{\fl}(\phi,1_\qb^-,2_q^+,3_\Qb^+,4_Q^-)\, ,
\nonumber
\eea
\bea
A_{4;1}(\phi,1_\qb^-,2_q^+,3_\Qb^-,4_Q^+) &=&
A_4^{\mathrm{lc}}(\phi,1_\qb^-,2_q^+,3_\Qb^-,4_Q^+)
\label{mpmp1}
\\ && \hskip0.0cm \null
- \frac{2}{N_c^2}\left[
A_4^{\mathrm{lc}}(\phi,1_\qb^-,2_q^+,3_\Qb^-,4_Q^+) +
A_4^{\mathrm{lc}}(\phi,1_\qb^-,2_q^+,4_\Qb^+,3_Q^-)\right]
\nonumber\\ && \hskip0.0cm \null
+ \frac{1}{N_c^2} A_4^{\mathrm{slc}}(\phi,1_\qb^-,2_q^+,4_\Qb^+,3_Q^-)
+ \frac{n_f}{N_c}
A_4^{\fl}(\phi,1_\qb^-,2_q^+,3_\Qb^-,4_Q^+)\, ,
\nonumber
\eea
\bea
A_{4;2}(\phi,1_\qb^-,2_q^+,3_\Qb^+,4_Q^-) &=&
A_4^{\mathrm{lc}}(\phi,1_\qb^-,2_q^+,4_\Qb^-,3_Q^+)
\label{mppm2}
\\ && \hskip0.0cm \null
+ \frac{1}{N_c^2}
\left[A_4^{\mathrm{lc}}(\phi,1_\qb^-,2_q^+,4_\Qb^-,3_Q^+) +
A_4^{\mathrm{lc}}(\phi,1_\qb^-,2_q^+,3_\Qb^+,4_Q^-)\right]
\nonumber\\ && \hskip0.0cm \null
+ \frac{1}{N_c^2} A_4^{\mathrm{slc}}(\phi,1_\qb^-,2_q^+,3_\Qb^+,4_Q^-)
- \frac{n_f}{N_c}
A_4^{\fl}(\phi,1_\qb^-,2_q^+,3_\Qb^+,4_Q^-)\, ,
\nonumber
\eea
\bea
A_{4;2}(\phi,1_\qb^-,2_q^+,3_\Qb^-,4_Q^+) &=&
A_4^{\mathrm{lc}}(\phi,1_\qb^-,2_q^+,4_\Qb^+,3_Q^-)
\label{mpmp2}
\\ && \hskip0.0cm \null
+ \frac{1}{N_c^2}\left[
A_4^{\mathrm{lc}}(\phi,1_\qb^-,2_q^+,4_\Qb^+,3_Q^-) +
A_4^{\mathrm{lc}}(\phi,1_\qb^-,2_q^+,3_\Qb^-,4_Q^+)\right]
\nonumber\\ && \hskip0.0cm \null
- \frac{1}{N_c^2} A_4^{\mathrm{slc}}(\phi,1_\qb^-,2_q^+,4_\Qb^+,3_Q^-)
- \frac{n_f}{N_c}
A_4^{\fl}(\phi,1_\qb^-,2_q^+,3_\Qb^-,4_Q^+)\,.
\nonumber
\eea
Here $A_4^{\mathrm{lc}}$, $A_4^{\mathrm{slc}}$, and $A_4^{\fl}$
describe respectively the leading-color, subleading-color and
fermion-loop primitive amplitudes.  The number of massless quark
flavors is denoted by $n_f$. The quantity
$A_4^{\mathrm{lc}}(\phi,1_\qb^-,2_q^+,4_\Qb^-,3_Q^+)$ refers to
the primitive amplitude
$A_4^{\mathrm{lc}}(\phi,1_\qb^-,2_q^+,3_\Qb^-,4_Q^+)$ with the
labels on legs 3 and 4 exchanged.  Sample Feynman diagrams
corresponding to these amplitudes are shown in
\fig{primitive_amplitudes}.


\FIGURE[t]{
\resizebox{0.65\textwidth}{!}{\includegraphics{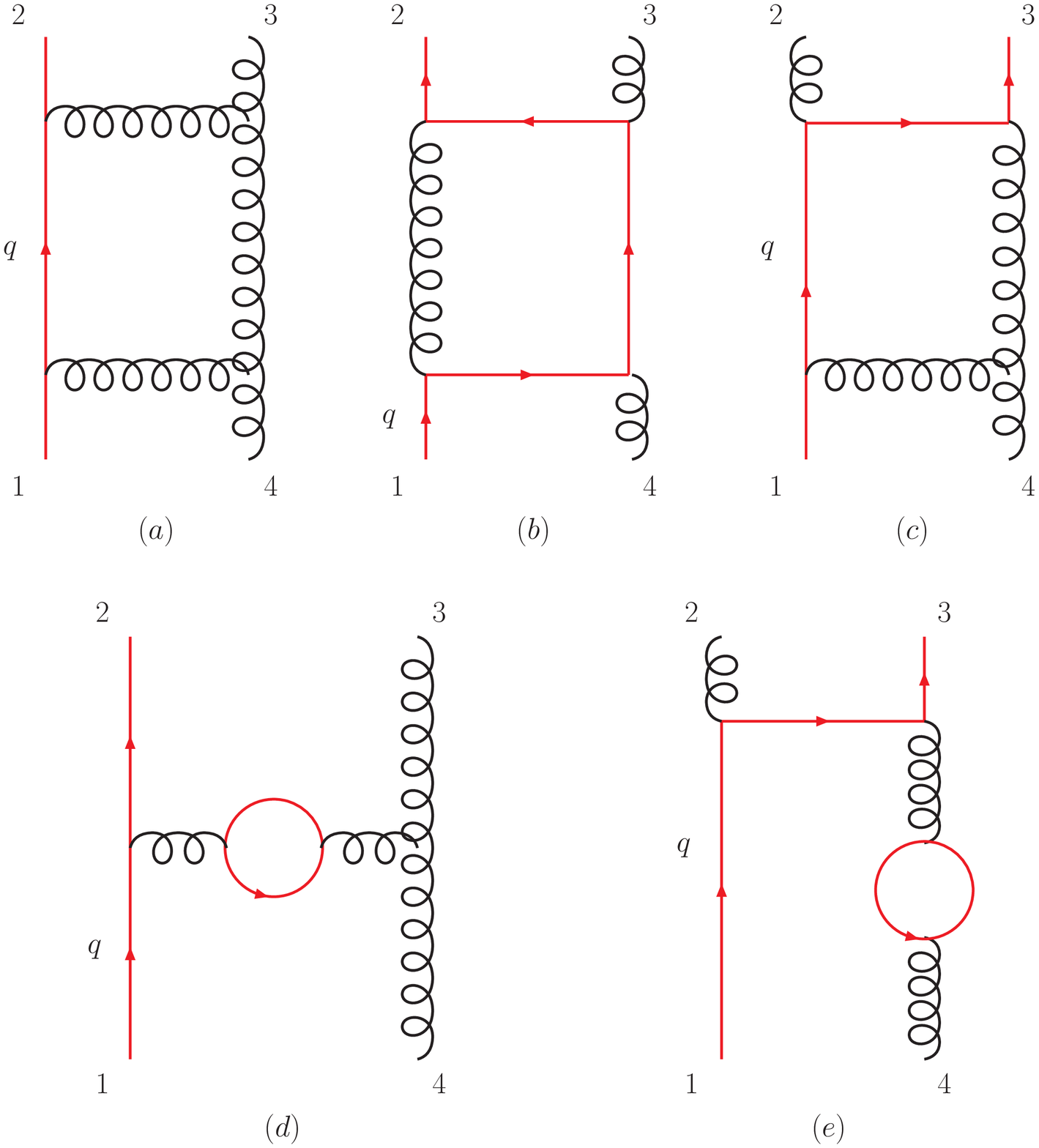}}
\caption{ Sample diagrams corresponding to the (a) $\qb qgg$ $L$,
(b) $\qb qgg$ $R$, (c) $\qb gqg$ $L$, (d) $\qb qgg$ fermion loop,
and (e) $\qb gqg$ fermion loop primitive amplitudes. The Higgs
field can attach to any gluon line in these diagrams.}
\label{primitive_amplitudes_qqgg} }

The virtual part of the unpolarized NLO cross section for $H\qb q\Qb Q$
requires the sum over both helicities and colors of the
interference of the tree and one-loop amplitudes.
The color sum is straightforward to work out from
\eqns{qqQQtreedecomp}{qqQQloopdecomp},
\bea
\sum_{\rm colors} \bigl[ {\cal A}_4^* {\cal A}_4 \bigr]_{\rm NLO}
&=& 2 C^2 \cg g^6 (N_c^2-1) \, N_c \, {\rm Re} \Bigl[
 {\cal A}_4^{\treenum\,*}(H,1_\qb,2_q,3_\Qb,4_Q)
  A_{4;1}(H,1_\qb,2_q,3_\Qb,4_Q) \Bigr] \,. \nonumber\\
\label{HqqQQcolorsum}
\eea
The same formula also applies with $H$ replaced by $A$.
The formula for the case of identical quarks $q=Q$ follows
from \eqn{identicalquarks}.

\subsubsection{$\phi\qb qgg$}

Similarly, the color decomposition for the $\phi\qb qgg$
amplitudes is identical to that for the process 
$e^+e^- \rightarrow \qb qgg$~\cite{Bern1997sc}. The tree amplitude is
given by
\be
{\cal A}_4^\treenum(\phi,1_\qb,2_q,3,4)
 = Cg^2 \, \sum_{\sigma\in S_2}\left(
T^{a_{\sigma(3)}}T^{a_{\sigma(4)}}\right)_{i_2}^{\,\,\,\bar{\imath}_1}
A_4^\treenum(\phi,1_\qb,2_q,\sigma(3),\sigma(4)) \,,
\label{qqggtreedecomp}
\ee
where the tree helicity amplitudes
$A_4^\treenum\left(\phi,1_\qb,2_q,3,4\right)$
are given in \eqn{qqggtrees}.

The one-loop amplitude is decomposed as
\bea
{\cal A}_4^\oneloopnum(\phi,1_\qb,2_q,3,4) &=& Cg^4\, \cg
\Bigg[ N_c\sum_{\sigma\in S_2}\left(
T^{a_{\sigma(3)}}T^{a_{\sigma(4)}}\right)_{i_2}^{\,\,\,\bar{\imath}_1}
A_{4;1}(\phi,1_\qb,2_q,\sigma(3),\sigma(4))
\nonumber\\ && \hskip1.3cm \null
+ \delta^{a_3 a_4} \, \delta_{i_2}^{\,\,\,\bar{\imath}_1}
A_{4;3}(\phi,1_\qb,2_q;3,4) \Bigg] \,.
\label{qqggloopdecomp}
\eea
The partial amplitudes $A_{4;1}$ and $A_{4;3}$ are given,
in a helicity-independent fashion, by
\bea A_{4;1}(\phi,1_\qb,2_q,3,4) &=& A_4^L(\phi,1_\qb,2_q,3,4) -
\frac{1}{N_c^2}A_4^R(\phi,1_\qb,2_q,3,4) \nonumber\\ && \hskip0.0cm
\null + \frac{n_f}{N_c} \, A_4^\fl(\phi,1_\qb,2_q,3,4) \,,
\eea
\bea A_{4;3}(\phi,1_\qb,2_q;3,4) &=& A_4^L(\phi,1_\qb,2_q,3,4) +
A_4^L(\phi,1_\qb,2_q,4,3) + A_4^L(\phi,1_\qb,3,2_q,4) \nonumber\\
&& \hskip0.0cm \null + A_4^L(\phi,1_\qb,4,2_q,3) +
A_4^R(\phi,1_\qb,2_q,3,4) + A_4^R(\phi,1_\qb,2_q,4,3) \,.
\nonumber\\
\eea
Here we choose to label the leading- and subleading-color
primitive amplitudes by ``$L$'' and ``$R$'' (corresponding to
fermion lines turning left or right upon entering the loop) to be
compatible with the notation used in ref.~\cite{Berger2006sh}.
Again $A_4^\fl$ denotes a fermion-loop contribution. Sample
Feynman diagrams corresponding to these primitive amplitudes are
shown in \fig{primitive_amplitudes_qqgg}.

The virtual part of the unpolarized NLO cross section for $H\qb
qgg$ requires the sum over both helicities and colors of the
interference of the tree and one-loop amplitudes. The color sum
can be expressed in the same form as that for $e^+e^- \to \qb
qgg$~\cite{Bern1997sc},
\bea
\sum_{\rm colors} \bigl[ {\cal A}_4^* {\cal A}_4 \bigr]_{\rm NLO}
&=& 2 C^2 \cg g^6 (N_c^2-1) \, {\rm Re} \biggl\{
 A_4^{\treenum\,*}(H,1_\qb,2_q,3,4)
\Bigl[ (N_c^2-1) A_{4;1}(H,1_\qb,2_q,3,4) \nonumber\\
&& \hskip3.7cm \null
    - A_{4;1}(H,1_\qb,2_q,4,3) + A_{4;3}(H,1_\qb,2_q;3,4) \Bigr] \biggr\}
\nonumber\\
&& \hskip0.3cm \null
\ +\ \{ 3 \lr 4 \} \,.
\label{Hqqggcolorsum}
\eea
The same formula also applies with $H$ replaced by $A$.


\subsection{Spinor helicity formalism}
\label{SpinorHelSubsection}

Primitive amplitudes depend only on kinematic variables.
They are functions of the external momenta of the Higgs boson,
$k_\phi = k_H$, and of the four partons, $k_i$, $i=1,\ldots,4$.
These momenta are all outgoing, by convention,
so momentum conservation and the on-shell conditions read,
\bea
&&k_\phi + \sum_{i=1}^4 k_i = 0\, ,
\label{momentum_conservation}\\
&&k_\phi^2 = k_H^2 = m_H^2\, , \qquad k_i^2 = 0\,.
\eea
A very convenient representation of the amplitudes is in terms of
spinor inner products, as reviewed {\it e.g.} in
refs.~\cite{Mangano1990by,Dixon1996wi}.  Let $u_\pm(k_i)$ be a
massless Weyl spinor of momentum $k_i$ and positive or negative
chirality. The corresponding two-component spinors are often
denoted $\lambda_i^\alpha$ and $\tilde\lambda_i^{\dot\alpha}$. The
spinor products are defined by
\bea
\spa{i}.{j} &=& \lambda_i^\alpha \lambda_{j\,\alpha}
             = \langle i^-|j^+\rangle = \bar{u}_-(k_i)u_+(k_j)\,,
             \label{defspa}\\
\spb{i}.{j} &=& \tilde\lambda_{i\,\dot\alpha} \tilde\lambda_j^{\dot\alpha}
             = \langle i^+|j^-\rangle = \bar{u}_+(k_i) u_-(k_j)\,.
             \label{defspb}
\eea
We use the convention $\spb{i}.{j} = \mathop{\rm
sgn}(k_i^0k_j^0)\spa{j}.{i}^*$, so that
\be
\spa{i}.{j}\spb{j}.{i} = 2k_i\cdot k_j \equiv s_{ij} \,.
\ee
For real momenta, and up to a complex phase, the spinor inner products
are square roots of the corresponding kinematic invariant
$s_{ij}\equiv (k_i+k_j)^2$.

Three-parton invariant masses are defined by,
\be
s_{ijl} \equiv (k_i+k_j+k_l)^2
= \spa{i}.{j}\spb{j}.{i} + \spa{j}.{l} \spb{l}.{j}
+ \spa{i}.{l} \spb{l}.{i}.
\ee
We also define the spinor strings,
\be
\langle a|i|b]\ =\ \spa{a}.{i}\spb{i}.{b} \,, \qquad\quad
\langle a|(i+j)|b]
\ =\ \spa{a}.{i}\spb{i}.{b} + \spa{a}.{j}\spb{j}.{b} \,.
\label{stringdef}
\ee
Strings involving the Higgs momentum, such as
$\langle a^-|k_H|b^-\rangle$, could also show up; however,
they can be eliminated in favor of strings such as in \eqn{stringdef}
using momentum conservation.

Besides momentum conservation, the other two spinor product identities
used to simplify expressions are antisymmetry,
\be
\spa{j}.{i} = - \spa{i}.{j} \,, \qquad\quad \spb{j}.{i} = - \spb{i}.{j} \,,
\ee
and the Schouten identity,
\bea
\spa{a}.{b}\spa{c}.{d} &=&
\spa{a}.{d}\spa{c}.{b} + \spa{a}.{c}\spa{b}.{d} \,, \\
\spb{a}.{b}\spb{c}.{d} &=&
\spb{a}.{d}\spb{c}.{b} + \spb{a}.{c}\spb{b}.{d} \,.
\eea
Composite spinors can appear in these products as well and they can
be handled similarly,
\bea
\langle a|P_{i\ldots j}|b]\spa{c}.{d} &=&
  \spa{a}.{d}\langle c|P_{i\ldots j}|b]
- \spa{a}.{c} \langle d|P_{i\ldots j}|b] \,, \\
\langle a|P_{i\ldots j}|b]\spb{c}.{d} &=&
   \langle a|P_{i\ldots j}|d]\spb{c}.{b}
 + \langle a|P_{i\ldots j}|c]\spb{b}.{d} \,, \\
\langle a|P_{i\ldots j}|b]\langle c|P_{i\ldots j}|d]
&=& \langle a|P_{i\ldots j}|d]\langle c|P_{i\ldots j}|b]
 - P_{i\ldots j}^2\spa{a}.{c}\spb{b}.{d} \,,
\eea
where $P_{i\ldots j}$ denotes an arbitrary momentum sum,
$P_{i\ldots j}^\mu \equiv \sum_{m=i}^j k_m^\mu$.
The spinor products defined above form the building blocks for
the primitive amplitudes and the basis for our calculations in the next
sections.


\section{Unitarity and recursive bootstrap method}
\label{BootstrapSection}

The method we employ in our paper combines the (generalized) unitarity
method~\cite{UnitarityMethod,UnitarityMethod2,Bern1997sc,Britto2004nc,%
Britto2005ha,Britto2006sj} with on-shell recursion
relations~\cite{Britto2005fq} operating at one
loop~\cite{Bern2005hs,Bern2005ji,Bern2005cq}. Using general
integral reduction
formulae~\cite{IntegralReductions,BDKIntegrals}, any
dimensionally-regulated one-loop amplitude ${\cal A}_n^{(1)}$ with
massless internal lines can be decomposed as,
\be {\cal A}_n^{(1)}\ =\ C_n + R_n \,, \label{CutRational}
\ee
where the {\it cut part} $C_n$ is a linear combination of scalar
integrals~\cite{IntegralReductions,BDKIntegrals},
\be
C_n\ =\ \sum_i d_i \, {\cal I}_4^i + \sum_i c_i \, {\cal I}_3^i
 + \sum_i b_i \, {\cal I}_2^i \,,
\label{IntegralBasis}
\ee
and $R_n$ denotes the {\it rational part}, which contains no
branch cuts (in four dimensions). The box integrals ${\cal
I}_4^i$, triangle integrals ${\cal I}_3^i$ and bubble integrals
${\cal I}_2^i$ are well known~(see for example
refs.~\cite{BDKPentagon,BGH,DuplancicNizic,EZ}). Hence determining
$C_n$ is equivalent to computing the coefficients of the
respective integrals, $d_i$, $c_i$ and $b_i$. These coefficients
are found by taking (generalized) four-dimensional unitarity cuts
of the amplitude in various channels. The rational part $R_n$ will
be calculated by utilizing on-shell recursion relations, which
requires only information about lower-point loop amplitudes and
tree amplitudes.  (For a review of the on-shell approach, see
ref.~\cite{Bern2007dw}.)

In the remainder of this section, we outline the various ingredients
in the method, and apply them to specific primitive amplitudes for
$\phi\qb q\Qb Q$ and $\phi\qb qgg$.  The results for these amplitudes
are collected in \sect{ResultsSection}.


\subsection{Generalized unitarity for box coefficients}
\label{GenUSubsection}

In this paper, the generalized unitarity method, along with
complex spinor
integration~\cite{Britto2004nc,Britto2005ha,Britto2006sj} is
employed in a twofold manner for the calculation of box functions
and bubble (single log) functions respectively.

In ref.~\cite{Britto2004nc} it was shown that using quadruple
cuts, it is possible to reduce the task of calculating the
coefficient of any box function into that of calculating a product
of four tree amplitudes. This method can be applied to box functions with
any number of external masses by taking advantage of the generally
non-vanishing behavior of the all-massless
three-point amplitudes when momenta are complexified.
A three-point amplitude with one leg carrying opposite helicity from the
other two can have either an MHV or an anti-MHV (sometimes also
denoted $\overline{\rm MHV}$) representation while satisfying
energy-momentum conservation. The complex momenta allow for all
four loop momenta of a box diagram to become on shell, freezing
the cut integral completely and simplifying it into an algebraic
product of tree amplitudes.


\FIGURE[t]{
\resizebox{0.65\textwidth}{!}{\includegraphics{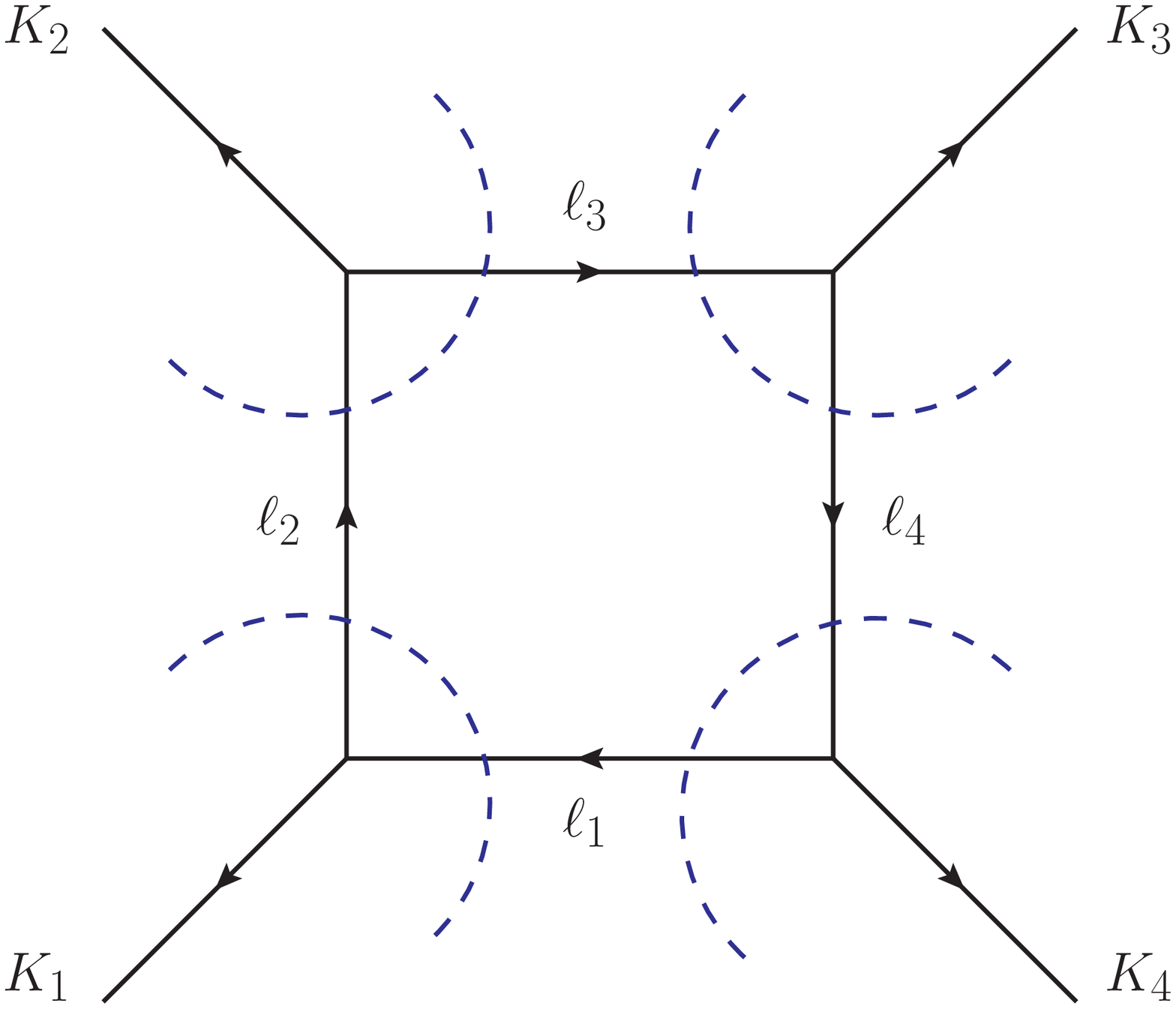}}
\caption{ Evaluation of a box function coefficient using quadruple
cuts.  The four on-shell conditions on the loop momentum reduce
the integration to an algebraic product of four tree amplitudes.}
\label{box_general_figure}}

For example, the coefficient of the box integral in
\fig{box_general_figure} is given by
\be
d = \frac{1}{\Delta_{\mathrm{LS}}{\cal I}_4} \,
\int d^4\ell_1\ \delta^{(+)}(\ell_1^2) \, \delta^{(+)}(\ell_2^2)
 \, \delta^{(+)}(\ell_3^2) \, \delta^{(+)}(\ell_4^2)
\ A_1^\treenum A_2^\treenum A_3^\treenum A_4^\treenum \,,
\ee
where $\ell_2 = \ell_1-K_1$, $\ell_3 = \ell_1-K_1-K_2$, $\ell_4=\ell_1+K_4$,
and each $A_i^\treenum$ corresponds to the tree amplitude in the
respective corner with massive (or massless)
momentum $K_i$ ($i=1\ldots 4$). The quadruple cut 
(or leading singularity) of the scalar box integral is given by
\be
\Delta_{\mathrm{LS}} {\cal I}_4 =
\int d^4\ell_1\ \delta^{(+)}(\ell_1^2) \, \delta^{(+)}(\ell_2^2)
 \, \delta^{(+)}(\ell_3^2) \, \delta^{(+)}(\ell_4^2) \,,
\ee
from which we obtain
\be
d = \frac{1}{2} \sum_{\sigma,h}
A_1^\treenum(\ell_1^\sigma) A_2^\treenum(\ell_1^\sigma)
A_3^\treenum(\ell_1^\sigma) A_4^\treenum(\ell_1^\sigma) \,,
\label{general_quad_cut_formula}
\ee
where the summation is over the two discrete solutions $\ell_1^\sigma$,
$\sigma=1,2$, of the loop-momentum localization constraints,
\be
\ell_1^2 = \ell_2^2 = \ell_3^2 = \ell_4^2 = 0 \,,
\label{quadruple_cut_eqs}
\ee
and over all possible helicities $h$ of internal particles propagating
in the loop.


\subsubsection{A two-mass box coefficient}
\label{TwoMassBoxCoefficientSubsubsection}

Let's consider, for example, the leading-color primitive amplitude
$A_4^{\mathrm{lc}}(\phi,1_\qb^-,2_q^+,3_\Qb^+,4_Q^-)$.
One of the box coefficients that needs to be determined
for this primitive amplitude
is that of the ``easy two mass'' box integral, with diagonally
opposite massive legs having mass $m_H^2$ and $s_{23}$.
Because all the two-mass boxes have $m_H^2$ as one of the two
masses, we denote this coefficient, using the other mass, as
$d^{2{\rm m}e}_{23}$.
This box integral is defined by the clustering of the five external
particles into the four legs of the box: $(1)(23)(4)(\phi)$.
The associated quadruple cut is shown in \fig{2me_example_figure}.


\FIGURE[t]{
\resizebox{0.6\textwidth}{!}{\includegraphics{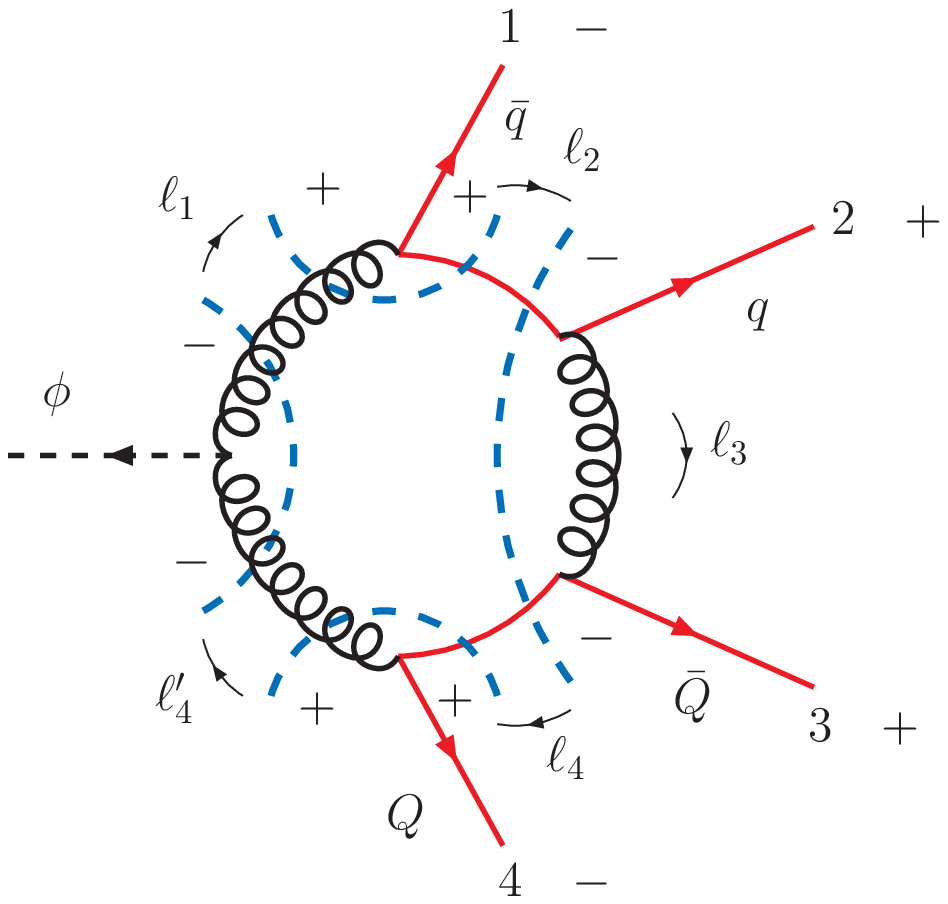}}
\caption{Quadruple cut for the evaluation of the easy-two-mass box
coefficient $d^{2{\rm m}e}_{23}$ of
$A_4^{\mathrm{lc}}(\phi,1_\qb^-,2_q^+,3_\Qb^+,4_Q^-)$.}
\label{2me_example_figure} }


The tree amplitude with $\phi$ and two gluons vanishes unless both
helicities are negative~\cite{Dixon2004za}.  It is easy to see
that this fact, plus fermion helicity conservation, forces the
unique assignment of intermediate helicities shown in
\fig{2me_example_figure}.  The three-point vertices containing
legs 1 and 4 are only nonvanishing for one of the two solutions to
the quadruple cut conditions~(\ref{quadruple_cut_eqs}).
\Eqn{general_quad_cut_formula} then becomes,
\be
d^{2{\rm m}e}_{23} =
\frac{1}{2} A_2^\treenum(\phi,\ell_1^-,{-\ell_4^\prime}^-)
A_3^\treenum(-\ell_1^+,1_\qb^-,\ell_{2q}^+)
A_4^\treenum(-\ell_{2\qb}^-,2_q^+,3_\Qb^+,\ell_{4Q}^-)
A_3^\treenum(-\ell_{4\Qb}^+,4_Q^-,{\ell_4^\prime}^+)\,.
\ee
Using the expressions for the tree amplitudes, we get
\be d^{2{\rm m}e}_{23} = \frac{i^4}{2} \times
\Bigl(-{\spa{\ell_1}.{(-\ell_4^\prime)}^2}\Bigr) \times
\frac{\spb{(-\ell_1)}.{\ell_2}^2}{\spb{1}.{\ell_2}} \times
\left(-\frac{\spb{2}.{3}^2}{\spb{(-\ell_2)}.{2}\spb{3}.{\ell_4}}\right)
\times \frac{\spb{\ell_4^\prime}.{(-\ell_4)}^2}{\spb{(-\ell_4)}.{4}}
\,. \label{quadcut23A} \ee

The positive chirality spinors for the three-point vertices containing
legs 1 and 4 are proportional,
\be
\lambda_{\ell_1}\ \propto\ \lambda_{\ell_2}
\ \propto\ \lambda_1 \,,\qquad\quad
\lambda_{\ell_4}\ \propto\ \lambda_{\ell_4^\prime}
\ \propto\ \lambda_4 \,.
\ee
Using this fact, along with momentum conservation relations, we can
eliminate all explicit loop momenta from \eqn{quadcut23A}:
\bea d^{2{\rm m}e}_{23} &=& \frac{1}{2} {\spb2.3}^2
\frac{(\spb{\ell_2}.{\ell_1}\spa{\ell_1}.{\ell_4^\prime}
       \spb{\ell_4^\prime}.{\ell_4})^2}
{\spb{1}.{\ell_2}\spb{\ell_2}.{2}\spb{3}.{\ell_4}\spb{\ell_4}.{4}}
\nn\\
&=& \frac{1}{2} {\spb2.3}^2
\frac{(\spb{\ell_2}.{1}\spa{1}.{4}\spb{4}.{\ell_4})^2}
{\spb{1}.{\ell_2}\spb{\ell_2}.{2}\spb{3}.{\ell_4}\spb{\ell_4}.{4}}
\nn\\
&=& \frac{1}{2} {\spa1.4}^2 {\spb2.3}^2 \frac{\spb{1}.{\ell_2}
\spa{\ell_2}.{4}\ \spa{1}.{\ell_4} \spb{\ell_4}.{4}}
{\spb{2}.{\ell_2} \spa{\ell_2}.{4}\ \spa{1}.{\ell_4}
\spb{\ell_4}.{3}}
\nn\\
&=& \frac{1}{2} {\spa1.4}^2 {\spb2.3}^2 \frac{\spba1.{(2+3)}.4\
\spab1.{(2+3)}.4} {\spb2.3\spa3.4\ \spa1.2\spb2.3}
\nn\\
&=& \frac{1}{2}\langle4|(2+3)|1] \langle 1|(2+3)|4]
\frac{\spa1.4^2}{\spa1.2\spa3.4} \,. \label{quadcut23B} \eea

In \sect{ResultsSection} we express the result, not in terms
of the scalar box integral ${\cal I}_4^{2\rm{m}e}$, but in terms
of the infrared-finite box function
$\Ls_{-1}^{2{\rm m}e}\left(s_{123},s_{234};s_{23},m_H^2\right)$.
These are related by \eqn{I2meexpr}.  After removing a factor of $\cg$
associated with \eqn{qqQQloopdecomp}, and using the identity
\be
s_{123} s_{234} - s_{23} m_H^2\ =\ \langle4|(2+3)|1] \langle 1|(2+3)|4] \,,
\ee
the coefficient of the box function
$\Ls_{-1}^{2{\rm m}e}\left(s_{123},s_{234};s_{23},m_H^2\right)$ in
$A_4^{\mathrm{lc}}(\phi,1_\qb^-,2_q^+,3_\Qb^+,4_Q^-)$ is
\be D^{2{\rm m}e}_{23}\ \equiv\ \frac{2 i}{s_{123} s_{234} -
s_{23} m_H^2} \, d^{2{\rm m}e}_{23} \ =\ i \,
\frac{\spa1.4^2}{\spa1.2\spa3.4} \ =\ -
A_4^\treenum(\phi,1_\qb^-,2_q^+,3_\Qb^+,4_Q^-) \,.
\label{D23final} \ee
It is a general feature of every primitive amplitude
presented in \sect{ResultsSection} that all easy-two-mass box functions
have coefficients equal to the negative of the
corresponding tree amplitude.
We therefore collect all the $\Ls_{-1}^{2{\rm m}e}$ functions
into ``$V$'' functions, which also contain the infrared and
ultraviolet poles in $\e$, since the latter have
to be proportional to the tree amplitude as well.


\subsubsection{Calculation of a one-mass box coefficient}

\FIGURE[t]{
\resizebox{0.6\textwidth}{!}{\includegraphics{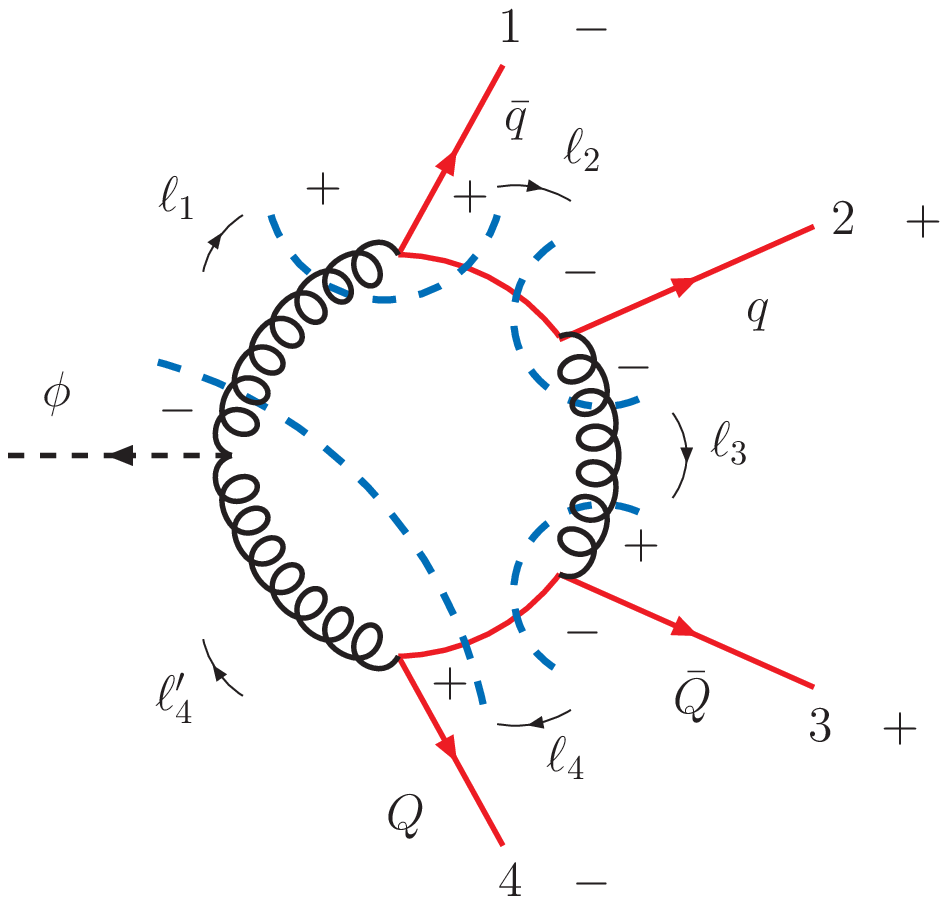}}
\caption{ Quadruple cut for the evaluation of the one-mass box
coefficient $d^{\mathrm{1m}}_{123}$ appearing in
$A_4^{\mathrm{lc}}(\phi,1_\qb^-,2_q^+,3_\Qb^+,4_Q^-)$.
\label{1m_example_figure} } }

As our second box example, we compute for the
same primitive amplitude the
coefficient of the one-mass box function
$\Ls_{-1}\left(s_{12},s_{23};s_{123}\right)$,
associated with the external leg clustering $(1)(2)(3)(4\phi)$.
The quadruple cut is depicted in \fig{1m_example_figure}.
We label the box integral coefficient by $d^{\mathrm{1m}}_{123}$.
Again the vanishing of a tree amplitude involving $\phi$,
in this case $A_3^\treenum(\phi,\ell_1^+,-\ell_{4\Qb}^+,4_Q^-)$,
along with fermion helicity conservation and the vanishing
of $A_4^\treenum(\qb^-,q^+,g^+,g^+)$, forces the unique
helicity assignment shown in the figure.
The cut evaluates to
\bea d^{\mathrm{1m}}_{123} &=& \frac{1}{2}
A_3^\treenum(\phi,\ell_1^-,-\ell_{4\Qb}^+,4_Q^-)
A_3^\treenum(-\ell_1^+,1_\qb^-,\ell_{2q}^+)
A_3^\treenum(-\ell_{2\qb}^-,2_q^+,\ell_3^-)
A_3^\treenum(-\ell_3^+,3_\Qb^+,\ell_{4Q}^-) \nonumber \\
&=& \frac{i^4}{2} \times
\left(-\frac{\spa{\ell_1}.{4}^2}{\spa{(-\ell_4)}.{4}}\right) \times
\frac{\spb{(-\ell_1)}.{\ell_2}^2}{\spb{1}.{\ell_2}} \times
\left(-\frac{\spa{\ell_3}.{(-\ell_2)}^2}{\spa{(-\ell_2)}.{2}}\right)
\times \frac{\spb{(-\ell_3)}.{3}^2}{\spb{3}.{\ell_4}} \,. \eea
After some spinor product manipulations similar to \eqn{quadcut23B},
we have
\be d^{\mathrm{1m}}_{123}\ =\ \frac{1}{2} s_{12}s_{23}
\frac{\spa1.4^2}{\spa1.2\spa3.4} \,, \ee
which yields for the coefficient of the
function $\Ls_{-1}\left(s_{12},s_{23};s_{123}\right)$,
\be D^{\mathrm{1m}}_{123}\ =\ \frac{2i}{s_{12}s_{23}}
d^{\mathrm{1m}}_{123} \ =\ i \, \frac{\spa1.4^2}{\spa1.2\spa3.4} \
=\ - A_4^\treenum(\phi,1_\qb^-,2_q^+,3_\Qb^+,4_Q^-) \,.
\label{D123final} \ee
The result is again proportional to the tree, up to a sign.
This property holds for all the one-mass box coefficients $\Ls_{-1}$
in the $\phi\qb q\Qb Q$ primitive amplitudes, apart from the
leading-color piece of the $({-}{+}{-}{+})$ helicity configuration.
However, in the $\phi\qb qgg$ primitive amplitudes it is typically
not true.


\subsection{Absence of three-mass triangles}
\label{TriSubsection}


\FIGURE[t]{
\resizebox{0.6\textwidth}{!}{\includegraphics{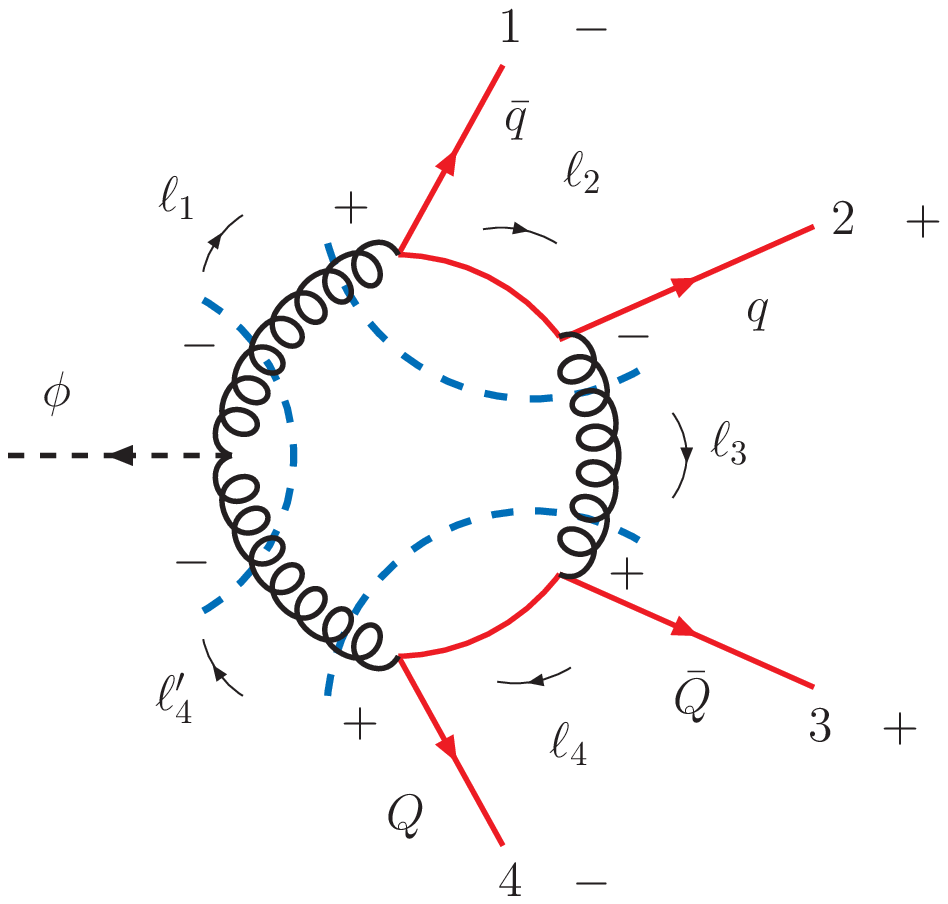}}
\caption{ Sample $\phi\qb q\Qb Q$ triple cut, illustrating the
vanishing coefficient for a three-mass triangle integral, caused
by the tree amplitude in the lower right corner associated with
the external invariant $s_{34}$. Consequently, hard-two-mass box
functions are absent from the amplitude as well.}
\label{tr3m_example_figure} }


The one-loop $\phi$ amplitudes that we consider in this paper have
the property that the triple cuts associated with three-mass
triangle integrals all vanish.  This general feature holds because
the tree-level four-parton amplitudes in QCD, $A_4^\treenum(1,2,3,4)$,
and those with an additional $\phi$ boson, $A_4^\treenum(\phi,1,2,3,4)$,
each require two negative helicities to be nonvanishing.
A nonvanishing product of three such amplitudes, as required
for a triple cut, implies six negative helicities.
Three of the negative helicities are associated with the three cut lines, so
there must be three external negative helicities. However, the
one-loop $\phi$ amplitudes we consider here only have two negative
helicities.  An example of a vanishing triple cut is shown
in~\fig{tr3m_example_figure}.  The same argument implies that all
``hard-two-mass'' boxes, with two adjacent massive legs, must
vanish: For any hard-two-mass box quadruple cut, one can remove
the cut between the two adjacent massless legs, relaxing the
quadruple cut into a (vanishing) triple cut.  Because the
$\phi\,+\,4$ parton amplitudes obviously contain no three-mass or
four-mass box integrals, only one-mass and easy-two-mass box
coefficients have to be computed here.

The amplitudes $\phi g^-g^-g^+g^+$~\cite{Badger2007si} and $\phi
g^-g^+g^-g^+$~\cite{Glover2008ffa} also contain only one-mass and
easy-two-mass boxes, and no three-mass triangles, for the same
reason, insufficiently many negative helicities in the triple
cuts. Interestingly, the amplitude $\phi g^-g^-g^-g^-$ computed in
ref.~\cite{BadgerGlover} also contains only one-mass and
easy-two-mass boxes, and no three-mass triangles, for the opposite
reason, a paucity of positive helicity gluons. On the other hand,
the primitive amplitudes for $\phi\qb^-q^+g^-g^-$ and $\phi
g^+g^-g^-g^-$, which have not yet been computed analytically, will
contain three-mass triangles and hard-two-mass boxes.

For the coefficients of two-mass and one-mass triangles, the above
triple-cut vanishing argument does not hold.  The existence of a
massless external leg implies that one of the tree amplitudes can
be an $\overline{\rm MHV}$ three-point amplitude, which contains
only one negative helicity, not two. However, the two-mass and
one-mass triangle integrals, given in \eqns{I31mdef}{I32mdef},
contain single log terms at order $1/\e$. Because of this, their
coefficients are completely determined by the known infrared poles
of the amplitude, so they do not have to be computed separately.
The remainder of the work to compute the cut part of the amplitude
involves determining the coefficients of bubble integrals.


\subsection{Unitarity and spinor integration for bubbles}
\label{BubSubsection}

In the case of the ordinary two-particle cuts used to determine
bubble coefficients, there are not enough constraints to fully
localize the cut integral.  The cut contains a residual
phase-space integral.  The method we use in this case was proposed
in refs.~\cite{Britto2005ha,Britto2006sj} and consists of writing
the cut loop momentum integral as an integral over spinor
variables $\ell\equiv|\ell\rangle$ and
$\tilde\ell\equiv|\ell]$~\cite{Cachazo2004kj}.
The integrand can be transformed into a total derivative in
$\tilde\ell$, leaving a single integral over $\ell$ which can be
evaluated by residue extraction.


\FIGURE[t]{
\resizebox{0.7\textwidth}{!}{\includegraphics{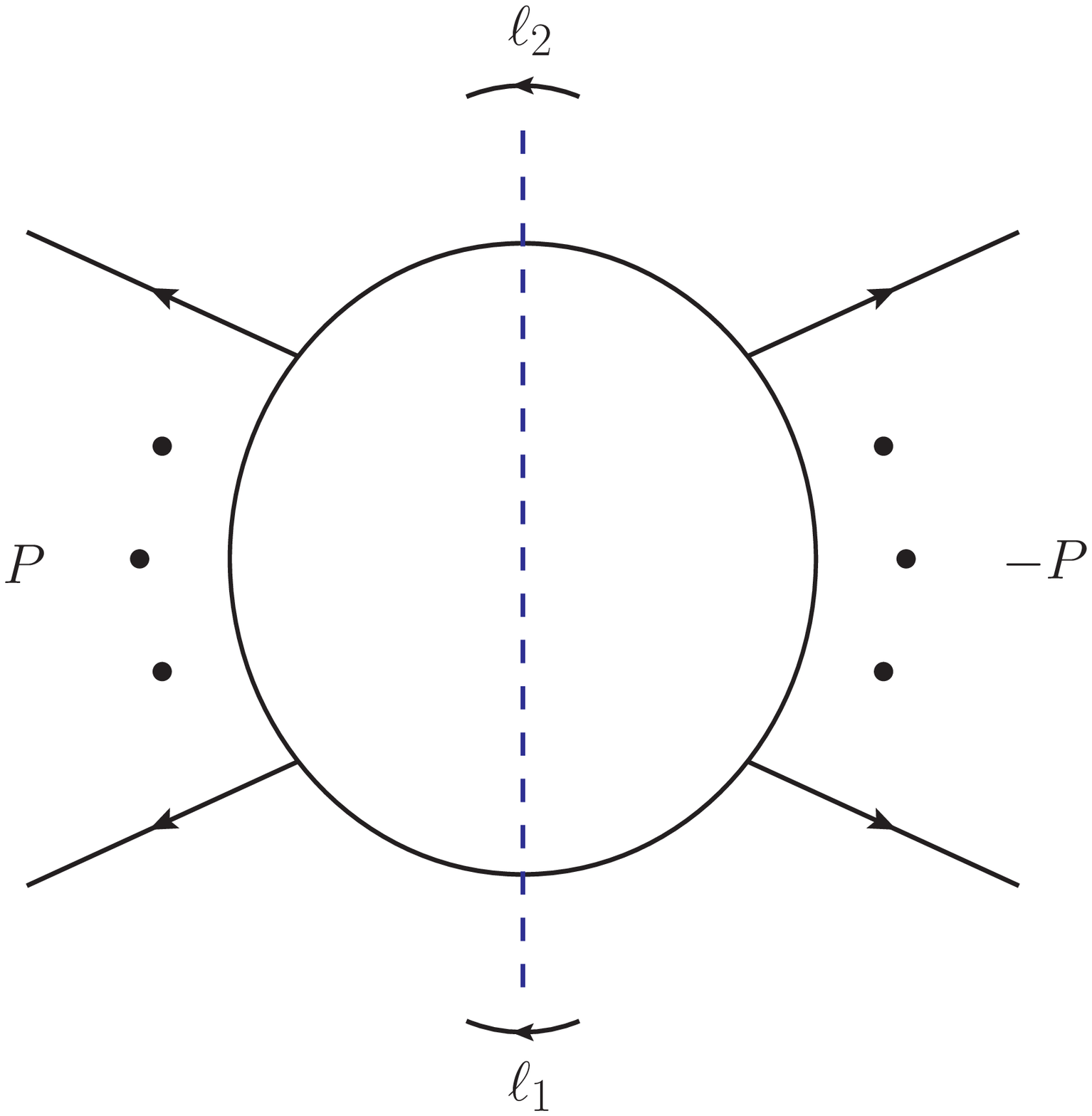}}
\caption{ Example of the evaluation of a bubble (single log)
function using an ordinary two-particle cut. Integration over the
loop momentum is required in this case.}
\label{bubble_general_figure} }


For the cut shown in
\fig{bubble_general_figure}, the coefficient of the logarithm will
be given by an integral of the form
\be
\int d^4\ell_1 \, \delta^{(+)}(\ell_1^2)\,\,
\delta^{(+)}(\ell_2^2)\,\, A_1^\treenum A_2^\treenum
\ =\ \int_0^\infty dt\,\,t \int
\spa{\ell}.{d\ell}[\ell\,d\ell]\,\,
\delta(P^2-t\langle\ell|P|\ell])\, f(\ell,\tilde{\ell})\,,
\label{2particlecut}
\ee
where $\ell$ can be either $\ell_1$ or $\ell_2$, whichever
is more convenient.
The function $f(\ell,\tilde{\ell})$ represents the product
of the two tree amplitudes and is in general a sum of terms of the
form
\be
\frac{\prod_i\langle a_i\,\ell\rangle \prod_j[b_j\,\ell]
\prod_k\langle\ell|R_k|\ell]} {\prod_i\langle c_i\,\ell\rangle
\prod_j[d_j\,\ell] \prod_k\langle\ell|Q_k|\ell]}\, ,\quad
\mathrm{with} \quad Q_k\neq P\, ,\quad Q_k^2\neq 0 \,.
\ee
After performing the integration over $t$ and partial fractioning
using Schouten identities, we can always
bring the remaining integrand into a form where we can take
advantage of the identity
\be
\spb{\ell}.{d\ell}
\left(\frac{\spb{\eta}.{\ell}^n}{\langle\ell|P|\ell]^{n+2}}\right)
= \spb{d\ell}.{\partial_\ell} \left(\frac{1}{n+1}
\frac{1}{\langle\ell|P|\eta]}
\frac{\spb{\eta}.{\ell}^{n+1}}{\langle\ell|P|\ell]^{n+1}}\right),
\label{total_derivative_id}
\ee
and convert it into a total derivative with respect to $|\ell]$. 
(In the special case of
$n=0$, the spinor $|\eta]$ appears only on the right-hand side of
\eqn{total_derivative_id}; hence it can be chosen arbitrarily.)
Then we can evaluate the integral over $|\ell\rangle$ by
calculating the residue for each pole. The case of multiple poles
was also examined in ref.~\cite{Britto2006sj}.

In this process one also detects the coefficients of box integrals
sharing the same cut. These are the terms that scale like
$1/\langle\ell |P|\ell]$ after partial fractioning of the integrand.
They can serve as an independent check of the box coefficients
determined by the quadruple cuts.

\subsubsection{Calculation of a bubble coefficient}

As an example, we compute the coefficient of the bubble integral
${\cal I}_2(s_{123})$, or equivalently, of the single logarithm
$\ln(-s_{123})$, in the primitive amplitude
$A_4^{\mathrm{lc}}(\phi,1_\qb^-,2_q^+,3_\Qb^+,4_Q^-)$.
The corresponding two-particle cut is shown in
\fig{bubble_example_figure}, and the cut integral is given by
\bea
&&\ i \int \mathrm{dLIPS}\,\,
A_3^\treenum(\phi,\ell_1^-,\ell_{2\Qb}^+,4_{Q}^-) \times
A_5^\treenum(-\ell_1^+,1_\qb^-,2_q^+,3_\Qb^+,-\ell_{2Q}^-)
\nonumber\\
&=& i^3 \int
\mathrm{dLIPS}\left(-\frac{\spa{\ell_1}.4^2}{\spa{\ell_2}.4}\right)
\times \frac{\spa1.{(-\ell_2)}^3 \spa2.3}{\spa1.2 \spa2.3
\spa3.{(-\ell_2)} \spa{(-\ell_2)}.{(-\ell_1)} \spa{(-\ell_1)}.1}
\nonumber\\
&=& \phantom{i^3} \int \mathrm{dLIPS} \,\,
\frac{\spa{\ell_2}.1^3\spa{\ell_1}.4^2}
{\spa1.2\spa{\ell_2}.3\spa{\ell_2}.{\ell_1}\spa{\ell_1}.1\spa{\ell_2}.4}
\,. \label{cut_integral}
\eea
Here $\mathrm{dLIPS}$ stands for the Lorentz-invariant phase space
measure.


\FIGURE[t]{
\resizebox{0.6\textwidth}{!}{\includegraphics{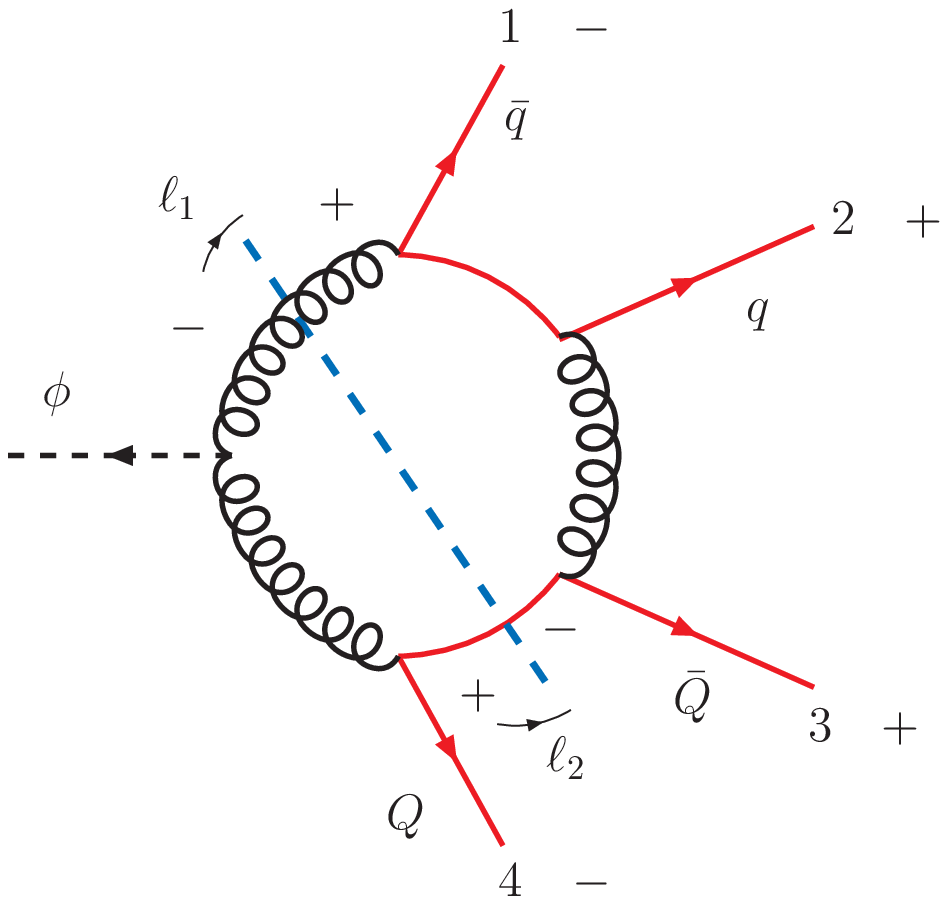}}
\caption{ Two-particle cut for the evaluation of the bubble function
coefficient $b_{123}$ of
$A_4^{\mathrm{lc}}(\phi,1_\qb^-,2_q^+,3_\Qb^+,4_Q^-)$.}
\label{bubble_example_figure} }


By multiplying both the numerator and denominator of the integrand
in \eqn{cut_integral} with factors of $\spb{\ell_2}.{\ell_1}$ we
can eliminate one of the loop momenta (in this case we choose to
eliminate $\ell_1$), leaving us with an expression that depends
only on the other loop momentum (in this case, $\ell_2$).
Furthermore, we rewrite the integral over the
Lorentz-invariant phase space as an integral over spinor variables,
as in \eqn{2particlecut},
\be
\int\mathrm{dLIPS} = \int_0^\infty t\,dt \int
\spa{\ell_2}.{d\ell_2}\spb{\ell_2}.{d\ell_2}
\delta\left(P^2-t\left\langle\ell_2|P|\ell_2\right]\right),
\ee
with $P\equiv P_{123} = k_1 + k_2 + k_3$.  We also need to track
factors of $\sqrt{t}$ from rescaling the $\ell_2$ spinors in the
integrand.  Performing these steps yields
\be
- \int_0^\infty t^2\,dt \int
\spa{\ell_2}.{d\ell_2}\spb{\ell_2}.{d\ell_2}
\delta\left(P^2 - t\left\langle\ell_2|P|\ell_2\right]\right)
\, \frac{\spa{\ell_2}.1^3\left\langle4|P|\ell_2\right]^2}{\spa1.2\spa{\ell_2}.3
P^2\left\langle 1|P|\ell_2\right]\spa{\ell_2}.4} \,.
\ee
We can readily integrate over $t$, eliminating the
$\delta$-function, to get
\be
- \int \spa{\ell_2}.{d\ell_2} \spb{\ell_2}.{d\ell_2} \frac
{P^2\spa{\ell_2}.1^3\left\langle 4|P|\ell_2\right]^2}
{\left\langle\ell_2|P|\ell_2\right]^3\spa1.2\spa{\ell_2}.3
\left\langle1|P|\ell_2\right] \spa{\ell_2}.4} \,.
\ee

The two-particle cut we have considered detects not only bubbles,
but also boxes that have cuts in this channel.  By using Schouten
identities we can rearrange terms so that we separate the
bubble contributions from the box contributions,
\bea
&&- \int \spa{\ell_2}.{d\ell_2} \spb{\ell_2}.{d\ell_2}\Bigg\{
\frac{P^2\spa{\ell_2}.1\left\langle
1|P|\ell_2\right]\spa{\ell_2}.4}
{\left\langle\ell_2|P|\ell_2\right]^3\spa1.2\spa{\ell_2}.3} -
\frac{2P^2\spa{\ell_2}.1\spa1.4}
{\left\langle\ell_2|P|\ell_2\right]^2\spa1.2\spa{\ell_2}.3}
\nonumber\\ && \hskip3.3cm \null
+ \frac{P^2\spa{\ell_2}.1\spa1.4^2}
{\left\langle\ell_2|P|\ell_2\right]\spa1.2\spa{\ell_2}.3\left\langle
1|P|\ell_2\right]\spa{\ell_2}.4} \Bigg\} \,.
\label{bub123A}
\eea
We identify the box contribution as the last term
in \eqn{bub123A}, with the $1/\left\langle\ell_2|P|\ell_2\right]$
dependence.  Because the box contributions are obtained
straightforwardly from the quadruple cuts, we can safely discard
them (or use them as an independent check, but we won't do so here).
The remaining part will be the coefficient of the bubble
${\cal I}_2(s_{123})$, given by
\be
b_{123} = - \int \spa{\ell_2}.{d\ell_2}
\spb{\ell_2}.{d\ell_2}\Bigg\{ \frac{P^2\spa{\ell_2}.1\left\langle
1|P|\ell_2\right]\spa{\ell_2}.4}
{\left\langle\ell_2|P|\ell_2\right]^3\spa1.2\spa{\ell_2}.3} -
\frac{2P^2\spa{\ell_2}.1\spa1.4}
{\left\langle\ell_2|P|\ell_2\right]^2\spa1.2\spa{\ell_2}.3}\Bigg\}
\,.
\label{bubble_C_123}
\ee
Transforming the integral over $\left|\ell_2\right]$ into a total
derivative using the general formula~(\ref{total_derivative_id}),
with $n = 0,1$ for the two terms in \eqn{bubble_C_123}, we obtain
\be
b_{123} = \int \spa{\ell_2}.{d\ell_2}
[d\ell_2\,\partial_{\ell_{2}}]\Bigg\{
\frac{1}{2}\frac{\left\langle 1|P|\ell_2\right]^2\spa{\ell_2}.4}
{\left\langle\ell_2|P|\ell_2\right]^2\spa1.2\spa{\ell_2}.3}
+ \frac{2P^2\spa{\ell_2}.1\spa1.4\spb{3}.{\ell_2}}
{\left\langle\ell_2|P|\ell_2\right]
\spa1.2\spa{\ell_2}.3\left\langle\ell_2|P|3\right]}\Bigg\}
\,.
\label{before_residues}
\ee
In this last step we have chosen the
value $\left|\eta\right] = \left|3\right]$ for the arbitrary
spinor $\left|\eta\right]$ appearing in the $n=0$ term after its
transformation into a total derivative.

At this point we are
ready to evaluate the integral over $\left|\ell_2\right\rangle$ by
computing the residues of the poles of the integrand.
We only have simple poles occurring for
$\left|\ell_2\right\rangle=\left|3\right\rangle$ in the first term,
and for $\left|\ell_2\right\rangle=P\!\left|3\right]$ in the
second term of~\eqn{before_residues}. Note that our choice for $\left|\eta\right]$
eliminates a pole for $\left|\ell_2\right\rangle=\left|3\right\rangle$ in the second term.
After substituting and simplifying,
using $\left\langle3|P|3\right] = s_{123}-s_{12}$, we get
for the coefficient of the single log $\ln(-s_{123})$
in $A_4^{\mathrm{lc}}(\phi,1_\qb^-,2_q^+,3_\Qb^+,4_Q^-)$,
\be
B_{123}\ =\ - i \, b_{123}
\ =\ -i \biggl[
\frac{1}{2} \frac{\spa1.2\spb2.3^2\spa3.4}{(s_{123}-s_{12})^2}
- 2\frac{\spb2.3\spa1.4}{s_{123}-s_{12}}
\biggr] \,,
\label{B123Final}
\ee
which is a rather simple final answer.

The procedure outlined
above for the computation of $B_{123}$ is a typical
example of the steps that have to be carried out to calculate any
bubble function coefficient. It has been automated and implemented
in \Maple\ and yields a fast analytical evaluation of these cuts.


\subsection{On-shell recursion}
\label{OnShellRecursionSection}

The (four-dimensional) unitarity technique can give us the
cut-containing parts of the amplitudes (terms associated with
functions with an imaginary part), but not the parts that are
rational functions of the kinematic variables.  However, these
terms can be determined from their analytic properties as well,
namely their factorization poles. On-shell recursion relations,
developed first at tree level~\cite{Britto2004ap,Britto2005fq}
and later at one loop~\cite{Bern2005hs,Bern2005ji,Bern2005cq,%
Berger2006ci,Berger2006vq,Berger2006cz}, exploit the known
factorization behavior and have greatly simplified the task of
calculating these terms.

At tree level, all amplitudes are rational functions and one can
consider~\cite{Britto2005fq} a complex shift of any two of the
external momenta $j$ and $l$ of an amplitude $A_n$, given by
\be
\tilde{\lambda}_j \rightarrow \tilde{\lambda}_j - z\tilde{\lambda}_l\,,
\quad
      \lambda_l\rightarrow\lambda_l + z\lambda_j \,.
\label{jlspinorshift}
\ee
This $[j,l\rangle$ shift preserves momentum conservation
as well as the massless conditions for the momenta $k_j$ and $k_l$,
which are now modified as follows,
\be
k_j^{\mu}\rightarrow k_j^{\mu}
-\frac{z}{2}\langle j|\gamma^{\mu}|l\rangle \,, \qquad\qquad
k_l^{\mu}\rightarrow k_l^{\mu}+\frac{z}{2}\langle
j|\gamma^{\mu}|l\rangle \,.
\label{jlmomentumshift}
\ee
The shifted amplitude $A_n(z)$ is an analytic
function of $z$ with only simple poles, which
are associated with factorizations of $A_n(z)$
onto lower-point tree amplitudes.

Provided that $A_n(z) \to 0$ as $z\to\infty$,
the integral of $A_n(z)/z$ over the contour at infinity vanishes.
This integral is also given by the sum over the residues at
the poles for finite $z$.  Therefore, the unshifted physical
amplitude we wish to compute is given by the residue at $z=0$,
\be
A_n = A_n(0)
= - \sum_{\mathrm{poles}\,\alpha}
\Res_{z=z_\alpha} \frac{A_n(z)}{z} \,.
\label{recursion_relation}
\ee
Each pole $z_\alpha$ in \eqn{recursion_relation} is associated
with a physical factorization channel of the amplitude.
Factorization allows the evaluation of the residue, leading to the
tree-level on-shell recursion
relation~\cite{Britto2004ap,Britto2005fq}
\be
A_n = \sum_h\sum_{r,s} A_L^h(z_{rs})
\frac{i}{P_{r\ldots s}^2} A_R^{-h}(z_{rs}) \,,
\label{tree_rec_diagrams}
\ee
where $h=\pm 1$ denotes the helicity
of the intermediate state carrying momentum $P_{r\ldots
s}$. The double sum over $r$ and $s$ is over
partitions of the external legs into two sets (contiguous
with respect to the color-ordering), for which the shifted legs
$j$ and $l$ lie on opposite sides of the pole ($j\in L$ and $l\in R$).
In the case of $\phi$ amplitudes, because $\phi$ is uncolored
(and we do not shift the $\phi$ leg), it can appear on either
the $L$ or $R$ side.
The tree amplitudes on each side are evaluated at the complex
momenta~(\ref{jlmomentumshift}), shifted by
$z=z_{rs}$, where
\be
z_{rs}\ =\ { P_{r\ldots s}^2
\over \langle j^- | P_{r\ldots s} | l^- \rangle }
\label{zrssolution}
\ee
is the solution to the condition $P^2_{r\ldots s}(z_{rs})=0$.

At the one-loop level the situation is more intricate.
\Fig{z_plane}(a) shows schematically the pole structure of a
typical tree amplitude $A_n^\treenum(z)$. As shown in
\fig{z_plane}(b), the shifted one-loop amplitude
$A_n^\oneloopnum(z)$ can in general have not just simple physical
poles, but also branch cuts in the complex plane, as well as
double poles~\cite{Bern2005hs}. In addition, it more frequently
has non-vanishing behavior at infinity.


\FIGURE[t]{
\resizebox{1.0\textwidth}{!}{\includegraphics{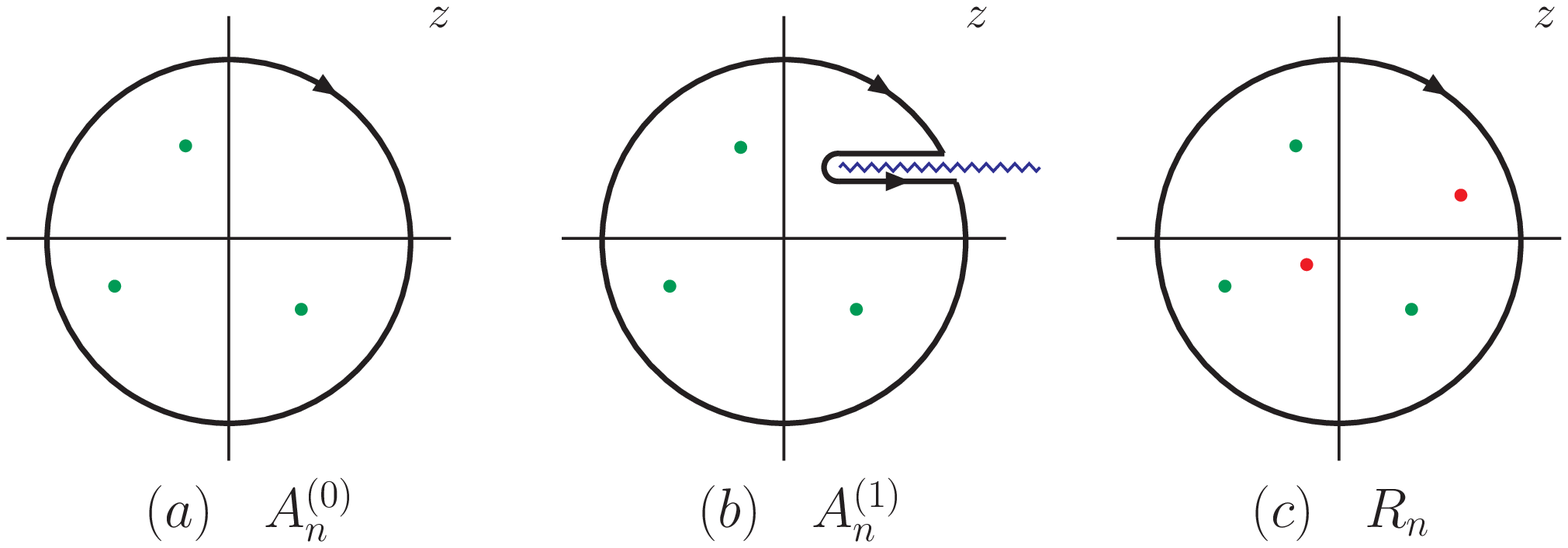}}
\caption{ Analytic structure in the $z$ plane of (a) tree amplitudes
$A_n^\treenum(z)$, (b) one-loop amplitudes $A_n^\oneloopnum(z)$,
and (c) the rational part of one-loop amplitudes $R_n(z)$.
The contour of integration is shown.  The (green) dots
in (a) and (b) represent physical poles.  The additional (red) dots
displayed in (c) represent spurious poles. }
\label{z_plane} }

Branch cuts would result in the need to evaluate
discontinuities along an integration contour, as shown
in \fig{z_plane}(b).
To avoid this, one can use the decomposition~(\ref{CutRational})
of the one-loop amplitude $A_n^\oneloopnum$ into a cut part
$C_n$ and a rational part $R_n$, and then
work with the shifted rational part
$R_n(z)$, instead of analyzing the behavior of the full shifted
one-loop amplitude $A_n^\oneloopnum(z)$.  Although
$R_n(z)$ is a rational function of $z$, only containing poles,
some of these poles are spurious.  The spurious poles are represented by
the additional (red) dots in \fig{z_plane}(c).
Unlike the physical poles, the spurious poles are not associated
with physical factorization channels.
Their contribution to the entire shifted amplitude $A_n^\oneloopnum(z)$
cancels between the shifted cut part $C_n(z)$ and the shifted
rational part $R_n(z)$.

One type of spurious pole arises from the existence of terms
such as $(\ln r)/(1-r)^2$ in $C_n$.  Here $r$ is a ratio of two kinematic
invariants that differ by a single massless leg.  For example,
\eqn{B123Final} displays a factor of $(s_{123}-s_{12})^2$
in the denominator of the coefficient of $\ln(-s_{123})$ in
the primitive amplitude
$A_4^{\mathrm{lc}}(\phi,1_\qb^-,2_q^+,3_\Qb^+,4_Q^-)$.
It corresponds to a term $(\ln r)/(1-r)^2$ with $r=s_{123}/s_{12}$.
Under many choices of shift~(\ref{jlspinorshift}), $r$ will
become a nontrivial function of $z$.  In this case a spurious pole,
located at the solution to $r(z)=1$,
will be generated for $C_n(z)$, and a compensating one for $R_n(z)$.

An analytic method for handling the contributions of spurious
poles was developed in a number of
papers~\cite{Bern2005cq,Berger2006ci,Berger2006vq,Berger2006cz}.
The method has also been applied to Higgs boson
amplitudes~\cite{Badger2007si,Glover2008ffa}. It consists of the
following approach:  We assume that the cut-containing pieces
$C_n$ have been obtained using methods such as those described in
the previous subsection.  Then, for a general shifted one-loop
amplitude we can write
\be
A_n^\oneloopnum(z)\ =\ C_n(z) + R_n(z) \,.
\ee
Our goal is to compute the rational terms $R_n$.
We can absorb spurious singularities
present in $R_n$ into the cut-containing pieces by rewriting
\be
A_n^\oneloopnum(z) = \widehat{C}_n(z) + \widehat{R}_n(z)\, ,
\ee
where the {\it completed-cut terms} $\widehat{C}_n(z)$
are free of spurious singularities, as are the
{\it remaining rational terms} $\widehat{R}_n(z)$.

To absorb all spurious singularities located at solutions
to $r(z)=1$, we make substitutions in $C_n$ of the form
\bea
\frac{\ln r}{(1 - r)^2} &\rightarrow&
\frac{\ln r + 1 - r}{(1 - r)^2}
\ \equiv\ \Ll_1(r) \,,
\label{L1replace} \\
\frac{\ln r}{(1 - r)^3} &\rightarrow&
\frac{\ln r - (r - 1/r)/2}{(1 - r)^3}
\ \equiv\ \Ll_2(r)\,,
\label{L2replace}
\eea
where $r$ represents the ratio of any two kinematic
invariants that differ by a single massless leg.
The amount by which the completed-cut terms have changed in
this process is given by the {\it rational completed-cut terms},
\be
\widehat{CR}_n(z) = \widehat{C}_n(z) - C_n(z)\, . \ee
Consequently,
\be
\widehat{R}_n(z)\ =\ R_n(z) - \widehat{CR}_n(z)\, .
\label{RhatDefinition}
\ee
One can then consider the contour integral at infinity
for $\widehat{R}_n(z)/z$.

Provided that all spurious poles are removed from
$\widehat{R}_n(z)$ by the
substitutions~(\ref{L1replace}) and (\ref{L2replace}),
this contour integral leads to an equation analogous
to the tree-level recursion relation~(\ref{recursion_relation}),
featuring residues only at physical poles,
\be
\widehat{R}_n = \widehat{R}_n(0)
= - \sum_{\mathrm{poles}\,\alpha}
\Res_{z=z_\alpha}\frac{\widehat{R}_n(z)}{z}
\,.
\label{Rhat_recursion_relation}
\ee
These residues can be split into two sets of terms, using
\eqn{RhatDefinition}.  The first set consists of the
{\it recursive diagrams}, associated with residues of $R_n(z)$;
it can be evaluated analogously to the tree-level recursive
diagrams~(\ref{tree_rec_diagrams}):
\bea
R_n^D &\equiv&
 - \sum_{{\rm poles}\ \alpha} \Res_{z=z_\alpha} {R_n(z)\over z}
\nonumber\\
 &=&
 \sum_{h} \sum_{r,s} \Biggl\{
R(k_r,\ldots,\hat k_j,\ldots,k_s,-\Ph^{-h})
\, {i\over P_{r\ldots s}^2} \,
A^\treenum(k_{s+1},\ldots,\hat k_l,\ldots,k_{r-1},\Ph^{h}) \nn\\
&& \null \hskip 0.6cm
+ A^\treenum(k_r,\ldots,\hat k_j,\ldots,k_s,-\Ph^{-h})
 \, {i\over P_{r\ldots s}^2} \,
R(k_{s+1},\ldots,\hat k_l,\ldots,k_{r-1},\Ph^{h}) \nn\\
&& \null \hskip 0.6cm
+ A^\treenum(k_r,\ldots,\hat k_j,\ldots,k_s,-\Ph^{-h})
{i R_\Fact(P_{r\ldots s}^2)\over P_{r\ldots s}^2}
A^\treenum(k_{s+1},\ldots,\hat k_l,\ldots,k_{r-1},\Ph^{h})
\Biggr\}
 \,.  \nn \\
&& \null \label{RationalRecursion}
\eea
The recursive diagrams are computed from the rational parts $R$ of
lower-point loop amplitudes, and lower-point tree amplitudes
$A^\treenum$, as well as the rational part of the factorization
function $R_\Fact$, which only enters for multi-particle
poles~\cite{Bern2005cq} (and not for collinear, two-particle
channels). Just as at tree level, $\phi$ can appear on either side
of the pole.

The second contribution from the physical poles
consists of the overlap terms,
\be O_n =
\sum_{\mathrm{poles}\,\alpha}\Res_{z=z_\alpha}
\frac{\widehat{CR}_n(z)}{z}
\,.
\label{overlap1}
\ee
They correct for the difference between $R_n(z)$ and
$\widehat{R}_n(z)$ in \eqn{RhatDefinition}.
In \sect{SpuriousRationalSubsubsection} we will describe a modification
of this procedure that can be used when the cut-completion described
above fails to remove all spurious poles.

Finally, we have to consider the potential contributions to the
integral from infinity, because
$\widehat{R}_n(z)=A_n^\oneloopnum(z)-\widehat{C}_n(z)$ may not
vanish as $z\to\infty$. In the case of the $H\qb q\Qb Q$
amplitudes there is always a shift that ensures a vanishing
behavior of $\widehat{R}_n(z)$ for large $z$, but there is no
guarantee that this is always the case. In fact, some of the
shifts used to compute the $H\qb qgg$ amplitudes have
non-vanishing large $z$ behavior. Usually it is straightforward to
compute the $z\to\infty$ limit of $A_n(z)$ and $\widehat{C}_n(z)$,
denoted by $\mathrm{Inf}A_n$ and $\mathrm{Inf}\,\widehat{C}_n$
respectively.  In some cases, a pair of shifts is
necessary~\cite{Berger2006ci} (the original shift plus an
auxiliary one), but that was not required for the amplitudes
computed here.  Putting together all the pieces, the full answer
is given by
\be
A_n^\oneloop = \widehat{C}_n + R_n^D + O_n
- \mathrm{Inf}\,\widehat{C}_n + \mathrm{Inf}A_n
\,.
\label{full_oneloop_recursion}
\ee


\subsubsection{Calculation of rational parts for
$A_4^{\mathrm{lc}}(\phi,1_\qb^-,2_q^+,3_\Qb^+,4_Q^-)$}
\label{SampleRationalSubsubsection}

To illustrate the calculation of the rational parts, we
consider the primitive amplitude
$A_4^{\mathrm{lc}}(\phi,1_\qb^-,2_q^+,3_\Qb^+,4_Q^-)$.
After completing the cut terms to form $\widehat{CR}_4$,
the first step is to choose a pair of legs $[j,l\rangle$
to shift according to \eqn{jlspinorshift}.
Then we compute the recursive diagrams $R_4^D$
and overlap terms $O_4$, as well as any contributions
from infinity (Inf terms) under this shift.

For the case of $A_4^{\mathrm{lc}}(\phi,1_\qb^-,2_q^+,3_\Qb^+,4_Q^-)$,
the only parts of the cut terms that need completing,
to remove spurious singularities, are the single log terms.
From the first term in \eqn{B123Final} we see
that the function $\Ll_1(\frac{-s_{123}}{-s_{12}})$,
as defined in \eqn{L1replace},
should be introduced to remove the singularity as $s_{123} \to s_{12}$.
Similarly, the coefficient of $\ln(-s_{234})$ (which
is related by symmetry to that of $\ln(-s_{123})$)
requires the function $\Ll_1(\frac{-s_{234}}{-s_{34}})$.
These functions are collected in \eqn{qqQQ_mppm_lc}.
From the rational parts of the $\Ll_1$ functions
we obtain $\widehat{CR}_4$,
\be
\widehat{CR}_4
= \frac{i}{2}
\frac{\spa3.4\spb2.3^2}{\spb1.2\left( s_{12} - s_{123} \right)} +
\frac{i}{2} \frac{\spa1.2\spb2.3^2}{\spb3.4\left( s_{34} - s_{234}
\right)} \,.
\label{CR_terms}
\ee

At this point there are no spurious poles left in $\widehat{C}_4$,
so we can proceed to choose a complex momentum
shift~(\ref{jlspinorshift}).  For
$A_4^{\mathrm{lc}}(1_\qb^-,2_q^+,3_\Qb^+,4_Q^-)$
we choose the $[4,2\rangle$ shift, namely
\be
\tilde\lambda_4 \rightarrow \tilde\lambda_4 - z\tilde\lambda_2
\,, \qquad\qquad
\lambda_2 \rightarrow \lambda_2 + z\lambda_4 \,,
\ee
or equivalently,
\be
|\widehat{4}] = \left|4\right] - z \left|2\right] \,, \qquad\qquad
|\widehat{2}\rangle = \left|2\right\rangle + z \left|4\right\rangle \,.
\ee
The contribution from infinity vanishes for this shift,
${\rm Inf}\,A_4 = 0$.  This behavior for the full amplitude
can be inferred from the corresponding behavior
of the known one-loop QCD amplitude found by deleting
$\phi$, $A_4^{\mathrm{lc}}(1_\qb^-,2_q^+,3_\Qb^+,4_Q^-)$.
The rational part of this amplitude is a constant times the tree
amplitude~\cite{KST}, and it is easy to see from \eqn{qqQQtrees}
that the tree amplitude vanishes under the $[4,2\rangle$ shift.
Injecting a finite amount of momentum through the field $\phi$
should not affect the large-$z$ behavior.  We confirm this assumption
{\it a posteriori} by checking factorization limits that are independent
of the ones used to construct the recursion relation.
It is also easy to verify that $\widehat{C}_4(z)$ vanishes at infinity;
{\it i.e.}, ${\rm Inf}\,\widehat{C}_4 = 0$.

Next we look at the recursive diagrams $R_4^D$.  The only diagrams that
give non-vanishing contributions are the ones shown in
\fig{RD_example_figure}.  We evaluate each of them separately.


\FIGURE[t]{
\resizebox{0.8\textwidth}{!}{\includegraphics{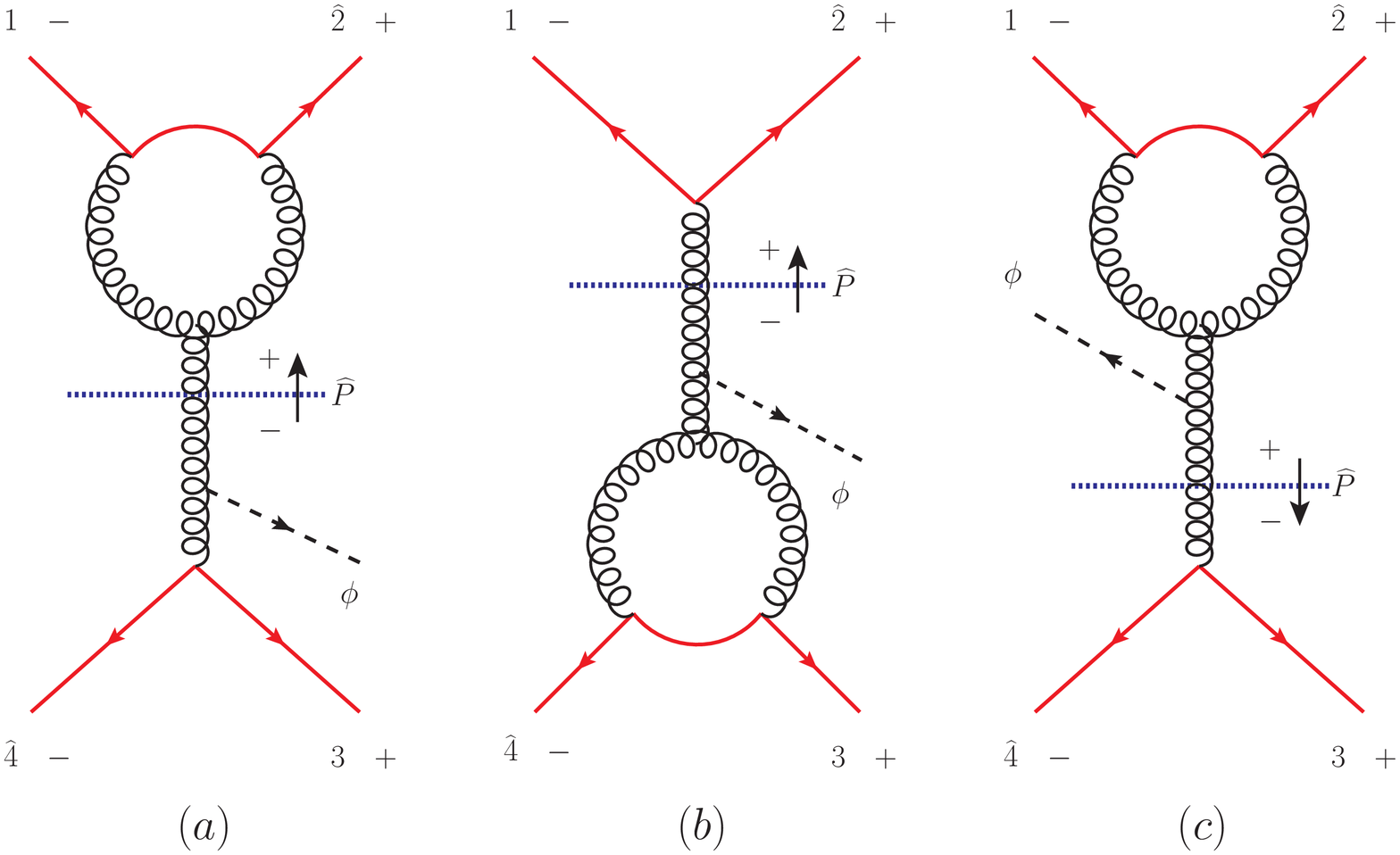}}
\caption{ Diagrams needed to evaluate the recursive diagrams
$R_4^D$ of $A_4^{\mathrm{lc}}(\phi,1_\qb^-,2_q^+,3_\Qb^+,4_Q^-)$.}
\label{RD_example_figure} }

Diagram (a) is given by,
\be
D_4^{(\mathrm{a})} =
R_3(1_\qb^-,\widehat{2}_q^+,-\widehat{P}^+) \,
\frac{i}{s_{12}} \,
A_3^\treenum(\phi,3_\Qb^+,\widehat{4}_Q^-,\widehat{P}^-) \,.
\ee
The tree amplitude
$A_3^\treenum(\phi,3_\Qb^+,\widehat{4}_Q^-,\widehat{P}^-)$
is a simple MHV $\phi$-amplitude.  The loop three-point rational part
$R_3(1_\qb^-,\widehat{2}_q^+,-\widehat{P}^+)$ can be extracted
from the rational part of a one-loop splitting amplitude for
$g \to \qb q$.  It is equal to the MHV tree amplitude
$A_3^\treenum(1_\qb^-,\widehat{2}_q^+,-\widehat{P}^+)$,
multiplied by the ($\Ord(\e^0)$) rational part of
of the loop splitting factor $r_S^{[1]}(\pm,\qb^\mp,q^\pm)$
defined in ref.~\cite{UnitarityMethod},
\be
r_S^{[1]}(\pm,\qb^\mp,q^\pm)\Bigr|_{\rm rat.}\ =\
{83\over18} - {\delta_R\over6} \,.
\ee
Here $\delta_R$ is a regularization-scheme dependent parameter, which fixes
the number of helicity states of the gluons running in the loop to
$(4 - 2 \delta_R \epsilon)$.  For the 't~Hooft-Veltman scheme~\cite{HV}
$\delta_R = 1$, while in the four-dimensional helicity (FDH)
scheme~\cite{BKStringBased,OtherFDH} $\delta_R = 0$.

Thus we get for diagram (a),
\be
D_4^{(\mathrm{a})}
\ =\ \left( - i \frac{[2\,(-\widehat{P})]^2}{\spb1.2} \right)
\left(\frac{83}{18}-\frac{\delta_R}{6}\right) \, \frac{i}{s_{12}}
\, \left( -i \frac{\langle\widehat{P}\,4\rangle^2}{\spa3.4} \right)
\,.
\ee
To remove the dependence on $\widehat{P}$ we use
the on-shell condition,
\be
\langle1\,\widehat{2}\rangle = 0
\quad \Leftrightarrow \quad
\spa1.2 + z\spa1.4 = 0
\quad \Leftrightarrow \quad
z = -\frac{\spa1.2}{\spa1.4} \,,
\ee
and
\be
\widehat{P}\ =\
|1]\langle1| + |\widehat{2}]\langle\widehat{2}|
\ =\ |1]\langle1| + |2]\langle2| + z|2]\langle4]
\ =\ P + z|2]\langle4] \,,
\ee
plus some simple spinor product algebra, to get
\be
D_4^{(\mathrm{a})}
\ =\ - i \frac{\spa1.4^2}{\spa1.2\spa3.4}
\left(\frac{83}{18} - \frac{\delta_R}{6}\right)
\ =\ A_4^{\treenum}(\phi,1_\qb^-,2_q^+,3_\Qb^+,4_Q^-)
 \times\left(\frac{83}{18} - \frac{\delta_R}{6}\right).
\ee

For diagram (b) we have the same on-shell condition as for (a). We
also need the rational part of the leading-color $\phi\qb qg$
amplitude $A_3^\oneloopnum(\phi,1_\qb^-,2_q^+,3^-)$. This can be
extracted from the $H\qb qg$ amplitude~\cite{Schmidt1997wr} and
the finite amplitude
$A_3^\oneloopnum(\phi,1_\qb^-,2_q^+,3^+)$~\cite{Berger2006sh}.
Then a calculation very similar to that for diagram (a) yields
\bea
D_4^{(\mathrm{b})} &=&
A_3^\treenum(1_\qb^-,\widehat{2}_q^+,-\widehat{P}^+) \,
\frac{i}{s_{12}} \,
R_3(\phi,3_\Qb^+,\widehat{4}_Q^-,\widehat{P}^-)
\nonumber\\
&=&  \left( - i \frac{[2\,(-\widehat{P})]^2}{\spb1.2} \right)
\, \frac{i}{s_{12}} \,
\left( - i \frac{\langle\widehat{P}\,4\rangle^2}{\spa3.4} \right)
\left(2 + \frac{83}{18} - \frac{\delta_R}{6}\right) \nonumber\\
&=& A_4^\treenum(\phi,1_\qb^-,2_q^+,3_\Qb^+,4_Q^-)
\times \left(\frac{119}{18} - \frac{\delta_R}{6}\right).
\eea

For diagram (c), the required one-loop $\phi$ amplitude is
$A_3^\oneloopnum(\phi,1_\qb^-,2_q^+,3^+)$~\cite{Berger2006sh}. The
on-shell condition becomes
\be [3\,\widehat{4}] = 0
\quad \Leftrightarrow \quad
\spb3.4 - z\spb3.2 = 0
\quad \Leftrightarrow \quad
z = \frac{\spb3.4}{\spb3.2}
\label{onshell34}
\ee
and
\be \widehat{P}
\ =\ |3]\langle3| + |\widehat{4}]\langle\widehat{4}|
\ =\ |3]\langle3| + |4]\langle4| - z|2]\langle4]
\ =\ P - z|2]\langle4] \,.
\ee
The diagram evaluates to
\bea
D_4^{(\mathrm{c})} &=&
R_3(\phi,1_\qb^-,\widehat{2}_q^+,\widehat{P}^+) \,
\frac{i}{s_{34}} \,
A_3^\treenum(3_\Qb^+,\widehat{4}_Q^-,-\widehat{P}^-)
\nonumber\\
&=&  \left( - i \frac{[2\,\widehat{P}]^2}{\spb1.2} \right)
\left( - 2
- \frac{1}{2} \frac{s_{1\widehat{2}}}{s_{\widehat{2}\widehat{P}}}
\right) \,
\frac{i}{s_{34}} \,
\left( -i \frac{\langle(-\widehat{P})\,4\rangle^2}{\spa3.4} \right)
\,.
\eea
By substituting $z$ with its value given by the on-shell condition
for $\widehat{P}$, \eqn{onshell34},
and using the Schouten identity, we find that
\bea
D_4^{(\mathrm{c})} &=& - i \frac{\spb2.3^2}
{\spb1.2\spb3.4} \,
\left( - 2 + \frac{1}{2}
\frac{\spb1.2\left\langle1|(2+4)|3\right]}{\spb2.3 s_{234}} \right)
\nonumber\\
&=& A_4^\treenum(\phi^\dagger,1_\qb^-,2_q^+,3_\Qb^+,4_Q^-) \,
\left( - 2 + \frac{1}{2}
\frac{\spb1.2\left\langle1|(2+4)|3\right]}{\spb2.3 s_{234}} \right).
\eea

In principle, there could be recursive diagrams associated with
the $s_{123}$ and $s_{341}$ channels.  However, these diagrams
vanish because on one side of the pole is a $\phi\qb q$ amplitude.
The amplitude ${\cal A}_2(\phi,1_\qb^-,2_q^+)$ vanishes by
angular momentum conservation, while the amplitude
${\cal A}_2(\phi,1_\qb^+,2_q^+)$ vanishes because the quarks are massless
and interact only with gluons, via chirality-preserving interactions.
Similarly, the $s_{23}$ and $s_{41}$ poles are absent
because there is no three-point amplitude containing two different
flavor quarks.  The sum of the recursive diagrams is
\be
R_4^D = D_4^{(\mathrm{a})} + D_4^{(\mathrm{b})}
+ D_4^{(\mathrm{c})}.
\label{RD_terms}
\ee

Next we evaluate the overlap terms $O_4$. They are given by
\be O_4 = \sum_{\mathrm{poles}\,\,\alpha}
\mathop{\mathrm{Res}}_{z=z_\alpha} \frac{\widehat{CR}_4(z)}{z}\, ,
\ee
where the sum is only over the physical poles.  In our
case, physical poles can arise only when the following
intermediate momenta go on shell (the same channels that admit
possible recursive diagrams):
\bea \widehat{P}_{12}^2 = 0 \ \ &\Leftrightarrow&\ \ z =
-\frac{\spa1.2}{\spa1.4} \,, \ \quad\quad\quad\quad
\widehat{P}_{23}^2 = 0 \ \ \Leftrightarrow\ \ z =
\frac{\spa2.3}{\spa3.4}\,,
\\
\widehat{P}_{34}^2 = 0 \ \ &\Leftrightarrow&\ \ z =
\frac{\spb3.4}{\spb3.2}\,, \ \qquad\quad\quad\quad
\widehat{P}_{41}^2 = 0 \ \ \Leftrightarrow\ \ z =
\frac{\spb1.4}{\spb1.2}\,,
\\
\widehat{P}_{123}^2 = 0 \ \ &\Leftrightarrow&\ \ z = -
\frac{s_{123}}{\left\langle4|(1+3)|2\right]} \,, \quad
\widehat{P}_{341}^2 = 0 \ \ \Leftrightarrow\ \ z =
\frac{s_{341}}{\left\langle4|(1+3)|2\right]} \,. \eea
However, we note that the $s_{23}$, $s_{41}$, $s_{123}$ and $s_{341}$
channels had no recursive diagrams.  This does not necessarily
imply the absence of an overlap diagram (in principle $\widehat{CR}_4$
could have a worse behavior than $R_4$ in a given channel),
but it is easy to check that \eqn{CR_terms} has no poles
in these channels.


\FIGURE[t]{
\resizebox{0.95\textwidth}{!}{\includegraphics{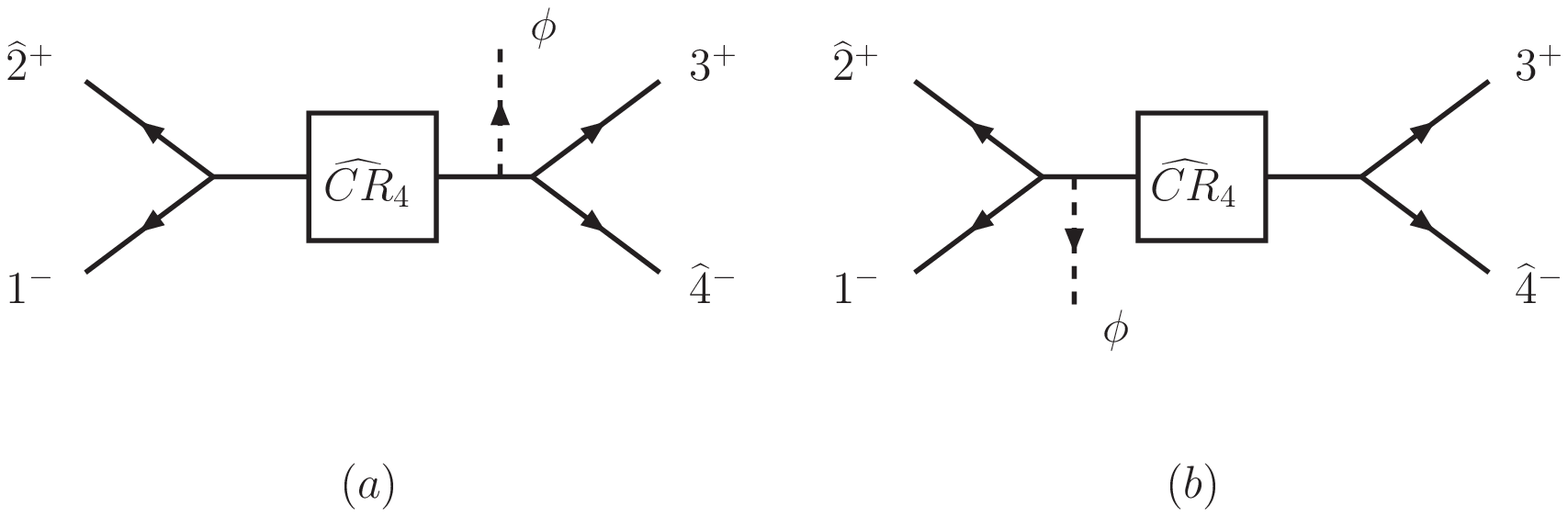}}
\caption{ Schematic diagrams corresponding to the overlap terms
$O_4$ of $A_4^{\mathrm{lc}}(\phi,1_\qb^-,2_q^+,3_\Qb^+,4_Q^-)$.
The diagrams are computed from the residues of $\widehat{CR}_4(z)/z$
at poles in the $z$ plane satisfying (a) $\widehat{P}_{12}^2=0$ and
(b) $\widehat{P}_{34}^2=0$. }
\label{O_example_figure} }

The two remaining cases correspond schematically to diagrams (a) and
(b) shown in \fig{O_example_figure}. We write
\be O_4 = O_4^{(\mathrm{a})} + O_4^{(\mathrm{b})} \,,
\ee
where diagram (a) corresponds to $\widehat{P}_{12}^2=0$
and diagram (b) to $\widehat{P}_{34}^2=0$.
It is simple to see from the $[4,2\rangle$ shift of $\widehat{CR}_4$
in \eqn{CR_terms} that $O_4^{(\mathrm{a})}$ also vanishes,
because there is no factor of $\spa1.2$ in the denominator of $\widehat{CR}_4$.
Thus the only non-vanishing overlap contribution comes from
$O_4^{(\mathrm{b})}$, due to the factor of $\spb3.4$ in the denominator
of $\widehat{CR}_4$. The residue is easily found to be
\be O_4 = O_4^{(\mathrm{b})} = \frac{i}{2}
\frac{\left\langle1|(2+4)|3\right]\spb2.3}{\spb3.4s_{234}}\, .
\label{overlap_terms} \ee

We are finally ready to assemble the remaining rational terms of
$A_4^{\mathrm{lc}}(\phi,1_\qb^-,2_q^+,3_\Qb^+,4_Q^-)$, given by
the sum
\be
\widehat{R}_4 = R_4^D + O_4 \,.
\label{rational_sum}
\ee
Using \eqns{RD_terms}{overlap_terms}, we get
\be
\widehat{R}_4 = A_4^{\treenum}(\phi,1_\qb^-,2_q^+,3_\Qb^+,4_Q^-)
 \times\left( \frac{101}{9} - \frac{\delta_R}{3} \right)
- 2 \, A_4^\treenum(\phi^\dagger,1_\qb^-,2_q^+,3_\Qb^+,4_Q^-) \,,
\label{rational_sum_final}
\ee
which is also recorded in \eqn{qqQQ_mppm_lc}.  Note that the
term with the unphysical pole in $s_{234}$ in the recursive
diagram $D_4^{(\mathrm{c})}$
cancels against the overlap diagram $O_4^{(\mathrm{b})}$
for the same channel.  Such recursive-overlap cancellations
are a common feature.


\subsubsection{Spurious poles in rational parts}
\label{SpuriousRationalSubsubsection}

The cut-completion process described in
\eqns{L1replace}{L2replace} removes certain types of spurious
singularities, namely those associated with terms like $(\ln
r)/(1-r)^n$ for $n>1$, where $r$ is a ratio of two momentum
invariants that differ by one massless external leg.  It ensures
that $\widehat{C}_n(z)$, and therefore $\widehat{R}_n(z)$, is free
of spurious singularities when $r(z)\rightarrow 1$, where $r(z)$
is the shifted value of $r$. In
refs.~\cite{Bern2005cq,Berger2006ci,Berger2006vq,Badger2007si} it
was found that this type of cut completion was sufficient to
remove all spurious poles in the $z$ plane for a large class of
QCD and $\phi$ amplitudes.  (Note that if a spurious denominator
factor is unaffected by the particular shift used,
\eqn{jlspinorshift}, then it will not produce a spurious pole in
the $z$ plane.)  However, poles of the type $r(z) = 1$ certainly
do not exhaust the set of potential spurious poles for a general
amplitude.  These poles correspond specifically to Gram
determinants associated with two-mass triangle integrals. In a
general amplitude, there are also poles associated with the Gram
determinants of a multitude of scalar box and triangle integrals
with varying numbers of external masses.  It would be very
difficult to construct a cut completion $\widehat{C}_n$ that
removed all spurious poles.

On the other hand, a completely general alternative method for
handling the spurious poles was sketched in
ref.~\cite{Bern2008ef}, and fully implemented numerically in the
\BlackHat\ program~\cite{Berger2008sj}. This method did not use
cut completion, but rather the original decomposition
$A_n^\oneloopnum=C_n+R_n$, and instead relied on the fact that the
residues of $C_n(z)/z$ and $R_n(z)/z$ cancel at every spurious
pole.  The contour integral of $R_n(z)/z$ at infinity requires the
sum over spurious pole residues of $R_n(z)/z$, but this sum can be
evaluated using the shifted cut part $C_n(z)$, as
\be
 - \sum_{{\rm spurious\ poles}\ \beta} \Res_{z=z_\beta} {R_n(z)\over z}
= \sum_{{\rm spurious\ poles}\ \beta} \Res_{z=z_\beta} {C_n(z)\over z} \,.
\label{spurpoleres}
\ee

Here we simply note that a hybrid approach is also feasible:
First one removes the spurious poles that can be easily removed
by cut completion, such as \eqns{L1replace}{L2replace}.
This procedure leads to the overlap terms~(\ref{overlap1})
in the usual way.  Then one considers the contour integral
at infinity of the remaining rational terms $\widehat{R}_n(z)/z$.
One evaluates the sum over residues of $\widehat{R}_n(z)/z$,
at the spurious poles $z_\gamma$
that were {\it not} removed by cut completion,
by using the fact that their residues still cancel against those of
the completed cut terms, $\widehat{C}_n(z)/z$:
\be
 - \sum_{{\rm spurious\ poles}\ \gamma}
   \Res_{z=z_\gamma} {\widehat{R}_n(z)\over z}
= \sum_{{\rm spurious\ poles}\ \gamma}
   \Res_{z=z_\gamma} {\widehat{C}_n(z)\over z} \,,
\label{completedspurpoleres}
\ee
where $\{ \gamma \}$ is a subset of $\{ \beta \}$.

Note that $\widehat{C}_n(z)$ in \eqn{completedspurpoleres}
includes rational terms as well as cut terms;
whereas $C_n(z)$ in \eqn{spurpoleres} is a pure cut function,
which nevertheless can have rational-function spurious-pole
residues after Taylor expansion around the pole.
In fact, it is only the rational-function part of the residue of
$\widehat{C}_n(z)/z$ that we require; the terms containing
logarithms, polylogarithms and $\pi^2$ factors are guaranteed
to cancel, because the residue of the rational function
$\widehat{R}_n(z)/z$ can have no such terms.

In our calculation of the $\phi\qb qgg$ amplitudes, after removing
the $r(z)=1$ spurious poles, we found that certain spurious poles
still remained, due to denominator factors in integral
coefficients of the form $\spa{i}.{j}$ or $\spb{i}.{j}$, where $i$
and $j$ are color {\it non-adjacent} legs. If $i$ and $j$ are
color-adjacent, then $\spa{i}.{j}$ and $\spb{i}.{j}$ denominator
factors represent physical singularities, corresponding (for real
momenta) to the region where $k_i$ and $k_j$ become collinear,
$s_{ij} \to 0$. However, physical collinear poles for
color-ordered primitive amplitudes only occur when $i$ and $j$ are
color-adjacent; in the non-adjacent case, the $\spa{i}.{j}$ and
$\spb{i}.{j}$ factors generate spurious singularities. They are
associated with the ``easy two mass'' box integral in which the
two diagonally opposite massless momenta are $k_i$ and $k_j$,
whose Gram determinant contains a factor of $s_{ij}=
\spa{i}.{j}\spb{j}.{i}$. The factors $\spa{i}.{j}$ and
$\spb{i}.{j}$ can also be seen in the denominators of the
solutions for the loop momenta in the quadruple cut for the
easy-two-mass box kinematics (see {\it e.g.} eq.~(2.7) of
ref.~\cite{Berger2008sj}). The same factors also persist in the
limiting case of a one-mass box.

Factors of $\spa{i}.{j}$ or $\spb{i}.{j}$ do appear in the
denominators of coefficients of scalar integrals in the QCD and
$\phi$ MHV $n$-gluon amplitudes with non-adjacent
negative-helicity gluons labeled $i$ and $j$, whose rational parts
were computed using on-shell recursion
relations~\cite{Berger2006vq,Glover2008ffa}. However, in these
cases the $[i,j\rangle$ shift was used, which leaves both
$\spa{i}.{j}$ and $\spb{i}.{j}$ unshifted, and therefore produces
no spurious pole in this channel.

In this work we encountered spurious poles associated with $\spa{i}.{j}$
factors in several $\phi\qb qgg$ amplitudes.  Here we illustrate
how to use the method described above for the specific example of the
leading-color primitive amplitude $A_4^L(\phi,1_\qb^-,2_q^+,3^+,4^-)$.
Using the methods of sections
\ref{GenUSubsection}--\ref{BubSubsection}, we have calculated the
cut-containing parts of $A_4^L(\phi,1_\qb^-,2_q^+,3^+,4^-)$, and
obtained the completed-cut terms $\widehat{C}_4$, which can be read
off \eqn{qqgg_mppm_L} by ignoring the purely rational terms in
the last few lines.  To compute the rational part of this amplitude
we chose to use a $\left[4,1\right\rangle$ shift.
Inspecting $\widehat{C}_4$, we see that several terms contain the
spurious denominator factor $\spa1.3$, which will potentially
lead to a spurious pole under the $\left[4,1\right\rangle$ shift.
These terms are,
\bea
&&i\,\frac{\spa1.4^3}{\spa1.2\spa3.4\spa1.3}\Big[
  \Ls_{-1}\left(s_{12},s_{23};s_{123}\right)
+ \Ls_{-1}\left(s_{34},s_{41};s_{341}\right)\Big] \nonumber \\
&&\null - \frac{i}{3}\frac{\spa1.2^2\spb2.3^3\spa3.4^2}{\spa1.3}
\frac{\Ll_2\left(\frac{-s_{123}}{-s_{12}}\right)}{s_{12}^3} +
\frac{i}{2}\frac{\spa1.2\spa3.4\spb2.3^2\spa1.4}{\spa1.3}
\frac{\Ll_1\left(\frac{-s_{123}}{-s_{12}}\right)}{s_{12}^2} \,.
\label{phiqqggmppmspur}
\eea
The spurious pole satisfies
$0 = \langle\widehat{1}\,3\rangle
   = \spa1.3 + z \spa4.3$,
or $z = \spa1.3/\spa3.4 \equiv z_{\mathrm{sp}}$.

Using \eqn{completedspurpoleres}, we need to compute
\be
\mathop{\mathrm{Res}}_{z=z_{\mathrm{sp}}}\frac{\widehat{C}_4(z)}{z} \,.
\ee
Now the $\Ls_{-1}$ functions in \eqn{phiqqggmppmspur} actually
vanish as $z\to z_{\mathrm{sp}}$.  This is because the relevant
scalar box integral in $D=6$ dimensions, which is nonsingular
as $s_{13} \to 0$, can be written as
\be
{\cal I}_4^{D=6}(s_{12},s_{23};s_{123})
= - i \, \cg \, \frac{\Ls_{-1}(s_{12},s_{23};s_{123})}{s_{13}} \,,
\label{Deq6integral}
\ee
and similarly for the other $\Ls_{-1}$ function.
Because ${\cal I}_4^{D=6}$ is smooth in this limit, the $\Ls_{-1}$
functions must contain a factor of $s_{13} = \spa1.3\spb3.1$
in the limit $s_{13} \to 0$. Thus the terms containing the
$\Ls_{-1}$ functions in \eqn{phiqqggmppmspur} do not contribute
to the residue.

After expanding the remaining logarithms and rational terms
in \eqn{phiqqggmppmspur} around $z=z_{\mathrm{sp}}$, we find
that the logarithmic part of the residue cancels, as expected.
Keeping the rational part of the residue, and simplifying, we get
\be
\mathop{\mathrm{Res}}_{z=z_{\mathrm{sp}}}\frac{\widehat{C}_4(z)}{z}
= -
\frac{i}{6}\frac{\spa3.4\spa1.4^2\spb2.3}{\spa2.3\spa1.3\langle
4|(1+3)|2]} +
i \, \frac{2}{3}\frac{\spa1.4\spb2.3\spa3.4}{\spb1.2\spa2.3\spa1.3}\, .
\ee
This term has to be added to the recursive diagrams, overlap terms
and $\widehat{CR}_4$ to complete the full rational terms of
$A_4^L(\phi,1_\qb^-,2_q^+,3^+,4^-)$. The full rational terms, as
well as the full amplitude, are now free of spurious singularities
as $\spa1.3 \rightarrow 0$.

The procedure outlined above can be performed in a systematic way
for any amplitude, since the locations of the possible spurious
poles under a chosen shift are known {\it a priori}, or they can be
inferred simply by inspecting the completed-cut terms,
$\widehat{C}_n$. Whenever, after absorbing spurious singularities
according to \eqns{L1replace}{L2replace}, we are left with
residual spurious poles, we can always compute their contribution
to the remaining rational terms by evaluating the corresponding
residues of $\widehat{C}_n(z)/z$ instead.


\section{The one-loop $H\qb q\Qb Q$ and $H\qb qgg$ amplitudes}
\label{ResultsSection}

In this section we present our main results for the one-loop $H\qb
q \Qb Q$ and $H\qb qg^\pm g^\mp$ amplitudes. First we outline how
to obtain all primitive $\phi$-amplitudes, using only a minimum
set of them. Then we give the full analytic expressions for these
amplitudes, followed by numerical results at a specific kinematic
point. We then show how to obtain the color- and helicity-summed
cross section for a pseudoscalar Higgs plus two quarks and two
gluons, using our results and those of ref.~\cite{Ellis2005qe}. As
another application, we show how to compute part of the virtual
one-loop color-singlet interference term between the gluon-fusion
and VBF Higgs production mechanisms. Finally, we mention the
various consistency checks we used to verify the correctness of
our expressions.


\subsection{Preliminaries}
\label{H4p_preliminaries}

We obtain the one-loop corrections to
${\cal A}_4(H,1_\qb,2_q,3_\Qb,4_Q)$ and
${\cal A}_4(H,1_\qb,2_q,3,4)$ by computing color-ordered
primitive $\phi$-amplitudes in a helicity basis,
following our discussion in \sect{ColorDecompSubsection}.
Once we have the complete set of $\phi$-amplitudes,
the $\phi^\dagger$-amplitudes are obtained by
parity, \eqn{phiphidaggerparity}.

Consider first ${\cal A}_4(\phi,1_\qb,2_q,3_\Qb,4_Q)$.
Because the quarks are massless, chirality is preserved along
a quark line.  By convention, all external legs are outgoing,
so the helicities of any quark-antiquark pair have to be opposite.
Thus we need only consider the four helicity configurations
${\cal A}_4(\phi,1_\qb^{-\lambda},2_q^\lambda,
3_\Qb^{-\Lambda},4_Q^\Lambda)$,
where $\lambda, \Lambda=\pm$ are the helicities of the $q$ and $Q$
quarks, respectively.

Suppose the anti-quark $\bar{q}$ (leg 1) has positive helicity.
We can obtain these cases from the cases where it has negative
helicity by using charge conjugation, which reverses both quark
lines ($q\lr\bar{q}$ and $Q\lr\bar{Q}$):
\bea
{\cal A}_4(\phi,1_\qb^+,2_q^-,3_\Qb^-,4_Q^+) &=&
{\cal A}_4(\phi,2_\qb^-,1_q^+,4_\Qb^+,3_Q^-)\,, \label{phi+--+} \\
{\cal A}_4(\phi,1_\qb^+,2_q^-,3_\Qb^+,4_Q^-) &=&
{\cal A}_4(\phi,2_\qb^-,1_q^+,4_\Qb^-,3_Q^+) \, . \label{phi+-+-}
\eea
Now taking the anti-quark $\bar{q}$ to have negative helicity,
we see that there are two independent
helicity configurations that we need to compute,
\be
{\cal A}_4(\phi,1_\qb^-,2_q^+,3_\Qb^+,4_Q^-)
\qquad\mathrm{and}\qquad
{\cal A}_4(\phi,1_\qb^-,2_q^+,3_\Qb^-,4_Q^+)\,.
\ee
Here ${\cal A}_4$ is shorthand for the three types of primitive
amplitude in this case (lc, slc, and $\fl$).

Recall from \eqn{phiphidaggerparity} that parity gives the
$\phi^\dagger$-amplitudes in terms of the $\phi$-amplitudes,
\be
{\cal A}_4(\phi^\dagger,1_\qb^\lambda,2_q^{-\lambda},
3_\Qb^\Lambda,4_Q^{-\Lambda})
=
\Bigl[
{\cal A}_4(\phi,1_\qb^{-\lambda},2_q^\lambda,3_\Qb^{-\Lambda},4_Q^\Lambda)
\Bigr] \bigg|_{\spa{i}.{j} \lr \spb{j}.{i} }
\,,
\label{phi2phidagger}
\ee
where the operation $\spa{i}.{j} \lr \spb{j}.{i}$ conjugates spinors
but does not reverse the sign of absorptive parts of loop integrals.

For the $H\qb qgg$ amplitude, there are two cases to consider,
depending on whether the helicities of the two gluons are the same
or opposite.  In the case that they are the same, say both positive,
we have, using the decomposition~(\ref{Hreconstruct}) and parity,
\be
{\cal A}_4(H,1_\qb^-,2_q^+,3^+,4^+)
= {\cal A}_4(\phi,1_\qb^-,2_q^+,3^+,4^+)
- \Bigl[ {\cal A}_4(\phi,1_\qb^+,2_q^-,3^-,4^-)
\Bigr] \bigg|_{\spa{i}.{j} \lr \spb{j}.{i} }
\,.
\label{Hqqggbothplus}
\ee
The amplitude ${\cal A}_4(\phi,1_\qb^-,2_q^+,3^+,4^+)$ vanishes at
tree level. For this reason, the one-loop amplitude is quite
simple~\cite{Berger2006sh} and is given below in
eqs.~(\ref{qqgg_mppp_L}), (\ref{qqgg_mppp_R}) and
(\ref{qqgg_mppp_fl}). However, the amplitude
$A_4(\phi,1_\qb^+,2_q^-,3^-,4^-)$ is next-to-maximally-helicity
violating (NMHV), and at one-loop it is considerably more complex.
(For example, the coefficients of the three-mass triangle
integrals are nonzero for this amplitude.) We will leave its
analytic computation for future work.

Instead we turn to the case of opposite-helicity gluons, which can
be decomposed as
\be 
{\cal A}_4(H,1_\qb^-,2_q^+,3^\pm,4^\mp)
= {\cal A}_4(\phi,1_\qb^-,2_q^+,3^\pm,4^\mp)
- \Bigl[ {\cal A}_4(\phi,1_\qb^+,2_q^-,3^\mp,4^\pm)
\Bigr] \bigg|_{\spa{i}.{j} \lr \spb{j}.{i} }
 \,.
\label{Hqqggopposite}
\ee
In this case the $\phi$ amplitudes are both MHV, and of a similar
complexity as the $\phi\qb q \Qb Q$ amplitudes.  Again, using
charge conjugation we can exchange the roles of anti-quark and quark,
so as to obtain the remaining $\phi$-amplitude helicity configurations,
in which the anti-quark $\bar{q}$ has positive helicity,
\be
 {\cal A}_4(1_{\bar{q}}^+,2_q^-,3^\pm,4^\mp)
= {\cal A}_4(2_{\bar{q}}^-,1_q^+,4^\mp,3^\pm).
\label{qqggchargeconj}
\ee

Using the color decompositions in \sect{ColorDecompSubsection},
the problem is reduced to computing the primitive amplitudes
$A_4^{\mathrm{lc}}$, $A_4^{\mathrm{slc}}$ and $A_4^{\fl}$ for
two four-quark helicity configurations, $1_\qb^- 2_q^+ 3_\Qb^\pm 4_Q^\mp$,
and $A_4^L$, $A_4^R$ and $A_4^\fl$ for two two-quark-two-gluon
helicity configurations, and two color orderings, namely
$1_\qb^- 2_q^+ 3^\pm 4^\mp$ and $1_\qb^- 3^\pm 2_q^+  4^\mp$.


\subsection{Full results}
\label{answers}

In section \ref{BootstrapSection} we showed in specific examples
how to compute various ingredients necessary to obtain the full
$A_4^{\mathrm{lc}}(\phi,1_\qb^-,2_q^+,3_\Qb^+,4_Q^-)$ and
$A_4^L(\phi,1_\qb^-,2_q^+,3^+,4^-)$ amplitudes.  We used the same
techniques for the quadruple cuts and ordinary two-particle cuts
in all channels, and for all other color components and
independent helicity configurations, in order to arrive at the full
results for the $\phi$-amplitudes.

It is worth noting that the computation of the $\phi\qb q\Qb Q$
primitive amplitudes in both helicity configurations was
significantly simpler than that of the $\phi\qb qgg$ amplitudes.
The expressions were more compact at each stage (due in part to the
higher symmetry of these amplitudes).
In addition, we had no remaining spurious poles for the shifts we
chose to perform.  Specifically, we used a
$\left[4,2\right\rangle$ shift for all the
$A_4(\phi,1_\qb^-,2_q^+,3_\Qb^+,4_Q^-)$ primitive amplitudes,
and encountered no contributions from $z\to\infty$ ({\it i.e.},
$\widehat{C}_4(z)$ as well as $A_4^\oneloopnum(z)$ vanish in this limit).
We used a $\left[1,3\right\rangle$ shift for
$A_4^{\mathrm{lc}}(\phi,1_\qb^-,2_q^+,3_\Qb^-,4_Q^+)$.
Here there was a contribution from $z\to\infty$, from
$\widehat{C}_4$.  We used a symmetry to obtain the slc and $\fl$
components of this helicity configuration from the previous one.

The $\phi\qb qgg$ amplitudes were more intricate.
For $A_4(\phi,1_\qb^-,2_q^+,3^+,4^-)$, we used the
$\left[4,1\right\rangle$ shift, and we had to compute residues of
spurious poles in the $L$, $R$ and fermion loop ($\fl$) amplitudes,
although there were no contributions from $z\to\infty$.
For $A_4(\phi,1_\qb^-,2_q^+,3^-,4^+)$, we used the
$\left[3,2\right\rangle$ shift.  There were not only spurious pole
residues in the $L$ and $R$ components,
but also a contribution from $z\to\infty$ (from $\widehat{C}_4$)
in the $L$ component.  We cross-checked our results for
the $L$ and $R$ components of $A_4(\phi,1_\qb^-,2_q^+,3^-,4^+)$
using the $\left[2,4\right\rangle$ shift.

Finally, for the $\phi\qb gqg$ $L$ amplitudes, we used a
$\left[4,1\right\rangle$ shift for
$A_4^L(\phi,1_\qb^-,2^+,3_q^+,4^-)$, and a $\left[2,1\right\rangle$
shift for $A_4^L(\phi,1_\qb^-,2^-,3_q^+,4^+)$.  There were neither
spurious pole contributions, nor contributions from $z\to\infty$.
The full results, after assembly and simplification, are presented below.


\subsubsection{$\boldsymbol{{\phi}\qb q\Qb Q}$}

For the $\phi\qb q\Qb Q$ $({-}{+}{+}{-})$ configuration we have
\bea
&&\null\hskip-0.3cm -i A_4^{\mathrm{lc}}(\phi,1_\qb^-,2_q^+,3_\Qb^+,4_Q^-)
\ =\
-i A_4^\treenum(\phi,1_\qb^-,2_q^+,3_\Qb^+,4_Q^-) \times V^{\mathrm{lc}}
\nonumber\\  && \hskip-0.3cm \null
- \frac{1}{2} \spa1.2\spa3.4\spb2.3^2
\left[\frac{\Ll_1\left(\frac{-s_{123}}{-s_{12}}\right)}{s_{12}^2}
+ \frac{\Ll_1\left(\frac{-s_{234}}{-s_{34}}\right)}{s_{34}^2}\right]
- 2 \spa1.4\spb2.3
\left[\frac{\Ll_0\left(\frac{-s_{123}}{-s_{12}}\right)}{s_{12}}
+ \frac{\Ll_0\left(\frac{-s_{234}}{-s_{34}}\right)}{s_{34}}\right]
\nonumber\\ && \hskip0.0cm \null
+ 2i \, A_4^\treenum(\phi^\dagger,1_\qb^-,2_q^+,3_\Qb^+,4_Q^-)\, ,
\label{qqQQ_mppm_lc}
\eea
with
\bea
V^{\mathrm{lc}} &=& - \frac{1}{\e^2} \left[
\left(\frac{\mu^2}{-s_{23}}\right)^\e +
\left(\frac{\mu^2}{-s_{41}}\right)^\e\right] + \frac{13}{6\e}\left[
\left(\frac{\mu^2}{-s_{12}}\right)^\e +
\left(\frac{\mu^2}{-s_{34}}\right)^\e \right]
\nonumber\\ && \hskip0.0cm \null
- \Ls_{-1}^{2{\rm m}e}\left(s_{123},s_{234};s_{23},m_H^2\right)
- \Ls_{-1}^{2{\rm m}e}\left(s_{341},s_{412};s_{41},m_H^2\right)
\nonumber\\ && \hskip0.0cm \null
- \Ls_{-1}\left(s_{23},s_{34};s_{234}\right)
- \Ls_{-1}\left(s_{34},s_{41};s_{341}\right)
\nonumber\\ && \hskip0.0cm \null
- \Ls_{-1}\left(s_{41},s_{12};s_{412}\right)
- \Ls_{-1}\left(s_{12},s_{23};s_{123}\right)
+ \frac{101}{9} - \frac{\delta_R}{3} \,,
\label{V_lc}
\eea
\bea
&&\null -i A_4^{\mathrm{slc}}(\phi,1_\qb^-,2_q^+,3_\Qb^+,4_Q^-)
\ =\ -i A_4^\treenum(\phi,1_\qb^-,2_q^+,3_\Qb^+,4_Q^-)\times V^{\mathrm{slc}}
\nn\\  && \hskip0.0cm \null
+ \frac{1}{2} \spa1.2 \! \spa3.4 \! \spb2.3^2
\left[\frac{\Ll_1\left(\frac{-s_{123}}{-s_{12}}\right)}{s_{12}^2}
    \! + \! \frac{\Ll_1\left(\frac{-s_{234}}{-s_{34}}\right)}{s_{34}^2}\right]
- \spa1.4 \! \spb2.3
\left[\frac{\Ll_0\left(\frac{-s_{123}}{-s_{12}}\right)}{s_{12}}
    \! + \! \frac{\Ll_0\left(\frac{-s_{234}}{-s_{34}}\right)}{s_{34}}\right]
\,,
\nn\\  && \hskip0.0cm \null{~}
\label{qqQQ_mppm_slc}
\eea
with
\bea
V^{\mathrm{slc}}
&=& - \frac{1}{\e^2} \left[
\left(\frac{\mu^2}{-s_{12}}\right)^\e +
\left(\frac{\mu^2}{-s_{34}}\right)^\e\right] - \frac{3}{2\e}\left[
\left(\frac{\mu^2}{-s_{12}}\right)^\e +
\left(\frac{\mu^2}{-s_{34}}\right)^\e \right]
\nonumber\\  && \hskip0.0cm \null
- \Ls_{-1}^{2{\rm m}e}\left(s_{412},s_{123};s_{12},m_H^2\right)
- \Ls_{-1}^{2{\rm m}e}\left(s_{234},s_{341};s_{34},m_H^2\right)
- 7 - \delta_R \,,~~~~~
\label{V_slc}
\eea
and
\be
A_4^{\fl}(\phi,1_\qb^-,2_q^+,3_\Qb^+,4_Q^-)
= A_4^\treenum(\phi,1_\qb^-,2_q^+,3_\Qb^+,4_Q^-)
\left\{-\frac{2}{3\e}\left[ \left(\frac{\mu^2}{-s_{12}}\right)^\e +
\left(\frac{\mu^2}{-s_{34}}\right)^\e\right] - \frac{20}{9}\right\} \,.
\label{qqQQ_mppm_mpmp_fl}
\ee

For the $\phi\qb q\Qb Q$ $({-}{+}{-}{+})$ case, we find
\bea
&&\null -i A_4^{\mathrm{lc}}(\phi,1_\qb^-,2_q^+,3_\Qb^-,4_Q^+)\ =\
-i A_4^\treenum(\phi,1_\qb^-,2_q^+,3_\Qb^-,4_Q^+)
\nonumber\\ && \hskip1.0cm \null
\times \left\{ V^{\mathrm{lc}}
+ \left[ 1 - \left(\frac{\spa1.4\spa2.3}{\spa1.3\spa2.4}\right)^2\right]
\Big[\Ls_{-1}\left(s_{23},s_{34};s_{234}\right) +
\Ls_{-1}\left(s_{41},s_{12};s_{412}\right) \Big]\right\}
\nonumber\\ && \hskip0.0cm \null
- \frac{1}{2} \frac{\spa1.2\spa3.4}{\spa2.4^2}
\left[s_{41}^2\frac{\Ll_1\left(\frac{-s_{412}}{-s_{12}}\right)}{s_{12}^2}
+ s_{23}^2\frac{\Ll_1\left(\frac{-s_{234}}{-s_{34}}\right)}{s_{34}^2}
- \ln\left(\textstyle{\frac{-s_{412}}{-s_{12}}}\right) -
\ln\left(\textstyle{\frac{-s_{234}}{-s_{34}}}\right) \right]
\nonumber\\ && \hskip0.0cm \null
+ 2 \frac{\spb2.4}{\spa2.4}\left\{\frac{\spa2.3}{\spb1.4}\left[s_{24}
\frac{\Ll_0\left(\frac{-s_{412}}{-s_{12}}\right)}{s_{12}}
+ \ln\left(\textstyle{\frac{-s_{412}}{-s_{12}}}\right)\right]
+ \frac{\spa1.4}{\spb2.3}
\left[s_{24}\frac{\Ll_0\left(\frac{-s_{234}}{-s_{34}}\right)}{s_{34}}
+ \ln\left(\textstyle{\frac{-s_{234}}{-s_{34}}}\right)\right]\right\}
\nonumber\\ && \hskip0.0cm \null
+ \frac{\spa1.4\spa2.3\spb2.4}{\spa2.4}
\left[\frac{\Ll_0\left(\frac{-s_{412}}{-s_{41}}\right)}{s_{41}}
+
\frac{\Ll_0\left(\frac{-s_{234}}{-s_{23}}\right)}{s_{23}}\right]
\nonumber\\ && \hskip0.0cm \null
- \frac{1}{2}\frac{1}{\spa2.4^2}
\left[\frac{\spa3.4\left(s_{24}-s_{41}\right)}{\spb1.2} +
\frac{\spa1.2\left(s_{24}-s_{23}\right)}{\spb3.4}\right]
+ 2i \, A_4^\treenum(\phi^\dagger,1_\qb^-,2_q^+,3_\Qb^-,4_Q^+) \,.
\label{qqQQ_mpmp_lc}
\eea
The slc and $\fl$ primitive amplitudes for the $\phi\qb q\Qb Q$ $({-}{+}{-}{+})$
are simply related to those for $({-}{+}{+}{-})$, because two of
the external legs can be exchanged at the cost of a minus sign,
\bea
A_4^{\mathrm{slc}}(\phi,1_\qb^-,2_q^+,3_\Qb^-,4_Q^+)
&=& - A_4^{\mathrm{slc}}(\phi,1_\qb^-,2_q^+,4_\Qb^+,3_Q^-) \,,\\
A_4^{\fl}(\phi,1_\qb^-,2_q^+,3_\Qb^-,4_Q^+)
&=& - A_4^{\fl}(\phi,1_\qb^-,2_q^+,4_\Qb^+,3_Q^-) \,.
\eea
These relations were used already in constructing the partial
amplitudes~(\ref{mppm1})--(\ref{mpmp2}).

The result for any color or helicity component of
${\cal A}_4^\oneloopnum(H,1_\qb,2_q,3_\Qb,4_Q)$ can be readily
obtained using eqs.~(\ref{phi+--+})--(\ref{phi2phidagger}).
%


\subsubsection{$\boldsymbol{{\phi}\qb qgg}$}

The results for the infrared- and ultraviolet-finite helicity
amplitude $\phi\qb qgg$ $({-}{+}{+}{+})$ can be extracted from
ref.~\cite{Berger2006sh}.  We give them here for completeness:
\bea
&&\null -i A_4^L(\phi,1_\qb^-,2_q^+,3^+,4^+)\ =\
  {1\over2} { \spa1.2 \spab1.{(3+4)}.2 \over \spa2.3 \spa3.4 \spa4.1 }
+ {1\over2} { \spa1.3 \spb3.4 \over \spa2.3 \spa3.4 }
\nonumber\\  && \hskip2.0cm \null
+ 2 \, { {\spab1.{(3+4)}.2}^2 \over \spa3.4 \spa4.1 \spab3.{(1+4)}.2 }
- 2 \, { {\spab1.{(2+3)}.4}^2 \spab2.{(1+3)}.4
   \over \spa1.2 \spa2.3 \, s_{123} \, \spab3.{(1+2)}.4 }~~~~~~~
\nonumber\\ && \hskip2.0cm \null
- 2 \, { {\spb2.4}^3 \, m_H^4
   \over \spb1.2 \, s_{412} \, \spab3.{(1+2)}.4 \, \spab3.{(1+4)}.2 }
- {1\over3} { \spa1.3 \spb3.4 \spa4.1 \over \spa1.2 {\spa3.4}^2 } \,,
\label{qqgg_mppp_L}
\eea
\be
-i A_4^R(\phi,1_\qb^-,2_q^+,3^+,4^+)\ =\
- {1\over2} \biggl[
   { \spab1.{(2+3)}.4 \over \spa2.3 \spa3.4 }
   + { \spa1.2 \spb2.3 \spa3.1 \over \spa2.3 \spa3.4 \spa4.1 }
 \biggr] \,,
\label{qqgg_mppp_R}
\ee
\be
-i A_4^{\fl}(\phi,1_\qb^-,2_q^+,3^+,4^+)\ =\
{1\over3} { \spa1.3 \spb3.4 \spa4.1 \over \spa1.2 {\spa3.4}^2 } \,.
\label{qqgg_mppp_fl}
\ee

For $\phi\qb qgg$ $({-}{+}{+}{-})$ we obtain,
\bea
&&\null -i A_4^L(\phi,1_\qb^-,2_q^+,3^+,4^-)\ =\
-i A_4^\treenum(\phi,1_\qb^-,2_q^+,3^+,4^-)
\nn\\
&&\null \hskip1.0cm \times \Big[ V_1^L - \Ls_{-1}\left(s_{23},s_{34};s_{234}\right)
-
\Ls_{-1}\left(s_{41},s_{12};s_{412}\right)\Big]
\nn\\
&&\null
+ \frac{\spa1.4^3}{\spa1.2\spa3.4\spa1.3}\Big[
  \Ls_{-1}\left(s_{12},s_{23};s_{123}\right)
+ \Ls_{-1}\left(s_{34},s_{41};s_{341}\right)\Big]
\nn\\
&&\null
+ \left[\frac{4}{3}\frac{\spa1.3^2 \langle
4|(1\!+\!2)|3]^3}{\spa1.2\spa3.4}-\spa1.2\spb2.3^2\spa3.4
\langle4|(1\!+\!2)|3] -
\frac{1}{3}\frac{\spa1.2^2\spb2.3^3\spa3.4^2}{\spa1.3}\right]
\frac{\Ll_2\left(\frac{-s_{123}}{-s_{12}}\right)}{s_{12}^3}
\nn\\
&&\null + \! \left[\frac{1}{2} \frac{\spa1.3^2 \! \spa2.4 \!
\langle4|(1\!+\!2)|3]^2}{\spa1.2\spa2.3\spa3.4}
 + \frac{\spa1.3 \! \spa1.4 \! \langle4|(1\!+\!2)|3]^2}{\spa1.2\spa3.4} +
\frac{1}{2}\frac{\spa1.2 \! \spa3.4 \! \spb2.3^2 \! \spa1.4}{\spa1.3}
\right]
\! \frac{\Ll_1\left(\frac{-s_{123}}{-s_{12}}\right)}{s_{12}^2}
\nn\\
&&\null -\frac{1}{2}\frac{\spa1.2\spa3.4\spa2.4\spb2.3^2}{\spa2.3}
\frac{\Ll_1\left(\frac{-s_{234}}{-s_{34}}\right)}{s_{34}^2} +
\frac{\spa1.4^2\langle 4|(1+2)|3]}{\spa1.2\spa3.4}
\frac{\Ll_0\left(\frac{-s_{123}}{-s_{12}}\right)}{s_{12}}
\nn\\
&&\null
- 2\,\frac{\spa1.4\spa2.4\spb2.3}{\spa2.3}\left[
\frac{\Ll_0\left(\frac{-s_{123}}{-s_{12}}\right)}{s_{12}} +
\frac{\Ll_0\left(\frac{-s_{234}}{-s_{34}}\right)}{s_{34}}\right]
\nn\\
&&\null
- \frac{5}{6}\left[
 - 2 \, i \, A_4^\treenum(\phi,1_\qb^-,2_q^+,3^+,4^-) +
\frac{\spa1.4^3}{\spa1.2\spa3.4\spa1.3}\right]
\ln\left(\textstyle{\frac{-s_{123}}{-s_{12}}}\right)
\nn\\
&&\null
+ \frac{5}{6}\frac{\spa1.4^2\langle
4|(1+2)|3]}{s_{12}\spa1.2\spa3.4} -
\frac{1}{6}\frac{\spa1.4^2\spb2.3\spa3.4}{\spa2.3\spa1.3\langle
4|(1+3)|2]}
+ \frac{2}{3}\frac{\spa1.4\spb2.3\spa3.4}{\spb1.2\spa2.3\spa1.3}
\nn\\
&&\null
- \frac{2}{3}\frac{\spa1.4\spa2.4\langle
4|(1+3)|2]}{s_{12}\spa2.3\spa3.4}
 + \frac{1}{3}\frac{\spb2.3\langle4|(1+3)|2]
\langle4|(2+3)|1]}{s_{123}\spb1.2^2\spa2.3}
\nn\\
&&\null
-\frac{1}{6}\frac{\langle 4|(2+3)|1]
(\langle 4|1|2]+2\,\langle4|3|2])
(2\,\langle 4|1|2]+\langle4|3|2])^2}
{s_{123}\spb1.2^2\spa2.3\spa3.4\langle 4|(1+3)|2]}
+ \frac{1}{2}\frac{\spb2.3\langle 2|(1+4)|3]}{\spb1.4\spa2.3\spb3.4}
\nn\\
&&\null
+\, \frac{1}{2}\frac{\langle 4|(1+3)|2]\langle
4|(1+2)|3]}{s_{123}\spa2.3\spb1.2} -
\frac{1}{2}\frac{\spa1.4^2\spb1.3}{s_{12}\spa2.3}
+ 2i\,A_4^\treenum(\phi^\dagger,1_\qb^-,2_q^+,3^+,4^-)\, ,
\label{qqgg_mppm_L}
\eea
with
\bea
V_1^L &=& - \frac{1}{\e^2} \left[
\left(\frac{\mu^2}{-s_{23}}\right)^\e +
\left(\frac{\mu^2}{-s_{34}}\right)^\e +
\left(\frac{\mu^2}{-s_{41}}\right)^\e\right] +
\frac{13}{6\e}\left(\frac{\mu^2}{-s_{12}}\right)^\e + \frac{119}{18}
- \frac{\delta_R}{6}
\nn\\
&&\null
- \Ls_{-1}^{2{\rm m}e}\left(s_{123},s_{234};s_{23},m_H^2\right)
- \Ls_{-1}^{2{\rm m}e}\left(s_{341},s_{412};s_{41},m_H^2\right)
\nn\\
&&\null
- \Ls_{-1}^{2{\rm m}e}\left(s_{412},s_{123};s_{12},m_H^2\right)
\,,
\label{V_1_L}
\eea
\bea
&&-i A_4^R(\phi,1_\qb^-,2_q^+,3^+,4^-)\ =\
-i A_4^\treenum(\phi,1_\qb^-,2_q^+,3^+,4^-)\,\times V^R
\nn\\
&&\null
+ \frac{\spa1.4^2}{\spa2.3\spa1.3}
\Bigr[ \Ls_{-1}\left(s_{12},s_{23};s_{123}\right)
+ \Ls_{-1}\left(s_{34},s_{41};s_{341}\right) \Big]
\nn\\
&&\null
-\,\frac{1}{2}\frac{\spa1.2^2\spb2.3^2\spa3.4^2}{\spa2.3\spa1.3}
\frac{\Ll_1\left(\frac{-s_{123}}{-s_{12}}\right)}{s_{12}^2} +
\frac{1}{2} \frac{\spa2.4^3\langle1|(3+4)|2]^2}{\spa1.2\spa2.3\spa3.4}
\frac{\Ll_1\left(\frac{-s_{234}}{-s_{34}}\right)}{s_{34}^2}
\nn\\
&&\null
- 2\,\frac{\spa1.2\spa3.4\spa1.4\spb2.3}{\spa2.3\spa1.3}
\frac{\Ll_0\left(\frac{-s_{123}}{-s_{12}}\right)}{s_{12}} -
2\,\frac{\spa1.4\spa2.4\spb2.3}{\spa2.3}
\frac{\Ll_0\left(\frac{-s_{234}}{-s_{34}}\right)}{s_{34}}
\nn\\
&&\null
- \frac{3}{2}\frac{\spa1.4^2}{\spa2.3\spa1.3}
\ln\left(\textstyle{\frac{-s_{123}}{-s_{12}}}\right)
- \frac{i}{2}\,A_4^\treenum(\phi,1_\qb^-,2_q^+,3^+,4^-)
\ln\left(\textstyle{\frac{-s_{234}}{-s_{34}}}\right)
\nn\\
&&\null
+\frac{1}{2}\Bigg[\frac{\spa1.4\spb2.3\spa3.4}{\spb1.2\spa2.3\spa1.3}
+ \frac{\spb2.3\spb1.3\langle 2|(1+4)|3]}{\spb3.4\spb1.4\langle
2|(3+4)|1]} - \frac{\langle 4|(1+3)|2]\langle
4|(1+2)|3]}{s_{123}\spa2.3\spb1.2}
\nn\\
&&\hskip1cm\null
+ \frac{\spa1.4^2\spa2.4^2(s_{21}+s_{23}+s_{24})}
{\spa1.2\spa2.3\spa3.4^2\langle2|(1+4)|3]}
- \frac{s_{341}^2\spb2.3\spa2.4^3}{s_{34}\spa2.3\spa3.4
\langle2|(1+4)|3]\langle 2|(3+4)|1]} \Bigg]\, ,
\nn\\
&&{~}
\label{qqgg_mppm_R}
\eea
with
\be
V^R\ =\ - \frac{1}{\e^2}\left( \frac{\mu^2}{-s_{12}}\right)^\e
- \frac{3}{2\e}\left( \frac{\mu^2}{-s_{12}}\right)^\e
- \frac{7}{2} - \frac{\delta_R}{2}
- \Ls_{-1}^{2{\rm m}e}\left(s_{234},s_{341};s_{34},m_H^2\right)
\,,
\ee
and
\bea
&&-i A_4^{\fl}(\phi,1_\qb^-,2_q^+,3^+,4^-)\ =\
-i A_4^\treenum(\phi,1_\qb^-,2_q^+,3^+,4^-)
\times\left[ - \frac{2}{3\e} \left(\frac{\mu^2}{-s_{12}}\right)^\e
- \frac{10}{9} \right]
\nn\\
&&\null
+\, \frac{1}{3}\left[ \frac{\spa1.2^2\spb2.3^3\spa3.4^2}{\spa1.3} -
\frac{\langle 4|(1+2)|3]^3\spa1.3^2}{\spa1.2\spa3.4}\right]
\frac{\Ll_2\left(\frac{-s_{123}}{-s_{12}}\right)}{s_{12}^3}
\nn\\
&&\null
-\frac{1}{3}\left[ \frac{\spa1.4^2}{\spa2.3\spa1.3} +
\frac{\spa1.4^2\spa2.4}{\spa1.2\spa2.3\spa3.4}\right]
\ln\left(\textstyle{\frac{-s_{123}}{-s_{12}}}\right)
- \frac{1}{2} \frac{{\spa1.4}^2 \spb2.3}{s_{12}\spa1.3}
\nn\\
&&\null
+ \frac{1}{6} \frac{\spa1.4
 ( \langle4|1|2] \langle4|(1+3)|2] - \langle4|3|2]^2 )}
  {\spb1.2 \spa3.4 \spa1.3 s_{123}}
+ \frac{1}{6} \frac{{\spa1.4}^3 s_{123}}
  {s_{12}\spa1.2\spa3.4\spa1.3} \,.
\label{qqgg_mppm_fl}
\eea
For the $\phi\qb qgg$ $({-}{+}{-}{+})$ case,
\bea
&& -i A_4^L(\phi,1_\qb^-,2_q^+,3^-,4^+)\ =\
-i A_4^\treenum(\phi,1_\qb^-,2_q^+,3^-,4^+)
\nn\\
&&\hskip1cm\null
\times \bigg[ V_1^L
- \frac{13}{6} \ln\left(\textstyle{\frac{-s_{412}}{-s_{12}}}\right)
- \Ls_{-1}\left(s_{34},s_{41};s_{341}\right)
- \Ls_{-1}\left(s_{12},s_{23};s_{123}\right) \bigg]
\nn\\
&&\null
+ \frac{\spa1.4^2\spa2.3^3}{\spa1.2\spa3.4\spa2.4^3}\Big[
\Ls_{-1}\left(s_{23},s_{34};s_{234}\right) +
\Ls_{-1}\left(s_{41},s_{12};s_{412}\right)
\Big]
\nn\\
&&\null
+ \frac{2}{3}\frac{\spa1.2^2\spa3.4^2\spb2.4^3}{\spa1.4}
\frac{\Ll_2\left(\frac{-s_{412}}{-s_{12}}\right)}{s_{12}^3}
- \frac{1}{2}\frac{\spa1.2\spa2.3\spa3.4\spb2.4^2}{\spa2.4}
\frac{\Ll_1\left(\frac{-s_{234}}{-s_{34}}\right)}{s_{34}^2}
\nn\\
&&\null
+ \left[\frac{1}{2}\frac{\spa1.4\langle 3|(1\!+\!2)|4]^2}{\spa2.4} -
\frac{1}{3}\frac{\spa1.3\spa1.4\langle 3|(1\!+\!2)|4]^2}{\spa1.2\spa3.4}
- \frac{2}{3}\frac{\spa1.3\spa1.2\spa3.4\spb2.4^2}{\spa1.4}\right]
\frac{\Ll_1\left(\frac{-s_{412}}{-s_{12}}\right)}{s_{12}^2}
\nn\\
&&\null
- \left[\frac{\spa1.2\spb2.4\langle3|(1+4)|2]^2}{\spa1.4\spb1.2} +
\frac{1}{2}\frac{\spa1.4\spa2.3^2\spb2.4^2}{\spa2.4}\right]
\frac{\Ll_1\left(\frac{-s_{412}}{-s_{41}}\right)}{s_{41}^2}
\nn\\
&&\null
- \left[\frac{\spb2.4\spa3.4\langle1|(2+3)|4]^2}{\spa1.4\spb3.4} +
\frac{1}{2}\frac{\spa1.4\spa2.3^2\spb2.4^2}{\spa2.4}\right]
\frac{\Ll_1\left(\frac{-s_{234}}{-s_{23}}\right)}{s_{23}^2}
\nn\\
&&\null
+ \left[3\,\frac{\spa1.3^2\langle 3|(1\!+\!2)|4]}{\spa1.2\spa3.4} +
2\,\frac{\spa1.3\langle 3|(1\!+\!2)|4]^2}{\spa1.2\spa3.4\spb1.4} +
\frac{1}{3}\frac{\spa1.3^2\spb2.4}{\spa1.4} - \frac{\langle
3|(1\!+\!2)|4]^2}{\spb1.4\spa2.4}\right]
\frac{\Ll_0\left(\frac{-s_{412}}{-s_{12}}\right)}{s_{12}}
\nn\\
&&\null
+ 3\,\frac{\spa2.3\spa1.3\spb2.4}{\spa2.4} \left[
\frac{\Ll_0\left(\frac{-s_{234}}{-s_{23}}\right)}{s_{23}} +
\frac{\Ll_0\left(\frac{-s_{412}}{-s_{41}}\right)}{s_{41}}
\right]
\nn\\
&&\null
+ \frac{\spa2.3\spb2.4\left(\spa1.2\spa3.4 + 2\spa1.4\spa2.3
\right)}{\spa2.4^2}
\frac{\Ll_0\left(\frac{-s_{234}}{-s_{34}}\right)}{s_{34}}
\nn\\
&&\null
- \left[\frac{1}{3}\frac{\spa1.3^3}{\spa1.2\!\spa3.4\!\spa1.4} +
\frac{1}{2}\frac{\spa2.3\spa1.3^2}{\spa1.2\!\spa2.4\!\spa3.4} +
\frac{\spa2.3^2\!\spb2.4}{\spa2.4^2\!\spb1.4} +
2\,\frac{\spa2.3^3\spa1.4\spb2.4}{\spa2.4^2\!\spa3.4\!\spa1.2\!\spb1.4}
\right]
\ln\left(\textstyle{\frac{-s_{412}}{-s_{12}}}\right)
\nn\\
&&\null
+ \frac{\spa1.2^2\spa3.4\spb2.4}{\spa2.4^2\spa1.4\spb3.4}
\ln\left(\textstyle{\frac{-s_{234}}{-s_{23}}}\right)
+ \frac{\spa3.4^2\spa1.2\spb2.4}{\spa2.4^2\spa1.4\spb1.2}
\ln\left(\textstyle{\frac{-s_{412}}{-s_{41}}}\right)
\nn\\
&&\null
- \frac{5}{6}\frac{\spa1.3^2\spb2.4}{s_{12}\spa1.4} -
\frac{1}{3}\frac{\spa1.3^2\langle 3|(1+2)|4]}{s_{12}\spa1.2\spa3.4}
-\frac{1}{3}\frac{\spa1.3\spb2.4(2\langle 3|4|2]+\langle
3|1|2])}{s_{412}\spa1.4\spb1.2}
\nn\\
&&\null
+ \frac{1}{2} \bigg[ \frac{\spb2.4\langle 3|(1+2)|4]
\langle3|(2+4)|1]}{s_{412}\spb1.4\spb1.2\spa2.4}
- \frac{\spa1.3^2\spb1.4}{s_{12}\spa2.4}
- \frac{\spa1.3\spb2.4\spa3.4}{\spa1.4\spb1.2\spa2.4}
+ \frac{\spa1.2\spb2.4^2}{\spb2.3\spb3.4\spa2.4} \bigg]
\nn\\
&&\null
+ \frac{\spa1.3\spb2.4\langle 1|(2+3)|4]}{s_{23}\spa1.4\spb3.4}
- \frac{\spa1.3\spb2.4\langle3|(1+4)|2]}{s_{41}\spa1.4\spb1.2}
- \frac{\spb2.4^2\spa3.4\spa2.3}{s_{41}\spa2.4\spb1.2}
\nn\\
&&\null
+ 2i\,A_4^\treenum(\phi^\dagger,1_\qb^-,2_q^+,3^-,4^+)\, ,
\label{qqgg_mpmp_L}
\eea
\bea
&& -i A_4^R(\phi,1_\qb^-,2_q^+,3^-,4^+)\ =\
-i A_4^\treenum(\phi,1_\qb^-,2_q^+,3^-,4^+)\,\times V^R
\nn\\
&&\null
+ \frac{\spa1.2^2\spa3.4^2}{\spa1.4\spa2.4^3}\Big[
\Ls_{-1}\left(s_{23},s_{34};s_{234}\right) +
\Ls_{-1}\left(s_{41},s_{12};s_{412}\right) \Big]
\nn\\
&&\null
- \frac{1}{2}\frac{\spa1.2^2\spa3.4^2\spb2.4^2}{\spa1.4\spa2.4}
\frac{\Ll_1\left(\frac{-s_{412}}{-s_{12}}\right)}{s_{12}^2}
+ \frac{\spb2.4\spa3.4\langle 1|(2+3)|4]^2}{\spa1.4\spb3.4}
\frac{\Ll_1\left(\frac{-s_{234}}{-s_{23}}\right)}{s_{23}^2}
\nn\\
&&\null
- \left[ \frac{\spa1.2\spa3.4\spb2.4^3}{\spb2.3} +
\frac{1}{2}\frac{\spa2.3^3\langle
1|(3+4)|2]^2}{\spa1.2\spa3.4\spa2.4}\right]
\frac{\Ll_1\left(\frac{-s_{234}}{-s_{34}}\right)}{s_{34}^2}
\nn\\
&&\null
- \frac{1}{2}\frac{\spa1.4\spa2.3^2\spb2.4^2}{\spa2.4}\left[
\frac{\Ll_1\left(\frac{-s_{412}}{-s_{41}}\right)}{s_{41}^2}
- \frac{\Ll_1\left(\frac{-s_{234}}{-s_{23}}\right)}{s_{23}^2}
\right]
- \frac{\spa1.2^2\spa3.4^2\spb2.4}{\spa1.4\spa2.4^2}
\frac{\Ll_0\left(\frac{-s_{412}}{-s_{12}}\right)}{s_{12}}
\nn\\
&&\null
+ \frac{\spa2.3\spb2.4(2\spa1.2\spa3.4 +
\spa1.4\spa2.3)}{\spa2.4^2}
\frac{\Ll_0\left(\frac{-s_{412}}{-s_{41}}\right)}{s_{41}}
- \frac{\spa1.2^2\spa3.4\spb2.4}{\spa2.4^2\spa1.4\spb3.4}
\ln\left(\textstyle{\frac{-s_{234}}{-s_{23}}} \right)
\nn\\
&&\null
+ \left[
\frac{\spa3.4\spa1.2\spb2.4}{\spa2.4^2\spb2.3} +
\frac{1}{2}\frac{\spa2.3\spa1.3^2}{\spa1.2\spa2.4\spa3.4}\right]
\ln\left(\textstyle{\frac{-s_{234}}{-s_{34}}} \right)
- \frac{1}{2}\frac{\spb2.4\langle 3|(1+2)|4]
\langle3|(1+4)|2]}{s_{41}s_{412}\spb1.2}
\nn\\
&&\null
- \frac{1}{2}\frac{\spb2.4^2\spa3.4\spa2.3}{s_{41}\spa2.4\spb1.2}
- \frac{1}{2}\frac{\spa1.3\spa2.3^2
\langle1|(3+4)|2]}{s_{34}\spa3.4\spa1.2\spa2.4}
+ \frac{1}{2}\frac{\spa2.3\spb2.4\langle1|(3+4)|2](s_{23}+s_{34})}
 {s_{34}s_{234}\spb2.3\spa2.4}
\nn\\
&&\null
+ \frac{1}{2}\frac{\spb2.4^2\langle
1|(2+3)|4]}{s_{234}\spb2.3\spb3.4}
- \frac{\spa1.2\spb2.4\spa3.4\langle
1|(2+3)|4]}{s_{23}\spa1.4\spb3.4\spa2.4} +
\frac{\spa1.3\spb2.4}{\spb2.3\spa2.4}\,,
\label{qqgg_mpmp_R}
\eea
and
\bea
&& -i A_4^{\fl}(\phi,1_\qb^-,2_q^+,3^-,4^+)\ =\
-i A_4^\treenum(\phi,1_\qb^-,2_q^+,3^-,4^+)
\times\left[ - \frac{2}{3\e}\left(\frac{\mu^2}{-s_{12}}\right)^\e
- \frac{10}{9} \right]
\nn\\
&&\null
- \frac{1}{3}\left[ \frac{\spa1.4^2\langle3|(1+2)|4]^3}{\spa1.2\spa3.4}
+ \frac{\spa1.2^2\spb2.4^3\spa3.4^2}{\spa1.4}\right]
\frac{\Ll_2\left(\frac{-s_{412}}{-s_{12}}\right)}{s_{12}^3}
\nn\\
&&\null
- \frac{i}{3}A_4^\treenum(\phi,1_\qb^-,2_q^+,3^-,4^+)
  \ln\left(\textstyle{\frac{-s_{412}}{-s_{12}}}\right)
+ \frac{1}{2} \frac{{\spa1.3}^2 \spb2.4}{s_{12}\spa1.4}
\nn\\
&&\null
+ \frac{1}{6} \frac{\spa1.3
 ( \langle3|1|2] \langle3|(1+4)|2] - \langle3|4|2]^2 )}
  {\spb1.2 \spa3.4 \spa1.4 s_{412}}
+ \frac{1}{6} \frac{{\spa1.3}^3 s_{412}}
  {s_{12}\spa1.2\spa3.4\spa1.4} \,.
\label{qqgg_mpmp_fl}
\eea


\subsubsection{$\boldsymbol{{\phi}\qb gqg}$}

For $\phi\qb gqg$, we have the ``reflection'' relation
$A_4^R(1_\qb,3,2_q,4) = A_4^L(1_\qb,4,2_q,3)$, so we do not need
to quote $A_4^R$ separately.

Again we take the results for the infrared- and ultraviolet-finite
helicity amplitude $\phi\qb gqg$ $({-}{+}{+}{+})$ from
ref.~\cite{Berger2006sh}:
\bea
-i A_4^L(\phi,1_\qb^-,2^+,3_q^+,4^+) &=&
{1\over2} \left[
{ \spa1.3 \spab1.{(3+4)}.2 \over \spa2.3\spa3.4\spa4.1 }
  + { {\spa1.3}^2 \spb3.4 \over \spa1.2\spa2.3\spa3.4 }
         \right]
\nn\\ &&\null
- 2 \, { {\spab1.{(2+3)}.4}^2 \over \spa1.2 \spa2.3 \, s_{123} } \,,
\label{phi_qgqg_mppp_L}\\
A_4^\fl(\phi,1_\qb^-,2^+,3_q^+,4^+) &=& 0.
\label{phi_qgqg_mppp_fl}
\eea

The results for $\phi\qb gqg$ $({-}{+}{+}{-})$ are given by
\bea
&&\null -i A_4^L(\phi,1_\qb^-,2^+,3_q^+,4^-)
\ =\
-i A_4^\treenum(\phi,1_\qb^-,2^+,3_q^+,4^-)
\nonumber\\ && \hskip1.0cm \null
\times \Bigg\{ V_2^L - \Ls_{-1}\left(s_{41},s_{12};s_{412}\right)
- \Ls_{-1}\left(s_{23},s_{34};s_{234}\right) \Bigg\}
\nonumber\\ && \hskip0.0cm \null
+ \frac{1}{2}\frac{\spa1.3^2\langle 4|(1+2)|3]^2}{\spa1.2\spa2.3}
\frac{\Ll_1\left(\frac{-s_{123}}{-s_{12}}\right)}{s_{12}^2}
- \frac{1}{2}\frac{\spa1.2\spa3.4^2\spb2.3^2}{\spa2.3}
\frac{\Ll_1\left(\frac{-s_{234}}{-s_{34}}\right)}{s_{34}^2}
\nonumber\\ && \hskip0.0cm \null
- 2 \, \frac{\spa1.4\spb2.3\spa3.4}{\spa2.3}\left[
\frac{\Ll_0\left(\frac{-s_{123}}{-s_{12}}\right)}{s_{12}} +
\frac{\Ll_0\left(\frac{-s_{234}}{-s_{34}}\right)}{s_{34}}\right]
\nonumber\\ && \hskip0.0cm \null
- \frac{i}{2}
A_4^\treenum(\phi,1_\qb^-,2^+,3_q^+,4^-)
\ln\left(\textstyle{\frac{-s_{123}}{-s_{12}}}
\right)
- \frac{1}{2}
\frac{s_{341}\spb2.3\spb1.3}{\spb3.4\spb1.4\langle2|(3+4)|1]}
- \frac{1}{2} \frac{\langle4|(1+3)|2]^2}{s_{123}\spb1.2\spa2.3}
\nonumber\\ && \hskip0.0cm \null
- \frac{\spa1.4\spb2.3\spa3.4}{s_{12}\spa2.3}
+ \frac{1}{2} \frac{\spa1.4^2(s_{13}+s_{23})}{s_{12}\spa1.2\spa2.3}
- \frac{1}{2}
\frac{\spb2.3\spa3.4\langle 2|(1+4)|3]}{\spa2.3\spb3.4\langle2|(3+4)|1]}
\,,
\eea
with
\bea
V_2^L &=& - \frac{1}{\e^2} \left[
\left(\frac{\mu^2}{-s_{34}}\right)^\e +
\left(\frac{\mu^2}{-s_{41}}\right)^\e \right] +
\frac{13}{6\e}\left(\frac{\mu^2}{-s_{123}}\right)^\e +
\frac{119}{18} - \frac{\delta_R}{6}
\nn\\
&&\null
- \Ls_{-1}^{2{\rm m}e}\left(s_{412},s_{123};s_{12},m_H^2\right)
- \Ls_{-1}^{2{\rm m}e}\left(s_{123},s_{234};s_{23},m_H^2\right)
\,,
\eea
and
\be
A_4^\fl(\phi,1_\qb^-,2^+,3_q^+,4^-)
\ =\ A_4^\treenum(\phi,1_\qb^-,2^+,3_q^+,4^-)
\left[ -{2\over3\e} \left(\frac{\mu^2}{-s_{123}}\right)^\e
- {10\over9} \right] \,.
\label{phi_qgqg_mppm_fl}
\ee

The results for $\phi\qb gqg$ $({-}{-}{+}{+})$ are
\bea
&&\null -i A_4^L(\phi,1_\qb^-,2^-,3_q^+,4^+)\ =\
-i A_4^\treenum(\phi,1_\qb^-,2^-,3_q^+,4^+)
\nn\\
&& \hskip1cm \null
\times
\Bigg\{ V_3^L - \Ls_{-1}\left(s_{34},s_{41};s_{341}\right)
- \Ls_{-1}\left(s_{12},s_{23};s_{123}\right) \Bigg\}
\nn\\ &&\null
- \frac{1}{2} \frac{\spa2.4^2\langle1|(2+3)|4]^2}{\spa3.4\spa1.4}
\frac{\Ll_1\left(\frac{-s_{234}}{-s_{23}}\right)}{s_{23}^2}
+ \frac{1}{2} \frac{\spa2.3^2\spb3.4^2\spa1.4}{\spa3.4}
\frac{\Ll_1\left(\frac{-s_{341}}{-s_{41}}\right)}{s_{41}^2}
\nn\\ &&\null
+ 2 \, \frac{\spa1.2\spa2.3\spb3.4}{\spa3.4} \left[
\frac{\Ll_0\left(\frac{-s_{234}}{-s_{23}}\right)}{s_{23}} +
\frac{\Ll_0\left(\frac{-s_{341}}{-s_{41}}\right)}{s_{41}}
\right]
- \frac{i}{2} A_4^\treenum(\phi,1_\qb^-,2^-,3_q^+,4^+)
\ln\left(\textstyle{\frac{-s_{234}}{-s_{23}}}\right)
\nn\\ &&\null
- \frac{1}{2} \frac{\langle 2|(1+3)|4]^2}{s_{341}\spb1.4\spa3.4}
+ \frac{1}{2} \frac{s_{123}\spb1.3\spb3.4}
         {\spb1.2\spb2.3\langle4|(2+3)|1]}
- \frac{1}{2} \frac{\spa1.2^2\spa2.4(s_{14}+s_{24}+s_{34})}
                   {\spa1.4\spa2.3\spa3.4\langle4|(1+2)|3]}
\nn\\ &&\null
- \frac{1}{2} \frac{s_{123}^2\spa2.4^2\spb3.4}
 {s_{23}\spa3.4\langle4|(2+3)|1]\langle 4|(1+2)|3]}
+ 2i\,A_4^\treenum(\phi^\dagger,1_\qb^-,2^-,3_q^+,4^+) \,,
\eea
with
\bea
V_3^L &=& - \frac{1}{\e^2} \left[
\left(\frac{\mu^2}{-s_{34}}\right)^\e +
\left(\frac{\mu^2}{-s_{41}}\right)^\e \right] -
\frac{3}{2\e}\left(\frac{\mu^2}{-s_{341}}\right)^\e - \frac{7}{2}
- \frac{\delta_R}{2}
\nn\\&&\null
- \Ls_{-1}^{2{\rm m}e}\left(s_{412},s_{123};s_{12},m_H^2\right)
- \Ls_{-1}^{2{\rm m}e}\left(s_{123},s_{234};s_{23},m_H^2\right)
\,,
\eea
and
\be
A_4^\fl(\phi,1_\qb^-,2^-,3_q^+,4^+)
\ =\ 0.
\label{phi_qgqg_mpmp_fl}
\ee

Using \eqns{phiphidaggerparity}{qqggchargeconj},
one can obtain any color or helicity component of the
${\cal A}_4^\oneloopnum(H,1_\qb,2_q,3^\pm,4^\mp)$ amplitudes.


\subsection{Numerical results}

In order to facilitate comparisons with future work, we present
here numerical values, at a single phase-space point, for the
bare, unrenormalized primitive amplitudes computed in the paper.
We choose the same kinematic point as in ref.~\cite{Ellis2005qe},
the configuration $H \to \bar{q}_1 q_2 \bar{Q}_3 Q_4$ in which the
Higgs (or $\phi$) and parton four-momenta take the values
\bea
k_\phi &=& (-1.0000000000,\ \, 0.00000000000,\ \ 0.00000000000,\ \ 0.00000000000),\nn\\
k_1 &=& (0.30674037867,-0.17738694693,-0.01664472021,-0.24969277974),\nn\\
k_2 &=& (0.34445032281,\ \ 0.14635282800,-0.10707762397,\ \ 0.29285022975),
~~~\label{PhaseSpacePoint}\\
k_3 &=& (0.22091667641,\ \ 0.08911915938,\ \ 0.19733901856,\ \ 0.04380941793),\nn\\
k_4 &=& (0.12789262211,-0.05808504045,-0.07361667438,-0.08696686795).\nn
\eea
We substitute $\mu=m_H=1$, and use the 't Hooft-Veltman
scheme~\cite{HV}, in which $\delta_R=1$ (in accord with
ref.~\cite{Ellis2005qe}).  Discussions of the conversion between
different dimensional regularization schemes can be found in
refs.~\cite{KST,OtherFDH,Ellis2005qe}. The dependence on
$\delta_R$ in our formulae agrees with the shift in $H\qb q\Qb Q$
and $H\qb qgg$ amplitudes between HV and FDH regularization
schemes that is quoted in ref.~\cite{Ellis2005qe}.

In tables~\ref{phiqqQQNumericalTable}, \ref{phiqqggNumericalTable}
and \ref{phiqgqgNumericalTable} we present numerical values for
the unrenormalized primitive amplitudes computed in the paper.
Note that the overall phases are convention-dependent.  Phase-independent
quantities can be constructed by dividing by the corresponding
tree amplitude.  The tree amplitude is identified as
$-1/2$ of the $\eps^{-2}$ slc entry in
table~\ref{phiqqQQNumericalTable}, the negative of the $\eps^{-2}$ $R$
entry in table~\ref{phiqqggNumericalTable}, and $-1/2$ of
the $\eps^{-2}$ $L$ entry in table~\ref{phiqgqgNumericalTable}.

\TABLE[t]{ \begin{tabular}{|c|c|c|c|}
\hline $\phi\qb q\Qb Q$ & $\e^{-2}$ & $\e^{-1}$ & $\e^0$\\
\hline \hline $({-}{+}{+}{-}) \quad \mathrm{lc}$ & $+0.10641628412$ & $+0.25970964611$ & $+1.94930173285$\\
& $-0.04813723405\,i$ & $+0.28524492960\,i$ & $+0.66145729341\,i$\\
\hline $({-}{+}{+}{-}) \quad \mathrm{slc}$ & $+0.10641628412$ & $+0.47949272770$ & $+0.82337543420$\\
& $-0.04813723405\,i$ & $+0.18582640802\,i$ & $+0.78352094637\,i$\\
\hline $({-}{+}{+}{-}) \quad \fl$ & $0$ & $+0.07094418941$ & $+0.33148585003$\\
& & $-0.03209148937\,i$ & $+0.11853569045\,i$\\
\hline
\hline $({-}{+}{-}{+}) \quad \mathrm{lc}$ & $-1.88930338066$ & $-1.08479447284$ & $+8.36061276059$\\
& $-0.26775736353\,i$ & $-6.20837676158\,i$ & $-4.87467657230\,i$\\
\hline $({-}{+}{-}{+}) \quad \mathrm{slc}$ & $-1.88930338066$ & $-4.98679990840$ & $-4.95070751222$\\
& $-0.26775736353\,i$ & $-6.76137989415\,i$ & $-19.34119382028\,i$\\
\hline $({-}{+}{-}{+}) \quad \fl$ & $0$ & $-1.25953558711$ & $-3.53445587012$\\
& & $-0.17850490902\,i$ & $-4.53733741427\,i$\\
\hline
\end{tabular}
\caption{Numerical values of $\phi\qb q\Qb Q$ primitive amplitudes at
kinematic point~(\ref{PhaseSpacePoint}). \label{phiqqQQNumericalTable} } }

\TABLE[t]{ \begin{tabular}{|c|c|c|c|}
\hline $\phi\qb qgg$ & $\e^{-2}$ & $\e^{-1}$ & $\e^0$ \\
\hline
\hline $({-}{+}{+}{-}) \quad L$ & $-0.06141673303$ & $+0.37791957375$ & $-0.34558862143$\\
& $-0.16247914884\,i$ & $-0.54354042568\,i$ & $-12.10809400189\,i$\\
\hline $({-}{+}{+}{-}) \quad R$ & $-0.02047224434$ & $+0.12098146140$ & $+0.89774344141$\\
& $-0.05415971628\,i$ & $-0.19438585924\,i$ & $+5.07992303199\,i$\\
\hline $({-}{+}{+}{-}) \quad \fl$ & $0$ & $-0.01364816290$ & $+0.08454550896$\\
& & $-0.03610647752\,i$ & $-0.11688473115\,i$\\
\hline
\hline $({-}{+}{-}{+}) \quad L$ & $-4.75526937444$ & $-43.39451947571$ & $-67.30255141380$\\
& $+10.54678423393\,i$ & $+7.81850845690\,i$ & $-40.68074759818\,i$\\
\hline $({-}{+}{-}{+}) \quad R$ & $-1.58508979148$ & $-14.85133091493$ & $-33.50442466808$\\
& $+3.51559474465\,i$ & $+3.46337394880\,i$ & $+5.37244309382\,i$\\
\hline $({-}{+}{-}{+}) \quad \fl$ & $0$ & $-1.05672652765$ & $-9.74999749954$\\
& & $+2.34372982976\,i$ & $+1.83728897650\,i$\\
\hline
\end{tabular}
\caption{Numerical values of $\phi\qb qgg$ primitive amplitudes at
kinematic point~(\ref{PhaseSpacePoint}). \label{phiqqggNumericalTable} } }

\TABLE[t]{ \begin{tabular}{|c|c|c|c|}
\hline $\phi\qb gqg$ & $\e^{-2}$ & $\e^{-1}$ & $\e^0$ \\
\hline
\hline $({-}{+}{+}{-}) \quad L$ & $-0.07267563934$ & $+0.04895177312$ & $-1.94503280260$\\
& $-0.06793690983\,i$ & $-0.38207096287\,i$ & $-4.71626523954\,i$\\
\hline $({-}{+}{+}{-}) \quad \fl$ & $0$ & $-0.02422521311$ & $+0.02361126144$\\
& & $-0.02264563661\,i$ & $-0.12053858081\,i$\\
\hline
\hline $({-}{-}{+}{+}) \quad L$ & $+23.05418438416$ & $+92.96105880288$ & $+154.70151920223$\\
& $+0.47135348735\,i$ & $+74.35776389434\,i$ & $+298.96823152311\,i$\\
\hline $({-}{-}{+}{+}) \quad \fl$ & $0$ & $0$ & $0$\\
\hline
\end{tabular}
\caption{Numerical values of $\phi\qb gqg$ primitive amplitudes at
kinematic point~(\ref{PhaseSpacePoint}). \label{phiqgqgNumericalTable} } }

In table~\ref{HqqQQNumericalTable} we give the numerical value of
the virtual correction to the color- and helicity-summed cross
section for the $H\qb q\Qb Q$ process, according to
\eqn{HqqQQcolorsum} but omitting an overall factor of $2 C^2 \cg
g^6 (N_c^2-1) N_c$. The result is constructed from the one-loop
amplitudes given in this paper and the tree
amplitudes~(\ref{qqQQtrees}). The dependence on the number of
colors $N_c$ and massless quark flavors $n_f$ is shown explicitly.
If one substitutes $N_c=3$ and $n_f=5$, adds the contributions,
and multiplies by $1/4 \times (N_c^2-1) N_c$, then the result
agrees with that for process $A$ in table I of
ref.~\cite{Ellis2005qe}. (The factor of $1/4$ arises because a
factor of $A^2 \equiv (2C)^2$ is extracted instead of $C^2$ in
ref.~\cite{Ellis2005qe}.)

\TABLE[t]{ \begin{tabular}{|c|c|c|c|}
\hline $H\qb q\Qb Q$ cross section & $\e^{-2}$ & $\e^{-1}$ & $\e^0$ \\
\hline
\hline 1 & $-12.9162958212$ & $-13.1670303819$ & 47.5186460764 \\
$1/N_c^2$ & 12.9162958212 & 75.7028593906 & 172.3194296444 \\
$n_f/N_c$ & 0 & $-8.61086 38808$ & $-27.9973052106$ \\
\hline
\end{tabular}
\caption{Numerical value of the one-loop correction to the
$H\qb q\Qb Q$ cross section at kinematic
point~(\ref{PhaseSpacePoint}), omitting
an overall factor of $2 C^2 \cg g^6 (N_c^2-1) N_c$
from \eqn{HqqQQcolorsum}.
 \label{HqqQQNumericalTable} } }


To convert the bare, unrenormalized amplitudes presented here
to renormalized ones in an $\overline{\rm MS}$-type subtraction
scheme, one should subtract the quantity
\be
 4 \, g^2 \, \frac{\cg}{2 \e}
 \biggl[ \frac{11}{3} N_c - \frac{2}{3} n_f \biggr]
 {\cal A}_4^\treenum
\label{MSrenom}
\ee
from the corresponding one-loop amplitude ${\cal A}_4^\oneloopnum$
in \eqn{qqQQloopdecomp} or (\ref{qqggloopdecomp}).
After this subtraction, the rational parts of the $1/\e$ poles
are purely infrared, and take the form of a sum over
contributions from each external parton,
\bea
&&\frac{g^2}{(4\pi)^2} \, 4
\, \biggl[ -\frac{3}{4} \biggl( N_c - \frac{1}{N_c} \biggr) \biggr]
   \frac{{\cal A}_4^\treenum}{\e}
\hskip4.1cm \hbox{for $\phi\qb q\Qb Q$},
\label{phiqqQQIRpole}\\
&&\null
\frac{g^2}{(4\pi)^2} \, 2
\, \biggl[  -\frac{3}{4} \biggl( N_c - \frac{1}{N_c} \biggr)
 -\frac{1}{2} \biggl( \frac{11}{3} N_c - \frac{2}{3} n_f \biggr)
\biggr] \frac{{\cal A}_4^\treenum}{\e}
\qquad  \hbox{for $\phi\qb qgg$},
\label{phiqqggIRpole}
\eea
in accordance with the general form of infrared singularities
of one-loop amplitudes~\cite{UniversalIR}.

A final contribution that needs to be included is the one-loop
correction to the $Hgg$ effective operator of \eqn{Lint}, which
shifts its coefficient from $C$ to
$C\times[1+11\alphas/(4\pi)]$~\cite{Chetyrkin1997un}. At the level
of the NLO virtual cross section, this effect can be taken into
account by an addition to \eqns{qqQQloopdecomp}{qqggloopdecomp} of
the form
\be
 11 \, \frac{g^2}{(4\pi)^2} {\cal A}_4^\treenum \,.
\label{Cshift}
\ee


\subsection{Pseudoscalar Higgs amplitudes and
cross section for $\boldsymbol{A\qb qgg}$}

As a byproduct of our calculation of the $\phi$-amplitudes, we can
obtain the respective amplitudes where the scalar Higgs boson $H$
has been replaced by a pseudoscalar Higgs boson, $A$.
Pseudoscalar Higgs bosons are present in many extensions of the
SM, such as the MSSM. Here we assume that we are in a kinematic
regime where the production of the $A$ boson plus jets can also be
treated in the large $m_t$ limit. As mentioned earlier, the
overall constant $C$ is different for the $A$
case~\cite{Berger2006sh}.  Otherwise, the only difference between
the two computations is that instead of taking the {\it sum} of
the $\phi$- and $\phi^\dagger$- components, the pseudoscalar
amplitudes are given by their {\it difference} divided by $i$,
according to \eqn{Areconstruct}.

Furthermore, we can combine our results with those of Ellis, Giele
and Zanderighi (EGZ)~\cite{Ellis2005qe}, to obtain the color- and
helicity-summed cross section for $A\qb qgg$. EGZ used a
semi-numerical method to compute the color- and helicity-summed
cross section for $H\qb qgg$, which we may write schematically as,
\be \sigma_{\mathrm{EGZ}} = \sum_{\lambda}\left[
{\cal A}_H^{\treenum*}(\lambda)
{\cal A}_H^\oneloopnum(\lambda)
 + {\cal A}_H^\treenum(\lambda)
   {\cal A}_H^{\oneloopnum *}(\lambda) \right]
= 2 \, \mathrm{Re} \left\{ \sum_{\lambda}\left[
{\cal A}_H^{\treenum*}(\lambda)
{\cal A}_H^\oneloopnum(\lambda)\right] \right\}.
\label{HqqggEGZ}
\ee
Here by ${\cal A}_H(\lambda)$ we denote a $H\qb qgg$ amplitude in a
helicity configuration $\lambda$, and the sum is over all possible
helicity configurations. In every summation in this subsection
there is also an implicit sum over colors which is given by
\eqn{Hqqggcolorsum}, but for simplicity we do not write it out here.
The helicity sum includes the two independent MHV
configurations $({-}{+}{+}{-})$ and $({-}{+}{-}{+})$ analyzed in
this paper, and their conjugate ones (obtained by parity).
Moreover, they include
configurations $({-}{+}{-}{-})$, $({-}{+}{+}{+})$ and their
conjugates.  These last configurations require NMHV $\phi$ amplitudes,
which we did not compute here, and which would be needed to provide
a full analytic description of the one-loop $H\qb qgg$ and $A\qb qgg$
helicity amplitudes.   However, we will see that because
the contribution of the latter configurations to the $H\qb qgg$
cross section is encoded in $\sigma_{\mathrm{EGZ}}$,
it is possible to use this result instead for the purpose of
computing the $A\qb qgg$ cross section.

Note that we can rewrite \eqn{HqqggEGZ} as
\be
\sigma_{\mathrm{EGZ}} = 4\mathrm{Re}\left\{
\sum_{\lambda^\prime}\left[ A_H^{\treenum
*}(\lambda^\prime)A_H^\oneloopnum(\lambda^\prime)\right] \right\},
\label{HqqggEGZ2}
\ee
with $\lambda^\prime$ now labelling each of the four helicity
configurations with fixed helicities for the antiquark-quark pair,
$\qb^- q^+$, namely $({-}{+}{\pm}{\pm})$.  Also,
from \eqn{HqqggEGZ2} we see that $\sigma_{\mathrm{EGZ}}$
contains the NMHV $H\qb qgg$ amplitudes in the quantity
\be \sigma_{\mathrm{EGZ}}^{\mathrm{NMHV}} \equiv
4 \, \mathrm{Re}\left\{ A_H^{\treenum
*}(\lambda_-)A_H^\oneloopnum(\lambda_-) + A_H^{\treenum
*}(\lambda_+)A_H^\oneloopnum(\lambda_+) \right\}\,\,
\subset\,\, \sigma_{\mathrm{EGZ}}, \label{Hqqgg_NMHV} \ee
with $\lambda_- \equiv ({-}{+}{-}{-})$ and $\lambda_+ \equiv
({-}{+}{+}{+})$. We can extract
$\sigma_{\mathrm{EGZ}}^{\mathrm{NMHV}}$ from \eqn{HqqggEGZ2},
knowing the $H\qb qgg$ MHV amplitudes inferred from our
formul\ae\, (\ref{qqgg_mppp_L})--(\ref{phi_qgqg_mpmp_fl}). Hence
we will consider $\sigma_{\mathrm{EGZ}}^{\mathrm{NMHV}}$ known
from now on, and we will use it in order to compute the color- and
helicity-summed cross section for $A\qb qgg$, $\sigma_A$.
Similarly to eqs.~(\ref{HqqggEGZ}) and (\ref{HqqggEGZ2}),
$\sigma_A$ is given by
\be
\sigma_A =
2 \, \mathrm{Re}\left\{ \sum_{\lambda}\left[ A_A^{\treenum
*}(\lambda)A_A^\oneloopnum(\lambda)\right] \right\}
= 4\,\mathrm{Re}\left\{ \sum_{\lambda^\prime}\left[ A_A^{\treenum
*}(\lambda^\prime)A_A^\oneloopnum(\lambda^\prime)\right] \right\},
\ee
with $A_A(\lambda)$ denoting the pseudoscalar $A\qb qgg$ amplitudes.

Because $\sigma_A$ is expressed as a sum over $\lambda^\prime$, we
focus on the four configurations $({-}{+}{+}{-})$,
$({-}{+}{-}{+})$ and $({-}{+}{-}{-})$, $({-}{+}{+}{+})$. We can
straightforwardly compute the terms coming from the first two
(MHV) configurations using our results from \sect{answers} and
\eqn{Areconstruct}, as mentioned in the beginning of this section.
For the last two (NMHV) configurations, we need to compute the
quantity
\be \sigma_A^{\mathrm{NMHV}} \equiv 4\mathrm{Re}\left\{
A_A^{\treenum *}(\lambda_-)A_A^\oneloopnum(\lambda_-) +
A_A^{\treenum *}(\lambda_+)A_A^\oneloopnum(\lambda_+) \right\}.
\label{Aqqgg_NMHV} \ee
Let's look at each amplitude in this expression separately.  The
tree amplitudes are simple, because
$A_A^\treenum = (A_\phi^\treenum - A_{\phi^\dagger}^\treenum)/i$, and
$A_{\phi^\dagger}^\treenum(\lambda_-) = A_\phi^\treenum(\lambda_+)= 0$.
Therefore, from \eqns{Hreconstruct}{Areconstruct}, we have
\begin{align}
&A_A^\treenum(\lambda_-) = \frac{1}{i} A_\phi^\treenum(\lambda_-) =
\frac{1}{i} A_H^\treenum(\lambda_-)\, ,\label{A0-}\\
&A_A^\treenum(\lambda_+) = - \frac{1}{i}
A_{\phi^\dagger}^\treenum(\lambda_+) = - \frac{1}{i}
A_H^\treenum(\lambda_+)\, .\label{A0+}
\end{align}
We choose to express the one-loop amplitudes
$A_A^\oneloopnum(\lambda_-)$ and $A_A^\oneloopnum(\lambda_+)$
using \eqns{Hreconstruct}{Areconstruct} in the following way
\begin{align}
&A_A^\oneloopnum(\lambda_-) = \frac{1}{i}\left[
A_\phi^\oneloopnum(\lambda_-) -
A_{\phi^\dagger}^\oneloopnum(\lambda_-)\right] = \frac{1}{i}\left[
A_H^\oneloopnum(\lambda_-) -
2A_{\phi^\dagger}^\oneloopnum(\lambda_-)\right],\label{A1-}\\
&A_A^\oneloopnum(\lambda_+) = \frac{1}{i}\left[
A_\phi^\oneloopnum(\lambda_+) -
A_{\phi^\dagger}^\oneloopnum(\lambda_+)\right] = \frac{1}{i}\left[ -
A_H^\oneloopnum(\lambda_+) +
2A_\phi^\oneloopnum(\lambda_+)\right].\label{A1+}
\end{align}

Substituting eqs.~(\ref{A0-})--(\ref{A1+}) into (\ref{Aqqgg_NMHV})
we find that
\bea
\sigma_A^{\mathrm{NMHV}} &=& 4 \, \mathrm{Re}\Bigg\{ \left(\frac{1}{i}
A_H^\treenum(\lambda_-)\right)^*\frac{1}{i}\left[
A_H^\oneloopnum(\lambda_-) -
2A_{\phi^\dagger}^\oneloopnum(\lambda_-)\right]
\nonumber \\
&&\hskip1cm\null
+ \left( - \frac{1}{i}
A_H^\treenum(\lambda_+)\right)^*\frac{1}{i}\left[ -
A_H^\oneloopnum(\lambda_+) + 2A_\phi^\oneloopnum(\lambda_+)\right]
\Bigg\} \nonumber \\
&=& 4 \, \mathrm{Re}\left\{ A_H^{\treenum
*}(\lambda_-)A_H^\oneloopnum(\lambda_-) + A_H^{\treenum
*}(\lambda_+)A_H^\oneloopnum(\lambda_+) \right\}
\nonumber \\
&&\hskip1cm\null
- 8 \, \mathrm{Re}\left\{ A_H^{\treenum*}(\lambda_-)
A_{\phi^\dagger}^\oneloopnum(\lambda_-)
+ A_H^{\treenum*}(\lambda_+)
A_\phi^\oneloopnum(\lambda_+)\right\} \nonumber \\
&=& \sigma_{\mathrm{EGZ}}^{\mathrm{NMHV}}
- 8 \, \mathrm{Re}\left\{
A_\phi^{\treenum*}(\lambda_-)
A_{\phi^\dagger}^\oneloopnum(\lambda_-)
+ A_{\phi^\dagger}^{\treenum*}(\lambda_+)
A_\phi^\oneloopnum(\lambda_+)\right\}.
\label{sigmaANMHV}
\eea
We notice that the first term in \eqn{sigmaANMHV} is given by
\eqn{Hqqgg_NMHV}, and the second term contains only tree and
finite one-loop helicity amplitudes with $\phi$ and
$\phi^\dagger$. The required NMHV tree amplitudes (see {\it e.g.}
refs.~\cite{Kauffman1996ix,DFM,Badger2004ty}) are given in
eqs.~(\ref{phiqqggmpmmtree}), (\ref{phidqqggmppptree}),
(\ref{phiqgqgmmpmtree}) and (\ref{phidqgqgmppptree}). The finite
one-loop amplitudes are given in
eqs.~(\ref{qqgg_mppp_L})--(\ref{qqgg_mppp_fl}) and
(\ref{phi_qgqg_mppp_L})--(\ref{phi_qgqg_mppp_fl}). Therefore, we
know all the ingredients necessary to obtain the full color- and
helicity-summed cross section for $A\qb qgg$, albeit only
semi-numerically. Our results of section \ref{answers} can be used
to convert $\sigma_{\mathrm{EGZ}}$ into $\sigma_A$.


\subsection{Interference with VBF production}

Our amplitudes for $H\qb q\Qb Q$ can be used to calculate
analytically part of the interference between the $qQ\rightarrow
HqQ$ gluon fusion process and the tree-level vector boson fusion
processes. Both these processes have the same initial and final
states. However, at tree level one has a color-octet exchange and
the other a color-singlet exchange.   So there is no interference
at tree level. (For identical quarks, the exchange term does
produce an interference, but it is extremely
small~\cite{Andersen2006ag}.) At one loop, however, the
color-singlet part of the one-loop correction to the gluon-fusion
$H\qb q\Qb Q$ amplitude can interfere with the tree-level VBF
amplitude, and we will provide an analytic formula for this
contribution.  This is only part of the virtual correction; the
other part comes from the interference between the tree-level
gluon fusion and one-loop VBF $H\qb q\Qb Q$ amplitudes. In
addition, there is a real correction. The sum of all three terms
has been computed numerically in
refs.~\cite{Andersen2007mp,Bredenstein2008tm} and it is quite
small.

We obtain the color-singlet part of $H\qb q\Qb Q$ from the
corresponding $\phi$-amplitude,
$A_{4;\mathrm{s}}(\phi,1_\qb,2_q,3_\Qb,4_Q)$, using
\eqn{Hreconstruct}.  From the color
decomposition~(\ref{qqQQloopdecomp}), contracted with
$\delta_{i_2}^{\,\,\,\bar{\imath}_1}$, we see that the
color-singlet part is given by
\be A_{4;\mathrm{s}}(\phi,1_\qb,2_q,3_\Qb,4_Q) =
A_{4;1}(\phi,1_\qb,2_q,3_\Qb,4_Q) +
A_{4;2}(\phi,1_\qb,2_q,3_\Qb,4_Q).
\ee
Using eqs.~(\ref{mppm1})--(\ref{mpmp2}) we get
\begin{align}
A_{4;\mathrm{s}}(\phi,1_\qb^-,2_q^+,3_\Qb^+,4_Q^-) &=
\frac{N_c^2-1}{N_c^2} \left[
A_4^{\mathrm{lc}}(\phi,1_\qb^-,2_q^+,3_\Qb^+,4_Q^-) +
A_4^{\mathrm{lc}}(\phi,1_\qb^-,2_q^+,4_\Qb^-,3_Q^+) \right],\\
A_{4;\mathrm{s}}(\phi,1_\qb^-,2_q^+,3_\Qb^-,4_Q^+) &=
\frac{N_c^2-1}{N_c^2} \left[
A_4^{\mathrm{lc}}(\phi,1_\qb^-,2_q^+,3_\Qb^-,4_Q^+) +
A_4^{\mathrm{lc}}(\phi,1_\qb^-,2_q^+,4_\Qb^+,3_Q^-) \right],
\end{align}
in terms of the leading-color primitive amplitude
$A_4^{\mathrm{lc}}(\phi,1_\qb,2_q,3_\Qb,4_Q)$. The relevant
tree-level $H\qb q\Qb Q$ VBF amplitudes involve only $ZZ$ fusion,
not $WW$ or $WZ$; they are given by
\begin{align} A_{4;\mathrm{VBF}}^{\treenum}(H,1_\qb^-,2_q^+,3_\Qb^+,4_Q^-) &=
2 i \frac{m_Z^2}{v}\frac{\spa1.4\spb2.3}{(s_{12}-m_Z^2)(s_{34}-m_Z^2)},\\
A_{4;\mathrm{VBF}}^{\treenum}(H,1_\qb^-,2_q^+,3_\Qb^-,4_Q^+) &=
A_{4;\mathrm{VBF}}^{\treenum}(H,1_\qb^-,2_q^+,4_\Qb^+,3_Q^-),
 \end{align}
with $v$ the vacuum expectation value of the Higgs field. Finally,
the color-singlet virtual interference between the two processes is
\be
2 \alphas^2 N_c^2 \, \mathrm{Re} \left[
A_{4;\mathrm{VBF}}^{\treenum*}(H,1_\qb,2_q,3_\Qb,4_Q)
A_{4;\mathrm{s}}(H,1_\qb,2_q,3_\Qb,4_Q) \right] \,.
\label{VBFinterf}
\ee
\Eqn{VBFinterf} is to be understood with an implicit summation
over all allowed polarization states of the external quarks.
We implemented~\eqn{VBFinterf} numerically at a few phase-space
points and obtained agreement~\cite{ASPrivate} with this part 
of the full interference computed in ref.~\cite{Andersen2007mp}.
We did not perform such a comparison against
ref.~\cite{Bredenstein2008tm}.


\subsection{Consistency checks}

It is important to verify that the methods and the results
presented in this paper yield the correct answers for the
$H\qb q\Qb Q$ and $H\qb qgg$ amplitudes. We have used
three types of independent and non-trivial checks on our
expressions. They are based on collinear limits that the
amplitudes should satisfy, symmetries under which they should
remain invariant, and numerical comparisons with previously
computed expressions. For $H\qb qgg$ only the first two types of
checks were possible, whereas for $H\qb q\Qb Q$ all of them were
performed, providing an even more solid check. We have found that
our amplitudes agree with all the checks, and we outline
the process further in the remainder of this section.

\subsubsection{Collinear behavior}

A powerful handle on the correctness of the amplitudes is their
collinear behavior. When two neighboring external legs become
collinear, an $n$-point amplitude has to correctly factorize onto
an $(n-1)$-point amplitude, multiplied by the corresponding splitting
amplitude for the two collinear legs. In the case of one-loop
amplitudes, the factorization is onto a sum of possible
factorizations with the loop belonging either to the splitting
amplitude, or to the remaining $(n-1)$-point
amplitude~\cite{UnitarityMethod}.
There is also a sum over the helicity $h$ of the intermediate state $P$
carrying momentum $k_P^2 \approx 0$.
For the $\phi\qb q\Qb Q$ amplitudes, in the limit that
momenta $k_1$ and $k_2$ become collinear the factorization is
onto a $\phi\Qb Qg$ amplitude,
\bea
A_4^{\oneloopnum}(\phi,1_\qb,2_q,3_\Qb,4_Q)
&\buildrel{1\parallel2}\over\longrightarrow& \sum_{h=\pm} \bigg[
A_3^{\oneloopnum}(\phi,3_\Qb,4_Q,P^h) \times
\mathrm{Split}^{\treenum}_{-h}(1_\qb,2_q;z) \nn\\
&&\hskip0.7cm\null + A_3^{\treenum}(\phi,3_\Qb,4_Q,P^h) \times
\mathrm{Split}^{\oneloopnum}_{-h}(1_\qb,2_q;z)\bigg] \,,
\label{factorization}
\eea
with $k_P=k_1+k_2$, $k_1 \approx z k_P$, and $k_2 \approx (1-z) k_P$.
The splitting amplitudes depend on the longitudinal momentum fraction
$z$, which is the momentum fraction carried by leg 1,
a real variable with $0<z<1$.  (It is unrelated to the complex variable
$z$ used for the shifts performed in the previous sections.)
Replacing $\phi$ with $H$ in \eqn{factorization}, we get the
collinear behavior of the Higgs amplitudes, while the $3\parallel
4$ collinear limit can be obtained by exchanging $q$ and $Q$. The
$1\parallel 2$ and $3\parallel 4$ limits are the only
collinear limits of $A_4^{\oneloopnum}(\phi,1_\qb,2_q,3_\Qb,4_Q)$ that
exhibit universal singular behavior; there is no
splitting amplitude for quarks of different flavor.

The extension of \eqn{factorization} to the $\phi\qb qgg$
amplitudes is straightforward. In this case, however, there are
additional factorization channels, including channels where a
gluon becomes collinear with an adjacent quark or gluon. For
example, for the $2\parallel 3$ and $3\parallel 4$ collinear
limits we have
\bea A_4^\oneloopnum(\phi,1_\qb,2_q,3,4)
&\buildrel{2\parallel3}\over\longrightarrow& \sum_{h=\pm} \bigg[
A_3^{\oneloopnum}(\phi,1_\qb,P_q^h,4) \times
\mathrm{Split}^{\treenum}_{-h}(2_q,3;z) \nn\\ &&\hskip0.7cm\null +
A_3^{\treenum}(\phi,1_\qb,P_q^h,4) \times
\mathrm{Split}^{\oneloopnum}_{-h}(2_q,3;z)\bigg],
\label{qqgg_factorization_23}\\
A_4^\oneloopnum(\phi,1_\qb,2_q,3,4)
&\buildrel{3\parallel4}\over\longrightarrow& \sum_{h=\pm} \bigg[
A_3^{\oneloopnum}(\phi,1_\qb,2_q,P^h) \times
\mathrm{Split}^{\treenum}_{-h}(3,4;z) \nn\\ &&\hskip0.7cm\null +
A_3^{\treenum}(\phi,1_\qb,2_q,P^h) \times
\mathrm{Split}^{\oneloopnum}_{-h}(3,4;z)\bigg],
\label{qqgg_factorization_34} \eea
and similarly for the $H\qb gqg$ amplitudes.
Eqs.~(\ref{factorization})--(\ref{qqgg_factorization_34}) apply
separately to the primitive amplitude components lc, slc, $L$, $R$
and $\fl$, after extracting the respective pieces of the one-loop
three-parton amplitudes~\cite{Berger2006sh} and splitting
amplitudes~\cite{UnitarityMethod}.

We have checked that our expressions factorize correctly according
to eqs.~(\ref{factorization})--(\ref{qqgg_factorization_34}) and
their analogues, for all possible non-trivial (singular)
collinear limits.

\subsubsection{Symmetries}

The one-loop amplitudes we computed in this paper are MHV
four-point amplitudes, with an equal number of
positive- and negative-helicity external legs.  As a consequence,
they have to satisfy certain non-trivial symmetries, that become
manifest only after assembling together all the pieces into a full
answer.

The symmetries of the four-quark $H\qb q\Qb Q$ primitive amplitudes
can be summarized in the following form:
\begin{itemize}
\item reflection or quark-exchange symmetry
\begin{align}
A_4^\oneloopnum(H,1_\qb^-,2_q^+,3_\Qb^+,4_Q^-) &=
A_4^\oneloopnum(H,4_\qb^-,3_q^+,2_\Qb^+,1_Q^-), \label{reflection1}\\
A_4^\oneloopnum(H,1_\qb^-,2_q^+,3_\Qb^-,4_Q^+) &=
A_4^\oneloopnum(H,3_\qb^-,4_q^+,1_\Qb^-,2_Q^+), \label{reflection2}
\end{align}
\item parity conjugation symmetry
\begin{align}
A_4^\oneloopnum(H,1_\qb^-,2_q^+,3_\Qb^+,4_Q^-) &=
A_4^\oneloopnum(H,2_\qb^-,1_q^+,4_\Qb^+,3_Q^-)\Big|_{\spa{i}.{j} \lr \spb{j}.{i}}
\,,
\label{conjugation1}\\
A_4^\oneloopnum(H,1_\qb^-,2_q^+,3_\Qb^-,4_Q^+) &=
A_4^\oneloopnum(H,2_\qb^-,1_q^+,4_\Qb^-,3_Q^+)\Big|_{\spa{i}.{j} \lr \spb{j}.{i}}
\,. \label{conjugation2}
\end{align}
\end{itemize}
The reflection symmetry properties
(\ref{reflection1})--(\ref{reflection2}) are
satisfied by the component $\phi$ and $\phi^\dagger$ amplitudes as
well, whereas the conjugation symmetry
(\ref{conjugation1})--(\ref{conjugation2}) only holds for the $H$
amplitudes.

For the two-quark-two-gluon $H\qb qgg$ and $H\qb gqg$ primitive
MHV amplitudes, although there is no reflection symmetry, the following
parity conjugation symmetries hold:
\begin{itemize}
\item $H\qb qgg$ conjugation symmetry
\begin{align}
A_4^\oneloopnum(H,1_\qb^-,2_q^+,3^+,4^-) =
-A_4^\oneloopnum(H,2_\qb^-,1_q^+,4^+,3^-)\Big|_{\spa{i}.{j} \lr \spb{j}.{i}}
\,, \label{qqgg_conjugation1}\\
A_4^\oneloopnum(H,1_\qb^-,2_q^+,3^-,4^+)
= -A_4^\oneloopnum(H,2_\qb^-,1_q^+,4^-,3^+)\Big|_{\spa{i}.{j} \lr \spb{j}.{i}}
\,,
\label{qqgg_conjugation2}
\end{align}
\item $H\qb gqg$ conjugation symmetry
\begin{align}
A_4^\oneloopnum(H,1_\qb^-,2^+,3_q^+,4^-) =
-A_4^\oneloopnum(H,3_\qb^-,4^+,1_q^+,2^-)\Big|_{\spa{i}.{j} \lr \spb{j}.{i}}
\,, \label{qgqg_conjugation1}\\
A_4^\oneloopnum(H,1_\qb^-,2^-,3_q^+,4^+)
= -A_4^\oneloopnum(H,3_\qb^-,4^-,1_q^+,2^+)\Big|_{\spa{i}.{j} \lr \spb{j}.{i}}
\,.
\label{qgqg_conjugation2}
\end{align}
\end{itemize}

The fact that the symmetries
(\ref{reflection1})--(\ref{qgqg_conjugation2}) have to be
respected provides a non-trivial check on the amplitudes. Since
our computation is done by obtaining separate, in general
non-symmetric, pieces of the amplitude ({\it e.g.}, the
coefficient of a single log in a particular channel), it is only
after putting them all together that the symmetry becomes
manifest. Therefore, it is the combination of many non-symmetric
ingredients that gives rise to a symmetric final answer.  An
error that spoils the symmetry can be detected by this check.
We have checked that our amplitudes obey all the required symmetries.


\subsubsection{Numerical comparison}

Ref.~\cite{Ellis2005qe} computed the virtual cross section for the
$H\rightarrow q\qb Q\Qb$ process to next-to-leading order accuracy
(one-loop diagrams, not counting the top quark loop vertex) using
a semi-numerical approach. They also obtained analytic expressions
for the aforementioned cross section summed over colors and
helicities.

Using our color- and helicity-decomposed $\phi$-amplitudes
presented in our paper, we have constructed the same quantity and
have compared with their analytical results numerically.
The result at the phase-space point~(\ref{PhaseSpacePoint}) was given
in table~\ref{HqqQQNumericalTable}, but we have also found
agreement with their analytical formulae for all the
randomly-generated phase-space points that we examined.


\section{Conclusions}
\label{ConclusionSection}

In this paper we have presented analytic results for the one-loop
amplitudes for a Higgs plus four quarks, and for a Higgs plus two
quarks and two opposite-helicity gluons.  We have obtained the
cut-containing and rational pieces of the answer separately, using
unitarity for the former and on-shell recursion for the latter. We
have also shown in specific examples how to compute the various
ingredients, and presented a way to deal with spurious poles
without fully eliminating them from the completed-cut terms. Our
expressions are relatively compact and in agreement with various
consistency checks as well as previous results. We believe that
they will provide a useful input for faster numerical programs
computing NLO cross sections relevant for the LHC, and will be an
important ingredient for future higher-point calculations.
Together with the NMHV $H\qb qgg$ case, and the remaining helicity
amplitudes for $Hgggg$ (beyond those already been
computed~\cite{BadgerGlover,Badger2007si,Glover2008ffa}) they
provide the one-loop corrections to Higgs plus four partons, and
can be used to compute the gluon fusion contribution to the
$H+2$~jets final state at the LHC, as well as its
interference with the vector boson fusion channel.  Further NLO
studies will be important for understanding SM backgrounds, and
enhancing the potential for the discovery of new physics in the
upcoming experiments at the LHC.

\acknowledgments The figures in this paper were made with {\sc
Jaxodraw}~\cite{Binosi2003yf}, based on {\sc
Axodraw}~\cite{Vermaseren1994je}. We are grateful to Jeppe
Andersen, Jeffrey Forshaw, Nigel Glover and Jennifer Smillie for
useful discussions, and to John Campbell and Keith Ellis for pointing
out typographical errors in earlier versions of this article.
Y.S. would like to thank Marvin Weinstein for assistance with \Maple.


\appendix

\section{Tree amplitudes}

In this appendix, we record various tree amplitudes
entering the main computations and results.

As mentioned in \sect{SampleRationalSubsubsection},
the $\phi\qb q$ and $\phi^\dagger \qb q$ amplitudes for massless quarks
vanish by fermion chirality conservation and angular momentum
conservation.   The $\phi gg$ and $\phi^\dagger gg$ tree amplitudes
are given by
\bea
A_2^\treenum(\phi,1^+,2^+) &=&  A_2^\treenum(\phi,1^\pm,2^\mp) = 0 \,,
\nn\\
A_2^\treenum(\phi,1^-,2^-) &=& - i \, {\spa1.2}^2 \,,
\label{phiggtrees}\\
A_2^\treenum(\phi^\dagger,1^-,2^-)
&=& A_2^\treenum(\phi^\dagger,1^\pm,2^\mp) = 0 \,,  \nn\\
A_2^\treenum(\phi^\dagger,1^+,2^+) &=& - i \, {\spb1.2}^2 \,. \nn
\eea
The $\phi\qb qg$ and $\phi^\dagger\qb qg$ amplitudes are
\bea
A_3^\treenum(\phi,1_\qb^-,2_q^+,3^+) &=& 0 \,, \nn\\
A_3^\treenum(\phi,1_\qb^-,2_q^+,3^-) &=&
- i \, {{\spa1.3}^2\over\spa1.2} \,,
\label{phiqqgtrees}\\
A_3^\treenum(\phi^\dagger,1_\qb^-,2_q^+,3^-) &=& 0 \,, \nn\\
A_3^\treenum(\phi^\dagger,1_\qb^-,2_q^+,3^+) &=&
- i \, {{\spb2.3}^2\over\spb1.2} \,, \nn
\eea
while the $\phi ggg$ amplitudes are
\bea
A_3^\treenum(\phi,1^+,2^+,3^+) &=& A_3^\treenum(\phi,1^-,2^+,3^+) = 0 \,,
\nn\\
A_3^\treenum(\phi,1^-,2^-,3^+) &=&
i \, {{\spa1.2}^3\over\spa2.3\spa3.1} \,, \nn\\
A_3^\treenum(\phi,1^-,2^-,3^-) &=&
- i \, {(m_H^2)^2\over\spb1.2\spb2.3\spb3.1} \,,
\label{phigggtrees}\\
A_3^\treenum(\phi^\dagger,1^-,2^-,3^-)
&=& A_3^\treenum(\phi,1^+,2^-,3^-) = 0 \,, \nn\\
A_3^\treenum(\phi^\dagger,1^+,2^+,3^-) &=&
- i \, {{\spb1.2}^3\over\spb2.3\spb3.1} \,, \nn\\
A_3^\treenum(\phi^\dagger,1^+,2^+,3^+) &=&
i \, {(m_H^2)^2\over\spa1.2\spa2.3\spa3.1} \,. \nn
\eea

The $\phi\qb q\Qb Q$ and $\phi^\dagger\qb q\Qb Q$ tree amplitudes
are given by
\bea
A_4^\treenum(\phi,1_\qb^-,2_q^+,3_\Qb^+,4_Q^-)
&=& -i\,\frac{\spa1.4^2}{\spa1.2\spa3.4} \,, \nn\\
A_4^\treenum(\phi,1_\qb^-,2_q^+,3_\Qb^-,4_Q^+)
&=& -\, A_4^\treenum(\phi,1_\qb^-,2_q^+,4_Q^+,3_\Qb^-) \,, \label{qqQQtrees}\\
A_4^\treenum(\phi^\dagger,1_\qb^-,2_q^+,3_\Qb^+,4_Q^-)
&=& -i\,\frac{\spb2.3^2}{\spb1.2\spb3.4}\, , \nn\\
A_4^\treenum(\phi^\dagger,1_\qb^-,2_q^+,3_\Qb^-,4_Q^+)
&=& -A_4^\treenum(\phi^\dagger,1_\qb^-,2_q^+,4_\Qb^+,3_Q^-) \,. \nn
\eea
For the case of $\phi\qb qgg$ and $\phi^\dagger\qb qgg$ we
have
\bea
A_4^\treenum(\phi,1_\qb^-,2_q^+,3^+,4^+) &=& 0\,, \nn\\
A_4^\treenum(\phi,1_\qb^-,2_q^+,3^+,4^-)
&=& -i\frac{\spa1.4^2\spa2.4}{\spa1.2\spa2.3\spa3.4}\,, \nn\\
A_4^\treenum(\phi,1_\qb^-,2_q^+,3^-,4^+)
&=& i\frac{\spa1.3^3}{\spa1.2\spa3.4\spa4.1} \,,\label{qqggtrees}\\
A_4^\treenum(\phi^\dagger,1_\qb^-,2_q^+,3^-,4^-) &=& 0\,, \nn\\
A_4^\treenum(\phi^\dagger,1_\qb^-,2_q^+,3^+,4^-)
&=& -i\frac{\spb2.3^2\spb1.3}{\spb1.2\spb3.4\spb4.1} \,, \nn\\
A_4^\treenum(\phi^\dagger,1_\qb^-,2_q^+,3^-,4^+)
&=& i\frac{\spb2.4^3}{\spb1.2\spb2.3\spb3.4} \,, \nn
\eea
and for the NMHV cases,
\bea
A_4^\treenum(\phi,1_\qb^-,2_q^+,3^-,4^-)
&=&
- i { {\spab3.{(1+4)}.2}^2 \spa4.1 \over \spb2.4 \, s_{412} }
    \biggl[ {1\over s_{12}} + {1 \over s_{41}} \biggr]
\nonumber\\
&& \hskip0.0cm \null
- i { {\spab4.{(1+3)}.2}^2 \spa1.3 \over \spb2.3 \, s_{12} \, s_{123} }
+ i { {\spab1.{(3+4)}.2}^2 \over \spa1.2\spb2.4\spb2.3\spb3.4 }
\,, \label{phiqqggmpmmtree}\\
A_4^\treenum(\phi^\dagger,1_\qb^-,2_q^+,3^+,4^+)
&=&
- i { {\spab1.{(2+3)}.4}^2 \spb2.3 \over \spa1.3 \, s_{123} }
    \biggl[ {1\over s_{12}} + {1 \over s_{23}} \biggr]
\nonumber\\
&& \hskip0.0cm \null
+ i { {\spab1.{(2+4)}.3}^2 \spb2.4 \over \spa1.4 \, s_{12} \, s_{412} }
- i { {\spab1.{(3+4)}.2}^2 \over \spb1.2\spa1.3\spa1.4\spa3.4 }
\,. \label{phidqqggmppptree}
\eea

The amplitudes for $\phi\qb gqg$ and $\phi^\dagger\qb gqg$ are,
\bea
A_4^\treenum(\phi,1_\qb^-,2^+,3_q^+,4^-)
&=& -i\frac{\spa1.4^2}{\spa1.2\spa2.3} \,, \nn\\
A_4^\treenum(\phi,1_\qb^-,2^-,3_q^+,4^+)
&=& -i\frac{\spa1.2^2}{\spa3.4\spa4.1} \,, \\
A_4^\treenum(\phi^\dagger,1_\qb^-,2^+,3_q^+,4^-)
&=& i\frac{\spb2.3^2}{\spb3.4\spb4.1} \,, \nn\\
A_4^\treenum(\phi^\dagger,1_\qb^-,2^-,3_q^+,4^+)
&=& i\frac{\spb3.4^2}{\spb1.2\spb2.3} \,, \nn
\eea
and for the NMHV cases,
\bea
A_4^\treenum(\phi,1_\qb^-,2^-,3_q^+,4^-)
&=&
- i \, { {\spab4.{(1+2)}.3}^2 \over \spb1.2\spb2.3 \, s_{123} }
- i \, { {\spab2.{(1+4)}.3}^2 \over \spb3.4\spb4.1 \, s_{341} } \,,
\label{phiqgqgmmpmtree}\\
A_4^\treenum(\phi^\dagger,1_\qb^-,2^+,3_q^+,4^+)
&=&
  i \, { {\spab1.{(2+3)}.4}^2 \over \spa1.2\spa2.3 \, s_{123} }
+ i \, { {\spab1.{(3+4)}.2}^2 \over \spa3.4\spa4.1 \, s_{341} } \,.
\label{phidqgqgmppptree}
\eea
Our $\phi\qb qgg$ and $\phi\qb gqg$ tree amplitudes, after
dividing by $i$, agree with the ones (implicit) in
refs.~\cite{Kauffman1996ix,Kauffman1998yg,Badger2004ty}. For the
$\phi^\dagger\qb qgg$ and $\phi^\dagger\qb gqg$ tree amplitudes,
after dividing by $i$, our amplitudes agree with the ones in
refs.~\cite{Kauffman1996ix,Kauffman1998yg}, but have the opposite
sign from those in ref.~\cite{Badger2004ty}.  (The relative sign
between $\phi$ and $\phi^\dagger$ amplitudes matters in
reconstructing the $H$ amplitudes.) Our $\phi\qb qg\ldots g$ and
$\phi^\dagger\qb qg\ldots g$ tree amplitudes are uniformly
opposite in sign to ref.~\cite{DFM}.


\section{$\Ll_i$ and $\Ls_{-1}$ functions and scalar integrals}

The definitions for the $\Ll_0$, $\Ll_1$,
$\Ll_2$ and $\Ls_{-1}$,
$\Ls_{-1}^{2{\rm m}e}$ functions used in the formul\ae\
presented above are
\be
\Ll_0(r)\ =\
\frac{\ln r}{1-r} \,, \qquad
\Ll_1(r)\ =\
\frac{\ln r + 1 - r}{(1-r)^2}\,, \qquad
\Ll_2(r)\ =\
\frac{\ln r - (r - 1/r)/2}{(1 - r)^3} \,,
\ee
\bea
&&\Ls_{-1}\left(s,t;m^2\right) =
\Li_2\left( 1 - \frac{s}{m^2}\right)
+ \Li_2\left( 1 - \frac{t}{m^2} \right)
+ \ln\left(\frac{-s}{-m^2}\right)\ln\left(\frac{-t}{-m^2}\right)
- \frac{\pi^2}{6} \,,~~~~~
\\ &&
  \Ls_{-1}^{2{\rm m}e}\left(s,t;m_1^2,m_3^2\right) = \null
- \Li_2\left( 1 - \frac{m_1^2}{s}\right)
- \Li_2\left( 1 - \frac{m_1^2}{t}\right)
- \Li_2\left( 1 - \frac{m_3^2}{s}\right)
\nn\\ && \hskip3.7cm \null
- \Li_2\left( 1 - \frac{m_3^2}{t}\right)
+ \Li_2\left( 1 - \frac{m_1^2 m_3^2}{st}\right)
- \frac{1}{2}\ln^2\left(\frac{-s}{-t}\right),
\eea
with
\be
\Li_2 (x) = - \int_{0}^{x} dz \, \frac{\ln(1-z)}{z} \,.
\ee
The one-mass and easy two-mass box integrals are related to the $\Ls_{-1}$
functions by,
\bea
{\cal I}_4^{1\rm{m}}(s,t;m^2) &=& {-2i \, \cg \over st} \biggl\{
-{1\over\e^2} \biggl[
\biggl({\mu^2\over-s}\biggr)^\e + \biggl({\mu^2\over-t}\biggr)^\e
- \biggl({\mu^2\over-m^2}\biggr)^\e \biggr]
\nn\\ && \hskip1.3cm \null
- \Ls_{-1}(s,t;m^2) \biggr\} \,,
\label{I1mexpr}\\
{\cal I}_4^{2\rm{m}e}(s,t;m_1^2,m_3^2)
&=& {-2i \, \cg \over st - m_1^2m_3^2} \biggl\{
-{1\over\e^2} \biggl[
\biggl({\mu^2\over-s}\biggr)^\e + \biggl({\mu^2\over-t}\biggr)^\e
- \biggl({\mu^2\over-m_1^2}\biggr)^\e
- \biggl({\mu^2\over-m_3^2}\biggr)^\e \biggr]
\nn\\ && \hskip2cm \null
- \Ls_{-1}^{2\rm{m}e}(s,t;m_1^2,m_3^2) \biggr\} \,.
\label{I2meexpr}
\eea

The one-mass and two-mass triangle integrals are related to
each other,
\bea
{\cal I}_3^{\mathrm{1m}}(s) &=&
{-i \cg \over \e^2} {1\over(-s)}  \biggl({\mu^2\over-s}\biggr)^\e \,,
\label{I31mdef}\\
{\cal I}_3^{\mathrm{2m}}(s_1,s_2) &=&
{-i \cg \over \e^2}  {1\over(-s_1)-(-s_2)}
\biggl[ \biggl({\mu^2\over-s_1}\biggr)^\e
      - \biggl({\mu^2\over-s_2}\biggr)^\e \biggr] \,,
\label{I32mdef}
\eea
and contain terms of the form $\ln(-s_i)/\e$; hence
their coefficients are dictated by the known infrared poles
of the amplitude.

The bubble integral is given by,
\be
{\cal I}_2(s) =
{i \cg \over \e(1-2\e)} \biggl({\mu^2\over-s}\biggr)^\e \,,
\label{I2def}
\ee
and contains a single logarithm, $\ln(-s)$, at order $\e^0$.



\begin{thebibliography}{99}

\bibitem{Higgs1966ev}
P.~W.~Higgs,
Phys.\ Rev.\  {\bf 145}, 1156 (1966).

\bibitem{Englert1964et}
F.~Englert and R.~Brout,
Phys.\ Rev.\ Lett.\  {\bf 13}, 321 (1964).

\bibitem{Guralnik1964eu}
G.~S.~Guralnik, C.~R.~Hagen and T.~W.~B.~Kibble,
Phys.\ Rev.\ Lett.\  {\bf 13}, 585 (1964).

\bibitem{Georgi1977gs}
H.~M.~Georgi, S.~L.~Glashow, M.~E.~Machacek and D.~V.~Nanopoulos,
Phys.\ Rev.\ Lett.\  {\bf 40}, 692 (1978).

\bibitem{Djouadi1991tka}
A.~Djouadi, M.~Spira and P.~M.~Zerwas,
Phys.\ Lett.\  B {\bf 264}, 440 (1991).

\bibitem{Dawson1990zj}
S.~Dawson,
Nucl.\ Phys.\  B {\bf 359}, 283 (1991).

\bibitem{Graudenz1992pv}
D.~Graudenz, M.~Spira and P.~M.~Zerwas,
Phys.\ Rev.\ Lett.\  {\bf 70}, 1372 (1993).

\bibitem{Spira1995rr}
M.~Spira, A.~Djouadi, D.~Graudenz and P.~M.~Zerwas,
Nucl.\ Phys.\  B {\bf 453}, 17 (1995)
[hep-ph/9504378].

\bibitem{Figy2003nv}
T.~Figy, C.~Oleari and D.~Zeppenfeld,
Phys.\ Rev.\  D {\bf 68}, 073005 (2003)
[hep-ph/0306109].

\bibitem{Figy2004pt}
T.~Figy and D.~Zeppenfeld,
Phys.\ Lett.\  B {\bf 591}, 297 (2004)
[hep-ph/0403297].

\bibitem{Berger2004pca}
E.~L.~Berger and J.~Campbell,
Phys.\ Rev.\  D {\bf 70}, 073011 (2004)
[hep-ph/0403194].

\bibitem{Wilczek1977zn}
F.~Wilczek,
Phys.\ Rev.\ Lett.\  {\bf 39}, 1304 (1977).

\bibitem{Shifman1978zn}
M.~A.~Shifman, A.~I.~Vainshtein and V.~I.~Zakharov,
Phys.\ Lett.\  B {\bf 78}, 443 (1978).

\bibitem{KLS98}
M.~Kr\"amer, E.~Laenen and M.~Spira,
Nucl.\ Phys.\  B {\bf 511}, 523 (1998)
[hep-ph/9611272].

\bibitem{Hautmann2002tu}
F.~Hautmann,
Phys.\ Lett.\  B {\bf 535}, 159 (2002)
[hep-ph/0203140].

\bibitem{DelDuca2003ba}
V.~Del Duca, W.~Kilgore, C.~Oleari, C.~R.~Schmidt and D.~Zeppenfeld,
Phys.\ Rev.\  D {\bf 67}, 073003 (2003)
[hep-ph/0301013].

\bibitem{Dawson1991au}
S.~Dawson and R.~P.~Kauffman,
Phys.\ Rev.\ Lett.\  {\bf 68}, 2273 (1992).

\bibitem{Kauffman1996ix}
R.~P.~Kauffman, S.~V.~Desai and D.~Risal,
Phys.\ Rev.\  D {\bf 55}, 4005 (1997)
[Erratum-ibid.\  D {\bf 58}, 119901 (1998)]
[hep-ph/9610541].

\bibitem{Kauffman1998yg}
R.~P.~Kauffman and S.~V.~Desai,
Phys.\ Rev.\  D {\bf 59}, 057504 (1999)
[hep-ph/9808286].

\bibitem{DDKOSZ}
V.~Del Duca, W.~Kilgore, C.~Oleari, C.~Schmidt and D.~Zeppenfeld,
Phys.\ Rev.\ Lett.\  {\bf 87}, 122001 (2001)
[hep-ph/0105129];
Nucl.\ Phys.\  B {\bf 616}, 367 (2001)
[hep-ph/0108030].

\bibitem{Ellis2005qe}
R.~K.~Ellis, W.~T.~Giele and G.~Zanderighi,
Phys.\ Rev.\  D {\bf 72}, 054018 (2005)
[Erratum-ibid.\  D {\bf 74}, 079902 (2006)]
[hep-ph/0506196].

\bibitem{Campbell2006xx}
J.~M.~Campbell, R.~K.~Ellis and G.~Zanderighi,
JHEP {\bf 0610}, 028 (2006)
[hep-ph/0608194].

\bibitem{DelDuca2004wt}
V.~Del Duca, A.~Frizzo and F.~Maltoni,
JHEP {\bf 0405}, 064 (2004)
[hep-ph/0404013].

\bibitem{Andersen2007mp}
J.~R.~Andersen, T.~Binoth, G.~Heinrich and J.~M.~Smillie,
JHEP {\bf 0802}, 057 (2008)
[0709.3513 [hep-ph]].

\bibitem{Bredenstein2008tm}
A.~Bredenstein, K.~Hagiwara and B.~J\"ager,
Phys.\ Rev.\  D {\bf 77}, 073004 (2008)
[0801.4231 [hep-ph]].

\bibitem{Anastasiou2004vj}
C.~Anastasiou and A.~Lazopoulos,
JHEP {\bf 0407}, 046 (2004)
[hep-ph/0404258].

\bibitem{Anastasiou2005cb}
C.~Anastasiou and A.~Daleo,
JHEP {\bf 0610}, 031 (2006)
[hep-ph/0511176].

\bibitem{Anastasiou2007qb}
C.~Anastasiou, S.~Beerli and A.~Daleo,
JHEP {\bf 0705}, 071 (2007)
[hep-ph/0703282].

\bibitem{Ellis2005zh}
R.~K.~Ellis, W.~T.~Giele and G.~Zanderighi,
Phys.\ Rev.\  D {\bf 73}, 014027 (2006)
[hep-ph/0508308].

\bibitem{Denner2005nn}
A.~Denner and S.~Dittmaier,
Nucl.\ Phys.\  B {\bf 734}, 62 (2006)
[hep-ph/0509141].

\bibitem{Ossola2006us}
G.~Ossola, C.~G.~Papadopoulos and R.~Pittau,
Nucl.\ Phys.\  B {\bf 763}, 147 (2007)
[hep-ph/0609007].

\bibitem{Ossola2007ax}
G.~Ossola, C.~G.~Papadopoulos and R.~Pittau,
JHEP {\bf 0803}, 042 (2008)
[0711.3596 [hep-ph]].

\bibitem{Ossola2008xq}
G.~Ossola, C.~G.~Papadopoulos and R.~Pittau,
JHEP {\bf 0805}, 004 (2008)
[0802.1876 [hep-ph]].

\bibitem{Mastrolia2008jb}
P.~Mastrolia, G.~Ossola, C.~G.~Papadopoulos and R.~Pittau,
JHEP {\bf 0806}, 030 (2008)
[0803.3964 [hep-ph]].

\bibitem{Draggiotis2009yb}
P.~Draggiotis, M.~V.~Garzelli, C.~G.~Papadopoulos and R.~Pittau,
JHEP {\bf 0904}, 072 (2009)
[0903.0356 [hep-ph]].

\bibitem{vanHameren2009dr}
A.~van Hameren, C.~G.~Papadopoulos and R.~Pittau,
0903.4665 [hep-ph].

\bibitem{Bern2007dw}
Z.~Bern, L.~J.~Dixon and D.~A.~Kosower,
Annals Phys.\  {\bf 322}, 1587 (2007)
[0704.2798 [hep-ph]].

\bibitem{Anastasiou2006jv}
C.~Anastasiou, R.~Britto, B.~Feng, Z.~Kunszt and P.~Mastrolia,
Phys.\ Lett.\  B {\bf 645}, 213 (2007)
[hep-ph/0609191].

\bibitem{Anastasiou2006gt}
C.~Anastasiou, R.~Britto, B.~Feng, Z.~Kunszt and P.~Mastrolia,
JHEP {\bf 0703}, 111 (2007)
[hep-ph/0612277].

\bibitem{Ellis2007br}
R.~K.~Ellis, W.~T.~Giele and Z.~Kunszt,
JHEP {\bf 0803}, 003 (2008)
[0708.2398 [hep-ph]].

\bibitem{Giele2008ve}
W.~T.~Giele, Z.~Kunszt and K.~Melnikov,
JHEP {\bf 0804}, 049 (2008)
[0801.2237 [hep-ph]].

\bibitem{Giele2008bc}
W.~T.~Giele and G.~Zanderighi,
JHEP {\bf 0806}, 038 (2008)
[0805.2152 [hep-ph]].

\bibitem{Ellis2008ir}
R.~K.~Ellis, W.~T.~Giele, Z.~Kunszt and K.~Melnikov,
0806.3467 [hep-ph].

\bibitem{Berger2008sj}
C.~F.~Berger {\it et al.},
Phys.\ Rev.\  D {\bf 78}, 036003 (2008)
[0803.4180 [hep-ph]].

\bibitem{Bredenstein2006ha}
A.~Bredenstein, A.~Denner, S.~Dittmaier and M.~M.~Weber,
JHEP {\bf 0702}, 080 (2007)
[hep-ph/0611234].

\bibitem{Ciccolini2007jr}
M.~Ciccolini, A.~Denner and S.~Dittmaier,
Phys.\ Rev.\ Lett.\  {\bf 99}, 161803 (2007)
[0707.0381 [hep-ph]].

\bibitem{Ciccolini2007ec}
M.~Ciccolini, A.~Denner and S.~Dittmaier,
Phys.\ Rev.\  D {\bf 77}, 013002 (2008)
[0710.4749 [hep-ph]].

\bibitem{Schmidt1997wr}
C.~R.~Schmidt,
Phys.\ Lett.\  B {\bf 413}, 391 (1997)
[hep-ph/9707448].

\bibitem{BadgerGlover}
S.~D.~Badger and E.~W.~N.~Glover,
Nucl.\ Phys.\ Proc.\ Suppl.\  {\bf 160}, 71 (2006)
[hep-ph/0607139].

\bibitem{Berger2006sh}
C.~F.~Berger, V.~Del Duca and L.~J.~Dixon,
Phys.\ Rev.\  D {\bf 74}, 094021 (2006)
[Erratum-ibid.\  D {\bf 76}, 099901 (2007)]
[hep-ph/0608180].

\bibitem{Badger2007si}
S.~D.~Badger, E.~W.~N.~Glover and K.~Risager,
JHEP {\bf 0707}, 066 (2007)
[0704.3914 [hep-ph]];
Acta Phys.\ Polon.\  B {\bf 38}, 2273 (2007)
[0705.0264 [hep-ph]].

\bibitem{Glover2008ffa}
E.~W.~N.~Glover, P.~Mastrolia and C.~Williams,
JHEP {\bf 0808}, 017 (2008)
[0804.4149 [hep-ph]].

\bibitem{Britto2004nc}
R.~Britto, F.~Cachazo and B.~Feng,
Nucl.\ Phys.\  B {\bf 725}, 275 (2005)
[hep-th/0412103].

\bibitem{Britto2005ha}
R.~Britto, E.~Buchbinder, F.~Cachazo and B.~Feng,
Phys.\ Rev.\  D {\bf 72}, 065012 (2005)
[hep-ph/0503132].

\bibitem{Britto2006sj}
R.~Britto, B.~Feng and P.~Mastrolia,
Phys.\ Rev.\  D {\bf 73}, 105004 (2006)
[hep-ph/0602178].

\bibitem{Bern2005cq}
Z.~Bern, L.~J.~Dixon and D.~A.~Kosower,
Phys.\ Rev.\  D {\bf 73}, 065013 (2006)
[hep-ph/0507005].

\bibitem{Berger2006ci}
C.~F.~Berger, Z.~Bern, L.~J.~Dixon, D.~Forde and D.~A.~Kosower,
Phys.\ Rev.\  D {\bf 74}, 036009 (2006)
[hep-ph/0604195].

\bibitem{Berger2006vq}
C.~F.~Berger, Z.~Bern, L.~J.~Dixon, D.~Forde and D.~A.~Kosower,
Phys.\ Rev.\  D {\bf 75}, 016006 (2007)
[hep-ph/0607014].

\bibitem{Bern2008ef}
Z.~Bern {\it et al.}  [NLO Multileg Working Group],
0803.0494 [hep-ph].

\bibitem{Andersen2006ag}
J.~R.~Andersen and J.~M.~Smillie,
Phys.\ Rev.\  D {\bf 75}, 037301 (2007)
[hep-ph/0611281].

\bibitem{ASPrivate}
J.~R.~Andersen and J.~M.~Smillie, private communication.

\bibitem{Forshaw2009fz}
J.~Forshaw, J.~Keates and S.~Marzani,
JHEP {\bf 0907}, 023 (2009)
[0905.1350 [hep-ph]].

\bibitem{Chetyrkin1997un}
K.~G.~Chetyrkin, B.~A.~Kniehl and M.~Steinhauser,
Nucl.\ Phys.\  B {\bf 510}, 61 (1998)
[hep-ph/9708255].

\bibitem{DFM}
V.~Del Duca, A.~Frizzo and F.~Maltoni,
JHEP {\bf 0405}, 064 (2004)
[hep-ph/0404013].

\bibitem{Dixon2004za}
L.~J.~Dixon, E.~W.~N.~Glover and V.~V.~Khoze,
JHEP {\bf 0412}, 015 (2004)
[hep-th/0411092].

\bibitem{Cachazo2004kj}
F.~Cachazo, P.~Svr\v{c}ek and E.~Witten,
JHEP {\bf 0409}, 006 (2004)
[hep-th/0403047].

\bibitem{Badger2004ty}
S.~D.~Badger, E.~W.~N.~Glover and V.~V.~Khoze,
JHEP {\bf 0503}, 023 (2005)
[hep-th/0412275].

\bibitem{TreeColor}
J.~E.~Paton and H.~M.~Chan,
Nucl.\ Phys.\ B {\bf 10}, 516 (1969);\\
%
P.~Cvitanovi\'c, P.~G.~Lauwers and P.~N.~Scharbach,
Nucl.\ Phys.\ B {\bf 186}, 165 (1981);\\
F.~A.~Berends and W.~Giele,
Nucl.\ Phys.\  B {\bf 294}, 700 (1987);\\
M.~L.~Mangano, S.~J.~Parke and Z.~Xu,
Nucl.\ Phys.\  B {\bf 298}, 653 (1988);\\
D.~Zeppenfeld,
Int.\ J.\ Mod.\ Phys.\  A {\bf 3}, 2175 (1988).

\bibitem{Mangano1990by}
M.~L.~Mangano and S.~J.~Parke,
Phys.\ Rept.\  {\bf 200}, 301 (1991)
[hep-th/0509223].

\bibitem{Dixon1996wi}
L.~J.~Dixon,
hep-ph/9601359.

\bibitem{Bern1996ka}
Z.~Bern, L.~J.~Dixon, D.~A.~Kosower and S.~Weinzierl,
Nucl.\ Phys.\  B {\bf 489} (1997) 3
[hep-ph/9610370].

\bibitem{Bern1997sc}
Z.~Bern, L.~J.~Dixon and D.~A.~Kosower,
Nucl.\ Phys.\  B {\bf 513}, 3 (1998)
[hep-ph/9708239].

\bibitem{UnitarityMethod}
Z.~Bern, L.~J.~Dixon, D.~C.~Dunbar and D.~A.~Kosower,
Nucl.\ Phys.\ B {\bf 425}, 217 (1994)
[hep-ph/9403226].

\bibitem{UnitarityMethod2}
Z.~Bern, L.~J.~Dixon, D.~C.~Dunbar and D.~A.~Kosower,
Nucl.\ Phys.\  B {\bf 435}, 59 (1995)
[hep-ph/9409265].

\bibitem{Britto2005fq}
R.~Britto, F.~Cachazo, B.~Feng and E.~Witten,
Phys.\ Rev.\ Lett.\  {\bf 94}, 181602 (2005)
[hep-th/0501052].

\bibitem{Bern2005hs}
Z.~Bern, L.~J.~Dixon and D.~A.~Kosower,
Phys.\ Rev.\  D {\bf 71}, 105013 (2005)
[hep-th/0501240].

\bibitem{Bern2005ji}
Z.~Bern, L.~J.~Dixon and D.~A.~Kosower,
Phys.\ Rev.\  D {\bf 72}, 125003 (2005)
[hep-ph/0505055].

\bibitem{IntegralReductions}
L.~M.~Brown and R.~P.~Feynman,
Phys.\ Rev.\  {\bf 85}, 231 (1952);\\
%
L.~M.~Brown, 
Nuovo Cim.\ {\bf 22}, 178 (1961);\\
%
B. Petersson,  J. Math. Phys.\ {\bf 6}, 1955 (1965);\\
%
G. K\"all\'en and J.~S.\ Toll, J. Math.\ Phys.\ {\bf 6}, 299 (1965);\\
%
D.~B.~Melrose,
Nuovo Cim.\  {\bf 40}, 181 (1965);\\
%
G.~Passarino and M.~J.~G.~Veltman,
Nucl.\ Phys.\  B {\bf 160}, 151 (1979);\\
%
W.~L.~van Neerven and J.~A.~M.~Vermaseren,
Phys.\ Lett.\  B {\bf 137}, 241 (1984).

\bibitem{BDKIntegrals}
Z.~Bern, L.~J.~Dixon and D.~A.~Kosower,
Phys.\ Lett.\  B {\bf 302}, 299 (1993)
[Erratum-ibid.\  B {\bf 318}, 649 (1993)]
[hep-ph/9212308].

\bibitem{BDKPentagon}
Z.~Bern, L.~J.~Dixon and D.~A.~Kosower,
Nucl.\ Phys.\  B {\bf 412}, 751 (1994)
[hep-ph/9306240].

\bibitem{BGH}
T.~Binoth, J.~P.~Guillet and G.~Heinrich,
Nucl.\ Phys.\  B {\bf 572}, 361 (2000)
[hep-ph/9911342].

\bibitem{DuplancicNizic}
G.~Duplan\v{c}i\'c and B.~Ni\v{z}i\'c,
Eur.\ Phys.\ J.\  C {\bf 35}, 105 (2004)
[hep-ph/0303184].

\bibitem{EZ}
R.~K.~Ellis and G.~Zanderighi,
JHEP {\bf 0802}, 002 (2008)
[0712.1851 [hep-ph]].

\bibitem{Britto2004ap}
R.~Britto, F.~Cachazo and B.~Feng,
Nucl.\ Phys.\  B {\bf 715}, 499 (2005)
[hep-th/0412308].

\bibitem{Berger2006cz}
C.~F.~Berger, Z.~Bern, L.~J.~Dixon, D.~Forde and D.~A.~Kosower,
Nucl.\ Phys.\ Proc.\ Suppl.\  {\bf 160}, 261 (2006)
[hep-ph/0610089].

\bibitem{KST}
Z.~Kunszt, A.~Signer and Z.~Tr\'ocs\'anyi,
Nucl.\ Phys.\  B {\bf 411}, 397 (1994)
[hep-ph/9305239].

\bibitem{HV}
G.~'t Hooft and M.~Veltman,
Nucl.\ Phys.\ B {\bf 44}, 189 (1972).

\bibitem{BKStringBased}
Z.~Bern and D.~A.~Kosower,
Nucl.\ Phys.\ B {\bf 379}, 451 (1992).

\bibitem{OtherFDH}
S.~Catani, M.~H.~Seymour and Z.~Tr\'ocs\'anyi,
Phys.\ Rev.\ D {\bf 55}, 6819 (1997)
[hep-ph/9610553];\\
%
Z.~Bern, A.~De Freitas, L.~J.~Dixon and H.~L.~Wong,
Phys.\ Rev.\ D {\bf 66}, 085002 (2002)
[hep-ph/0202271].

\bibitem{UniversalIR}
W.~T.~Giele and E.~W.~N.~Glover,
Phys.\ Rev.\ D {\bf 46}, 1980 (1992);\\
Z.~Kunszt, A.~Signer and Z.~Tr\'ocs\'anyi,
Nucl.\ Phys.\ B {\bf 420}, 550 (1994)
[hep-ph/9401294];\\
S.~Catani,
Phys.\ Lett.\ B {\bf 427}, 161 (1998)
[hep-ph/9802439].

\bibitem{Binosi2003yf}
D.~Binosi and L.~Theussl,
Comput.\ Phys.\ Commun.\  {\bf 161}, 76 (2004)
[hep-ph/0309015].

\bibitem{Vermaseren1994je}
J.~A.~M.~Vermaseren,
Comput.\ Phys.\ Commun.\  {\bf 83}, 45 (1994).

\end{thebibliography}
\end{document}